\documentclass[fleqn,usenatbib,useAMS]{mnras}

\usepackage{newtxtext,newtxmath}
\usepackage{mathrsfs}
\usepackage[utf8]{inputenc}
\usepackage{graphicx}                           
\usepackage{bm}
\usepackage{multirow}
\usepackage{multicol}        

\usepackage{amsmath}	
\usepackage{pdflscape}	
\usepackage[T1]{fontenc}
\usepackage{ae,aecompl}
\usepackage{ulem}
\usepackage[dvipsnames]{xcolor}
\usepackage{color}

\DeclareRobustCommand{\VAN}[3]{#2}
\let\VANthebibliography\thebibliography
\def\thebibliography{\DeclareRobustCommand{\VAN}[3]{##3}\VANthebibliography}

\title[]{AI-assisted reconstruction of cosmic velocity field from redshift-space spatial distribution of halos}

\author[Ziyong Wu et al.]{
Ziyong Wu$^{1,2}$,
Liang Xiao$^{4,5}$\thanks{xiaoliang5@mail2.sysu.edu.cn},
Xu Xiao$^{4}$, 
Jie Wang$^{7,8}$, 
Xi Kang$^{3,1}$,
Yang Wang$^{6}$,
Xin Wang$^{4,5}$,
\newauthor{
Le Zhang$^{4,5,6}$\thanks{zhangle7@mail.sysu.edu.cn},
Xiao-Dong Li$^{4,5}$\thanks{lixiaod25@mail.sysu.edu.cn}
}
\\
$^{1}$Purple Mountain Observatory, Chinese Academy of Sciences, 10 Yuanhua Road, Nanjing 210033, China\\
$^{2}$School of Astronomy and Space Sciences, University of Science and Technology of China, Hefei 230026, China,\\
$^{3}$Institute for Astronomy, The school of Physics, Zhejiang University, Hangzhou 310037, China\\
$^{4}$School of Physics and Astronomy, Sun Yat-Sen University, Guangzhou 510297, P. R. China\\
$^{5}$CSST Science Center for the Guangdong–Hong Kong–Macau Greater Bay Area, SYSU, Zhuhai 519082, P. R. China\\
$^{6}$Peng Cheng Laboratory, No. 2, Xingke 1st Street, Shenzhen 518000, P. R. China\\
$^{7}$National Astronomical Observatories, Chinese Academy of Sciences, 20A Datun Road, Beijing 100101, China\\
$^{8}$University of Chinese Academy of Sciences, Beĳing 100049, China
}

\date{Accepted XXX. Received YYY; in original form ZZZ}

\pubyear{2022}

\begin{document}
\label{firstpage}
\pagerange{\pageref{firstpage}--\pageref{lastpage}}
\maketitle

\begin{abstract}

The peculiar velocities of dark matter halos are crucial to study many issues in cosmology and galaxy evolution. In this study, by using the state-of-the-art deep learning technique, a UNet-based neural network, we propose to reconstruct the peculiar velocity field from the redshift-space distribution of dark matter halos. Through a point-to-point comparison and examination of various statistical properties, we demonstrate that, the reconstructed velocity field is in generally good agreement with the ground truth. The power spectra of various velocity field components, including velocity magnitude, divergence and vorticity, can be successfully recovered when $k\lesssim 1.1$ $h/\rm Mpc$ (the Nyquist frequency of the simulations) at about 80\% accuracy. This approach is very promising and presents an alternative method to correct the redshift-space distortions using the measured 3D spatial information of halos. Additionally, for the reconstruction of the momentum field of halos, UNet achieves similar good results. Hence the applications in various aspects of cosmology are very broad, such as correcting redshift errors and improving measurements in the structure of the cosmic web, the kinetic Sunyaev-Zel'dovich effect, BAO reconstruction, etc.

\end{abstract}


\begingroup
\let\clearpage\relax
\tableofcontents
\endgroup
\newpage

\section{Introduction}
The Large Scale Structure (LSS) of the Universe is crucial to our study of the expansion and structure formation history of the Universe.
In the next decade, the stage IV surveys such as  DESI\footnote{https://desi.lbl.gov/},
EUCLID\footnote{http://sci.esa.int/euclid/}, LSST\footnote{http://sci.esa.int/euclid/},
WFIRST\footnote{https://wfirst.gsfc.nasa.gov/},CSST,Roman and Subaru will map the Universe with extraordinary precision on an unprecedented large volume,  deepening the understanding of dark energy, dark matter, gravity, the Hubble constant, the neutrino mass, and the initial condition of the Universe.

Due to the initial inhomogeneity, the peculiar velocity field of the universe is generated together with the density field during the process of structure formation, and thus contains enormous information about LSS.  Accurate observations or reconstruction of the cosmic velocity field will greatly help us to quantify and understand the redshift spatial distortions~\citep{jackson1972critique, kaiser1987clustering}, baryon acoustic oscillations~\citep{Eisenstein:2005su,Eisenstein2007BAOReconstruction}, the Alcock-Paczynski effect~\citep{ap,Li2014,Li2015,Li2016,KR2018}, the cosmic web~\citep{1986Bardeen,hahn2007properties,forero2009dynamical,hoffman2012kinematic,forero2014cosmic,Fang2019}, the kinematic Sunyaev-Zeldovich effect ~\citep{1972CoASP...4..173S,1980MNRAS.190..413S}, the integrated Sachs Wolfe effect~\citep{1967ApJ...147...73S, 1968Natur.217..511R, 1996PhRvL..76..575C}, and to constrain cosmology parameters~\citep{2015MNRAS.450..317C,2021MNRAS.507.1557L}

Observationally, however, measuring the peculiar velocity of galaxies is extremely difficult, mainly
as it requires redshift-independent distance estimates that can only be made by distance indicators such as type Ia Supernovae~\citep{1993ApJ...413L.105P,SNIaflow...1997ApJ...488L...1R, SNIaflow...2004MNRAS.355.1378R,SNIaflow...2012MNRAS.420..447T,SNIaflow...2016ApJ...827...60M},
the Tully-Fisher relation \citep{TullyFisher...1977A&A....54..661T, TullyFisher...2006ApJ...653..861M,TullyFisher...2008AJ....135.1738M}
the "fundamental plane" relation \citep{FundPlan...1987ApJ...313...42D,FundPlan...1987ApJ...313...59D,FundPlan...2007ApJS..172..599S}.
Therefore, it is essential to extract cosmological information by directly inferring the velocity field from observable measurements of large-scale structure, such as the distribution of dark matter halos.
Here, the difficulty lies in the complexity arising from the nonlinear evolution of the structure and the gravitational collapse, and many studies have been made in this direction ~\citep{VelocityRecon...1991ApJ...379....6N,
  VelocityRecon...1992ApJ...390L..61B,
  1994ApJ...421L...1N,
  VelocityRecon...1995ApJ...449..446Z,
  VelocityRecon...1997MNRAS.285..793C,
  VelocityRecon...1999MNRAS.309..543B,
  VelocityRecon...2000MNRAS.316..464K,
  VelocityRecon...2002MNRAS.335...53B,
  VelocityRecon...2005ApJ...635L.113M,
  VelocityRecon...2008MNRAS.383.1292L,
  VelocityRecon...2008MNRAS.391.1796B,
  VelocityRecon...2012MNRAS.425.2422K,
  VelocityRecon...2012MNRAS.420.1809W,
VelocityRecon...2015MNRAS.449.3407J,2017MNRAS.467.3993A,2019A&A...625A..64J}, etc.

Recent tremendous advances in machine learning algorithms, especially those based on deep neural networks, provide us with a great opportunity to extract useful information from complex data.
In more recent years, deep learning-based techniques have been applied to almost all areas of cosmology and astrophysics~\citep{Mehta:2018dln,Jennings:2018eko,Carleo:2019ptp,Ntampaka:2019udw}, such as weak gravitational lensing~\citep{Schmelzle:2017vwd,
Gupta:2018eev,Springer:2018aak,Fluri:2019qtp,Jeffrey:2019fag,Merten:2018bgr,Peel:2018aei,Tewes:2018she}, the Cosmic Microwave Background \citep{Caldeira:2018ojb,Rodriguez:2018mjb,Perraudin:2018rbt,Munchmeyer:2019kng,Mishra:2019sep},
the LSS including estimating cosmological parameters from the distribution of matter~\citep{Ravanbakhsh:2017bbi,Lucie-Smith:2018smo,Li2020...ML...2020SCPMA..63k0412P,2021JCAP...09..039L}, identifying dark matter halos 
and reconstruct the initial conditions of the universe using machine learning \citep{Modi:2018cfi,Berger:2018aey,Lucie-Smith:2019hdl,Ramanah:2019cbm}, mapping rough cosmology to fine one~\citep{He:2018ggn,Li_2021}, extracting line intensity maps~\citep{Pfeffer:2019pca}, foreground removal in 21cm intensity
mapping~\citep{Makinen:2020gvh}, augmenting N-body simulations with gas~\citep{Troster:2019mys}, 
a mapping between the 3D galaxy distribution in hydrodynamic simulations and its underlying dark matter distribution~\citep{Zhang:2019ryt}, modelling small-scale galaxy formation physics in large cosmological volumes~\citep{2021MNRAS.507.1021N}, reconstructing the baryon acoustic oscillations~\citep{AIBAORecon...2020arXiv200210218M} and reconstructing the initial linear-regime matter density field~\citep{2022arXiv220712511S}, searching for gravitational waves~\citep{Dreissigacker:2019edy,Gebhard:2019ldz} and cosmic reionization~\citep{LaPlante:2018pst,Gillet:2018fgb,Hassan:2018bbm,Chardin:2019euc,Hassan:2019cal}, as well as supernovae ~\citep{Lochner:2016hbn,Moss:2018tug,Ishida:2018uqu,Li:2019ybe,Muthukrishna:2019wpf}.

For velocity reconstruction, the pioneering work~\citep{Wu:2021jsy} shows that a UNet network can reconstruct the nonlinear velocity field of dark matter particles with high precision down to a scale of 2 $h^{-1}\rm Mpc$. When pushing down to highly non-linear scales of $k\lesssim 1.4 ~h^{-1}\rm Mpc$, they could achieve 90\% accuracy in reconstructing the power spectra of the velocity and momentum fields of the magnitude, the divergence and the vorticity components.
This demonstrates that, compared with the traditional perturbation-based theory, deep learning methods would be more effective and have a great advantage in reconstructing the cosmic velocity field at nonlinear scales.



Moreover, the scientific community widely agrees that dark matter halos and subhaloes serve as a reliable indicator of the distribution of galaxies. Consequently, employing these haloes as tracers in the study is closely linked to the actual observations. Additionally, taking into account the halo bias combined with the RSD effects amplifies the similarity between the simulated data in this study and the actual observations. Several pioneering studies have explored this area, such as a measurement of kinetic Sunyaev-Zel’dovich effect using individual cluster velocities, which are reconstructed from the distribution of galaxies surrounding them~\citep{2022A&A...662A..48T}. Meanwhile,~\cite{2021ApJ...913...76H,2022arXiv221206439G} demonstrate a density field or  peculiar velocity field reconstructed from galaxy distributions. However, reconstructing the velocity field from observed halos is technically more difficult because the halos are located only at density peaks, making them much sparser than simulation particles.

Therefore, in this study, we propose a modified UNet model dedicated to the reconstruction of the velocity field of {\it dark matter halos (and subhalos)}. From our simulation tests, this proposed method can reconstruct the peculiar velocities of each individual halos on high accuracy. It turns out that both the velocity and real-space density fields down to the non-linear scales can be well inferred from a redshift-space measurement alone. Therefore, this study is a major step toward applying the deep leaning technique to real observational data, which is extremely important for cosmology.



The layout of this paper is as follows. In Sect.~\ref{sect:method}, we introduce the simulation data used in this study and detail the architecture choice of the neural network and the training procedure, as well as the validation tests in velocity reconstruction. Results for our network are presented in Sect.~\ref{sect:result}, and finally the conclusion and discussion are present in Sect.~\ref{sect:conclusion}. 

For a comparison to the velocity reconstruction we discuss in this paper, we present reconstruction results for the  corresponding momentum field (the number-weighted velocity) of halos in Appendix~\ref{app:m}, obatined with the same UNet model.

\section{Method}
\label{sect:method}

\subsection{Dataset}
To train and validate our deep learning framework, the  training and tests data sets are based on the dark matter halos/subhalos of the BigMultiDark (BigMD) Planck
simulation\footnote{http://www.cosmosim.org}, which is the high-resolution N-body simulation described in~\cite{2016MNRAS.457.4340K} and was performed with \texttt{GADGET-2}~\citep{2005MNRAS.364.1105S}. The simulation was created in a box of $2.5 h^{-1}$ Gpc on each side, with $3840^3$ dark matter particles and the mass resolution of $M_{\rm DM}=2.4 \times 10^{10} h^{-1} \rm M_{\odot}$. The initial conditions are generated with Zeldovich approximation at $z_{\rm init}=100$. The simulation provides 79 redshift snapshots in the range of 0-8.8. For the analysis, we use the ROCKSTAR (Robust Overdensity Calculation using K-Space Topologically Adaptive Refinement) halo finder~\citep{2013ApJ...762..109B} to identify spherical dark matter halos/subhalos in the simulation, based on adaptive hierarchical refinement of friends-of-friends groups in six phase-space dimensions and one time dimension. ROCKSTAR provides halo mass using spherical overdensities of a virial structure.

The cosmology we assume in this study is the standard flat $\Lambda$CDM, compatible with Planck 2013 results~\citep{Planck:2018vyg}, with the fiducial parameters of \{$\Omega_m,\Omega_b, h, n_s, \sigma_8 \} = \{0.307, 0.048, 0.678, 0.96, 0.823\}$.

We construct the redshift-space halo/subhalo catalogue at $z=0$ with the number density of $10^{-3} ({\rm Mpc}/h)^3$ fixed, to be compatible with current spectral observations. The redshift-space position ${\bf s}$ is related to the real-space position ${\bf r}$ for a distant observer along the line of sight by 
\begin{equation}\label{eq:rsd}
    \bm{s} = \bm{r}+ \frac{ \bm{v}\cdot \hat{z} }{aH(a)}\,,
\end{equation}
where $\bm{v}$ is the peculiar velocity, $a$ is  scale factor and $H$ is the Hubble parameter, and the unit vector $\hat{z}$ denotes the line-of-sight direction. For simplicity, we have defined the $z$ direction of the simulation box as the line of sight direction. Based on the catalogue samples, we compute the density field and velocity field in the mesh cells by assigning the particle mass to a $900^3$ mesh using the CIC (Cloud-in-Cell) scheme, with a cell resolution of $(2.78 h^{-1} \rm Mpc)^3$.

\subsection{Input Preprocessing}
\label{sect:ip}
Our framework consists of the prepossessing of the input dataset and there are some points need to be clarified, as detailed below. 

\begin{enumerate}
\item[1)] 
In our UNet model, the input is a 6-channel 3D number density map of halos (and subhalos) in redshift space. For each channel, the map contains only halos in a certain mass range. To do so, we sort the halo (and subhalo) sample by mass in descending order and split it into 6 mass intervals, with the bin edges: $\log_{10}(M/M_\odot)\in[ 15.09,13.52,13.12,12.84,12.62,12.43,12.30]$, corresponding to binning the halo mass in percentiles of [5, 15, 30, 50, 75, 100]. The reason for doing so is that, 1) the correlation between velocity and halo density may depend on the mass of dark matter halos. Thus, in practice, a useful approach for neural network learning is to classify dark matter halos based on their mass; 2) in observations we can have approximately estimated mass of halos.

\item[2)] 
Based on the limitations in size of GPU memory, training time and model size, we have to divide the large box of side length 2500 ${\rm Mpc}/h$ into 8000 smaller boxes, each with side 125 ${\rm Mpc}/h$ ($45^3$ meshgrid points in CIC).
500 of them are used for training, 300 for validation, and a total of 3375 for testing. Both the input and output for the neural network model are for subboxes, each consisting of $45^3$ voxels. During the testing phase, instead of randomly selecting small boxes for velocity field prediction, we utilize the halo density fields of $5^3$ adjacent small boxes to predict the velocity fields of the corresponding 125 small boxes. These 125 small boxes are then created into a single large box, and we compute statistics based on the velocities at all lattice points within this large box. We repeat this process 27 times to create 27 such large boxes, which are not adjacent to each other to reduce correlation. Each of these large boxes is referred to as a test sample. We then calculate the mean and associated errors of the predictions from these test samples. Upon testing, we observe that the final results obtained from these 27 test samples had converged. 


\item[3)]
Given that the box division procedure and physically small boxes lead to loss of large-scale velocity modes, we thus use the linear perturbation theory to compensate for such loss in the training data. The velocity field $\bm{v}$ in the linear regime are directly related to the density field $\delta$ through 
\begin{equation}
\bm{v}(\bm{k}) =a f H \frac{i \bm{k}}{k^2} \delta(\bm{k})\,,
\label{eq:linear}
\end{equation}
where both fields are expressed in Fourier space, and $f=d \ln D / d \ln a$ denotes the growth rate with $D$ the growth factor.
The whole BigMD simulation box is used to calculate the linear velocity field, which is then partitioned into smaller boxes measuring $375^3~({\rm Mpc}/h)^3$ each, resulting in a grid size of 13.89 Mpc. The larger volume of the linear velocity field accounts for the possibility that motion in a larger volume may affect the motion of matter in the $141.67^3~({\rm Mpc}/h)^3$ box. The linear velocity grid is then convolved and concatenated with the feature maps of the density field (after convolution and pooling their feature maps have the same number of grid cells, so the two fields can be concatenated). Although this consideration is a bit complicated, we have observed that it indeed improves the accuracy of the power spectra of the velocity, velocity divergence and velocity curl, and have therefore adopted it in our study. The grid size and physical box size in each data set are illustrated in Tab.~\ref{tab:physicsize}.


\begin{table}
 \centering
 \caption{Grid size and physical box size in each data set.}
  \begin{tabular}{c|cc} 
  \hline
  field  & grid size & box size $({\rm Mpc}/h)^3$  \\
  \hline
   density  & $51^3$ & $141.67^3$   \\
  \hline
        velocity  & $45^3$ & $125^3$   \\
        \hline
        linear velocity  & $27^3$ & $375^3$   \\
        \hline
 \end{tabular}\label{tab:physicsize}
\end{table}

\item[4)]
To ensure that the training results of the model achieve good rotational invariance, we use a data augmentation method, in which the input data to the model for training are randomly rotated by one of 8 rotational transformations. There are 8 rotational transformations in total, as the $z$-direction (due to RSD effects in line of sight) behaves differently from the other two axes, which involve changes to the x- and y-coordinates and flipping the sign of the x- and y-coordinates. Our training set contains only a small fraction of the simulated boxes (500 subboxes), which can reduce the potential correlation between the training and test sets. However, the correlation between the samples within the training set can be considered as random noise. With a sufficiently large training set, this noise would have a negligible effect on the results during the training process. The data augmentation technique expands the effective size of the training dataset by a factor of 8, resulting in a total of $500\times 8=4000$ subboxes, further mitigating the noise and enhancing the model's performance.

\item[5)]
Instead of learning the 3D velocity vector field directly, we decompose the velocity vector into two parts: magnitude and direction. The predicted velocity field is then reconstructed using these two parts.


\item[6)] 
Since the dynamic range of the velocity field is very wide, to improve the accuracy and the convergence speed, the velocity magnitude in the output is normalized, where the normalization factor $c$ is chosen by  $c=1/200$ for $v\geq 60~{\rm km/s}$ and $c=1/12$ otherwise. Therefore, the output of the velocity magnitude contains two parts, the large velocity one and the small one, labelled as $v_{\rm large}$ and $v_{\rm smaller}$, respectively. Here the normalization factors and the strategy of velocity splitting are both empirical. This approach originates from our observation that small velocities tend to vanish during training due to their minimal weighting in sparse fields. With the loss function, we fine-tuned these two normalization coefficients to obtain the current values. Although there may be better normalization methods, our results indicate that this simple approach is effective. As scuh, the training set contains velocities that are split into two channels, with one channel having a non-zero value while the other must be zero. Consequently, by summing these two channels, we obtain the final velocity magnitude. In reality, one of the two output predictions of the velocity channel will not be exactly zero. However, as the training improves, the output value of the channel that should be zero will become very close to zero. Therefore, in the output, the velocity of each channel is multiplied by the corresponding normalization factor, and then the two resulting fields are simply added up to obtain the final velocity field.

\item[7)]
In addition to the velocity field, we have also trained the model to account for the momentum field as an output and the results are summarized in Appendix~\ref{app:m}.  
\end{enumerate}




\subsection{Neural Network Model}
\label{sec:maths} 

\begin{figure*}
	\includegraphics[width=500pt]{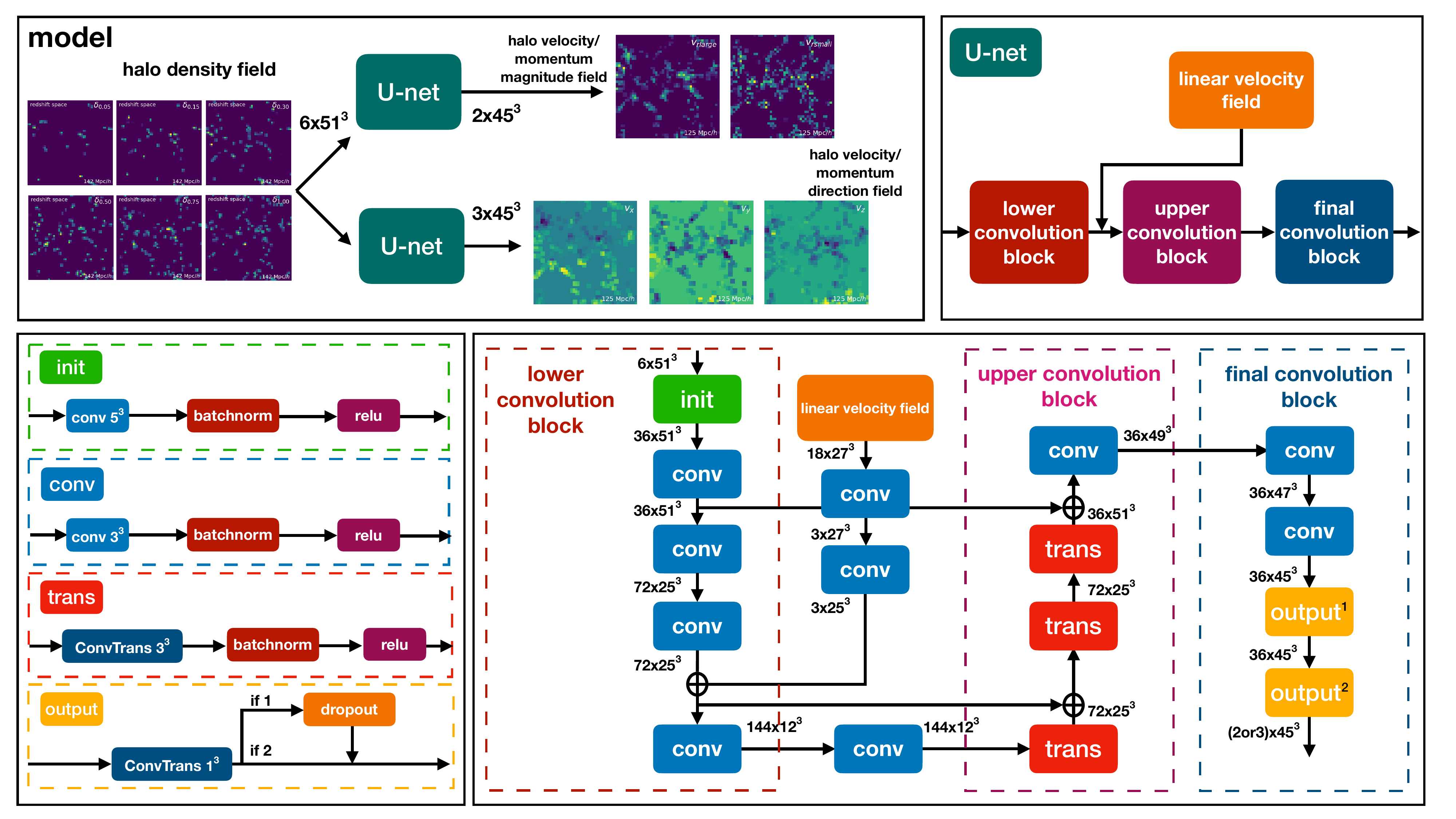}
    \caption{UNet neural network architecture and training scheme used for the velocity (momentum) reconstruction.
     Starting  with a 6-channel $51^3$-voxel input layer  that corresponds to the number density field (a side length of 142 ${\rm Mpc}/h$) of halos for the six different mass intervals (over the mass range of $10^{12.3}$--$10^{15.1} M_\odot$) in redshift space, our model is consisted of two U-net neural network architecture for reconstruction of velocity magnitude and direction, where one contains two channels corresponding to the large and small velocity fields ($v_{\rm large}, v_{\rm small}$), and the other consists of three channels corresponding to the three velocity directions ($v_x, v_y, v_z$) (upper left). This U-net architecture essentially consists of the upper, lower and final convolution blocks, together with a compensation for the linear velocity field (upper right). The dimension of each output field is $45^3$, corresponding to a box volume of $125^3$ $(h^{-1}{\rm Mpc})^3$. The lower-right part shows the   details of the components given in the three-block structure of the UNet. The layers of "init", "conv", "trans" and "output" are detailed on the lower-left part. In the final convolution block, a dropout layer between the convolution transform and batchnorm layers is used to enhance the UNet performance and prevent overfitting, where the dropuout value is chosen as 0.3.}
    \label{fig:model}
\end{figure*}

\begin{figure*}
    \includegraphics[width=0.4\textwidth]{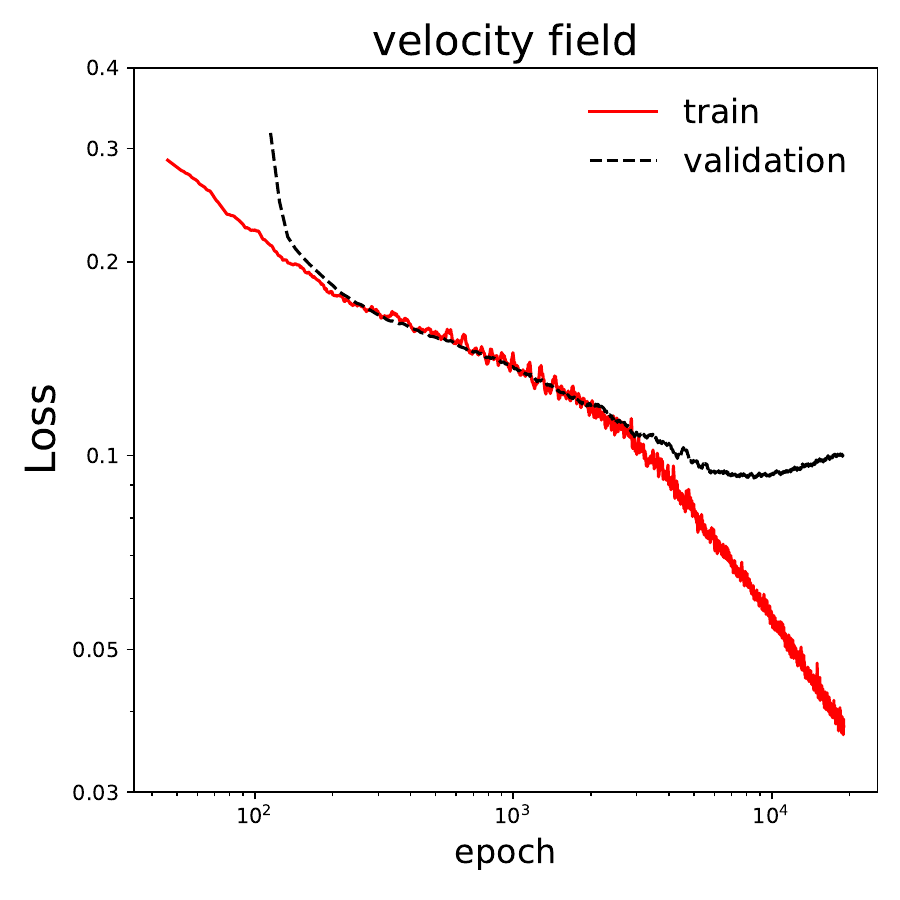}
    \includegraphics[width=0.4\textwidth]{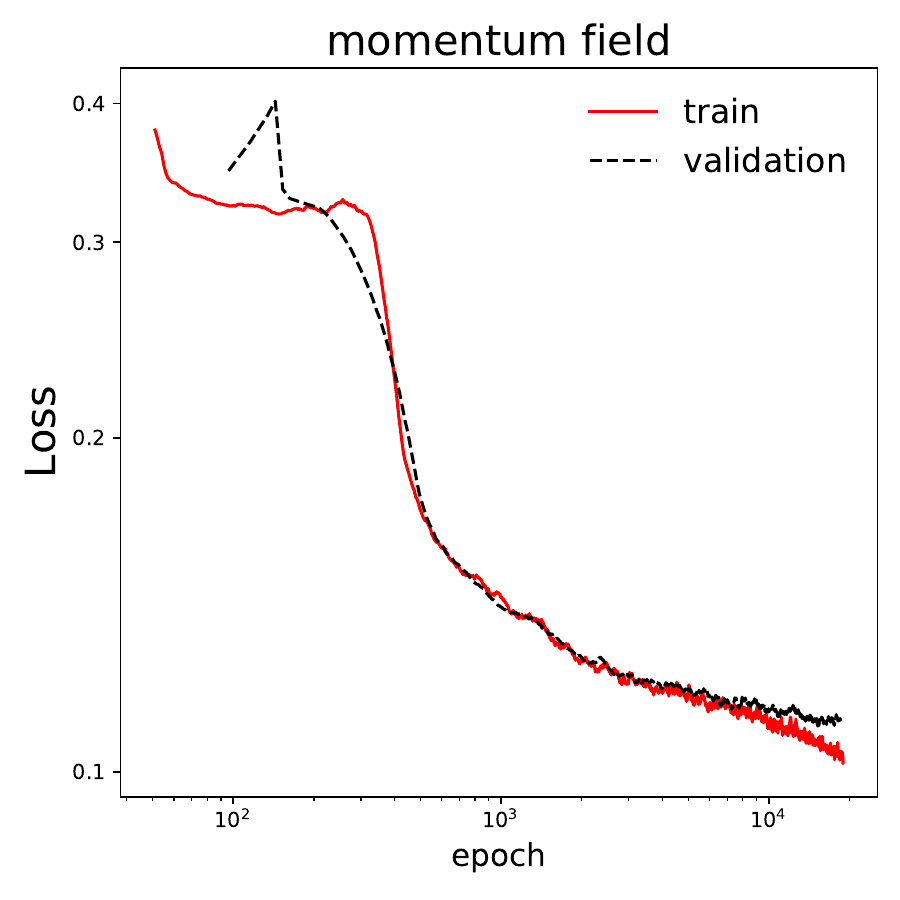}\\
    \caption{Loss of the training (black) and validation (red) sets. The UNnet model achieves convergence after 20,000 epochs of training for both velocity field (left) and momentum field (right) predictions.}
    \label{fig:loss}
\end{figure*}

\begin{figure*}
    \includegraphics[width=500pt]{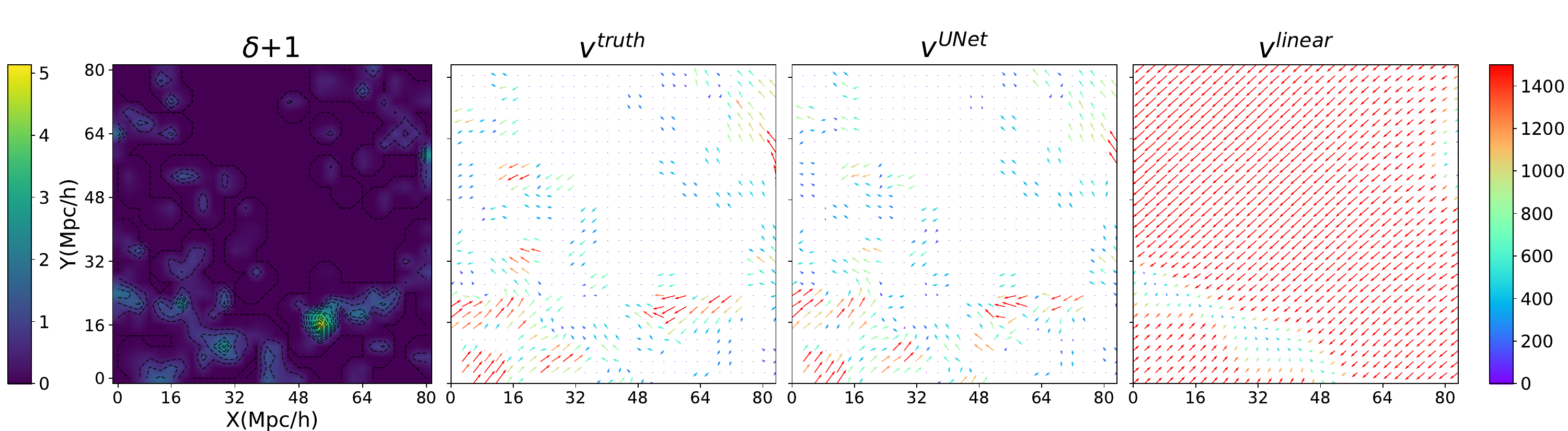}
    \caption{A point-wise comparison was performed between the true velocity field, the UNet-reconstructed velocity field, and the velocity field reconstructed by the linear theory in the test sets. The slice had a volume of $83\times 83\times 28$ $h^{-3}{\rm Mpc}^3$. From left to right, the fields of halo number density, the true velocity, the UNet-reconstructed velocity, and the linear velocity were displayed. The length and orientation of the colored arrows in the velocity fields represented the velocity magnitude and direction. As demonstrated, at the scale we plotted, non-linear effects dominated in this area, making it difficult to reconstruct the velocity using only linear methods.}
    \label{fig:linear_v_point}
\end{figure*}

Motivated by the neural network model~\citep{Wu:2021jsy}, we use a modified UNet neural network architecture for model construction. The architecture of our neural network and its components are shown in Fig.~\ref{fig:model}. The input is the 6-channel number density field of halos, each channel corresponding to number density field for a certain mass range of halos. As mentioned in Sect.~\ref{sect:ip}, as the velocity field is decomposed into the two parts: velocity magnitude and the velocity direction, we build two structurally similar neural networks to deal with them separately. The network ends with the output layer of 2+3 fields, three of which correspond to the components of velocity direction ($\hat{v}_x, \hat{v}_y, \hat{v}_z$) and two to the velocity magnitude ($v_{\rm large}$, $v_{\rm smaller}$). A complete reconstruction of the 3D velocity field is finally achieved by combining all of the output field components.



More specifically, the detail of the UNet network are shown on the bottom-left panel in Fig.\ref{fig:model}. The colored plates represent different operations in the neural network, which are connected from the inputs to the outputs by means of arrow lines. The size of the input, the intermediate and the output fields (number of channels $\times$ spatial pixels) is specified. Also, the size and the number of 3D convolution kernels ("conv") are also labelled. Moreover, the combination of padding schemes gives the desired dimensionality after each convolution. Note that, 1) To modify the size of the field, we use the stride=2 parameter, which allows to reduce or enlarge the field by half or double its original size, respectively. 2) the "init" 3D convolutional layers allow for a sufficiently large receptive field, enabling the network to quickly learn the large-scale information in the beginning, 3) the "output" convolutional layers increase complexity, followed by the dropout layers to avoid overfitting and to change the number of channels at the end, and 4) the inclusion of the batch normalization (BN) layer in neural networks has the potential to accelerate the convergence of training and prevent overfitting, while incorporating the rectified linear unit (ReLU) implemented as an activation layer after convolutional layers, could enhance the nonlinearity of the network.

With the trained UNet, the velocity field of halos is predicted by feeding the number density field of halos in the redshift space, and the relevant statistics such as clustering information can be measured straightforwardly.  

A crucial ingredient in our model is a three-block structure: an lower convolution block (red), a upper convolution block (pink), and a final one (blue). The advantages of these blocks are capable of passing the initial information to the deep network structure. In parallel, to avoid the bias in small-box simulations, the linear theory-predicted velocity field is used as an additional input in the lower block, compensating for the lack of information on large-scale velocities. 
The velocity information at the boundary of the density field is indeed strongly influenced by the density distribution outside the box. To avoid spurious signals caused by boundary effects, the final convolution block (as shown in the lower left panel of Fig.1) is specifically designed to learn the velocity field at the center of the number density field rather than the whole velocity field of the same size as the density field.

\subsection{Loss Function}
The objective of the training our UNet is to minimize a loss function between prediction, $\bm{v}$, and simulation truth, $\bm{v}^{\rm true}$ of each voxel. Specifically, to account for the contributions from the velocity magnitude ($v\equiv|\bm{v}|$) and the velocity direction (unit vector $\hat{\bm{v}}\equiv \bm{v}/v$), we choose the following loss function with two terms, 
\begin{equation}
 \mathcal{L} = \frac{1}{N} \sum_{i=1}^N \left[\frac{2}{5} (v_i-v^{\rm true}_i)^2 +  \frac{3}{5}\left(1-\cos\phi_i \right) \right]\,,
 \label{eq:loss}
\end{equation}
where $\cos\phi_i\equiv  \hat{\bm{v}}_i\cdot \hat{\bm{v}}_i^{\rm true}$, and the index $i$ denotes the $i$-th voxel. As observed, the first term is responsible for $v$, and corresponds to the standard and simple mean error (MSE) loss that is essentially equivalent to the maximum likelihood solution under a Gaussian assumption with constant variance. The second term naturally measures the deviation between the reconstructed and the true values of $\hat{\bm{v}}$. The coefficients of these two terms can be regarded as normalization factors and are determined by the number of channels, i.e., 2 for magnitude ($v_{\rm large}$,  $v_{\rm small}$), and 3 for direction ($\hat{v}_x$, $\hat{v}_y$, $\hat{v}_z$). Empirically, such loss function is effective, and has proven to be stable and effective during our training process, providing good results in the velocity (momentum) reconstruction. We trained our UNet using the most popular algorithm Adam~\citep{2014arXiv1412.6980K} for training deep neural networks, which can iteratively decrease the training loss by calculating its gradient with respect to model parameters and performing a small step along the direction with the maximum decrease.

Based on Fig.~\ref{fig:loss}, it is evident that both the velocity model and momentum model converge after 20,000 epochs of training. While the velocity model exhibits slight overfitting, the validation loss never exceeds 1.1 times its minimum until the final epochs, suggesting that our results would not be significantly affected by the issue of overfitting.

\section{Results}
\label{sect:result}
In this section, we evaluate the performance of the trained UNet model and present our results based on predictions from 27 large boxes. Each of these large boxes comprises 125 adjacent small boxes, with the same volume as the training sets (side length 125 ${\rm Mpc}/h$), selected from the test sets. To form each large box, we padded the 125 small test boxes together. Consequently, the total number of test simulation boxes is 3325. Therefore, the box volume for each test set we have performed the analysis on is $625^3 (h^{-1}\rm Mpc)^3$. We chose this because measurements on large boxes would have a better statistical behavior, reducing the statistical errors.  To ensure reliable test results, these test simulation boxes were not used for the model training and refining the model structure/training parameters.

First we shall describe the statistics we will use throughout 
the paper. The 2-point correlation function is one of the most commonly used statistics to characterize a homogeneous density  field,
\begin{equation}
\xi(\bm{r})=\left\langle\delta\left(\bm{x}\right) \delta\left(\bm{x}+\bm{r}\right)\right\rangle\,,
\end{equation}
where $\delta(\bm{x})$ is the density contrast field, $\bm{x}$ denote for any point, $\bm{r}$ is a separation vector, and $\langle\cdot\rangle$ stands for the ensemble mean, computed with a spatial mean over $\bm{x}$ in practice.


The power spectrum of $\delta\left(\bm{x}\right)$ is just related to $\xi(\bm{r})$ by the Fourier transform, i.e.,   
\begin{equation}
P(\bm{k})=\int \xi(\bm{r}) \mathrm{e}^{i \bm{k} \cdot \bm{r}} {\rm d}^3 \bm{r}\,,
\end{equation}
where $\bm{k}$ is the 3D wavevector of the plane wave, with the magnitude $k\equiv |\bm{k}|$ (the wavenumber) related to the wavelength $\lambda$ by $k=2\pi/\lambda$.

Similar to the scalar field $\delta$, we can also define power spectra for velocity and momentum vector fields of interest. Via the Helmholtz decomposition, the velocity field, $\bm{v}$, is completely described by its divergence, $\theta \equiv \nabla \cdot \bm{v} $ and its vorticity, $\bm{\omega} = \nabla \times \bm{v}$, which, in Fourier space, become purely radial and transversal velocity modes, respectively, defined by  $\theta(\bm{k}) = i \bm{k} \cdot \bm{v}(\bm{k})$ and $\bm{\omega}(\bm{k}) = i \bm{k} \times \bm{v}(\bm{k})$. The power spectra of the velocity, divergence, vorticity and velocity magnitude are given by 
\begin{equation}
\begin{aligned}
\langle \theta (\bm{k}) \theta^* (\bm{k}') \rangle =& (2\pi)^3 P_{\theta}(k) \delta ( \bm{k}  -  \textbf{k}')\,,\\
\langle \omega^i (\bm{k}) \omega^{*j} (\bm{k}') \rangle =& (2\pi)^3 \frac{1}{2} \bigg( \delta^{ij}-\frac{k^ik^j}{k^2} \bigg) P_{\omega}(\bm{k}) \delta ( \bm{k}  -  \bm{k}')\,,\\
\langle \bm{v} (\bm{k}) \cdot \bm{v}^* (\bm{k}') \rangle =& (2\pi)^3 P_{v}(\bm{k}) \delta ( \bm{k}  -  \bm{k}')\,,
\end{aligned}\label{eq:v}
\end{equation}
where indices $i,j$ denote the components in the Fourier space coordinates. 

In the linear perturbation theory, the continuity equation leads to $\theta=-\mathcal{H}f\delta$, where $\mathcal{H} =aH$ is the conformal Hubble parameter,  $a$ denotes the cosmic scale factor and $f$ is the linear growth rate defined by $f = d\ln D/d\ln a$, with $D$ being the linear density growth factor. In a $\Lambda$CDM model, $f \approx \Omega_{\mathrm{m}}^{0.55}$~\citep{2005PhRvD..72d3529L}, with a good approximation.

\subsection{Analysis on Velocity Field}
First we shall describe the metrics that we will use throughout this section for evaluating the reconstruction accuracy. For an arbitrary reconstructed field of halos from UNet, denoted by the shorthand notation $f$, where $f \in\{\theta, \bm{\omega}, \bm{v}\}$ for velocity, we use the following metrics to describe the {\it relative deviation} and {\it correlation coefficient}, to compare a reconstructed field ($f$) with the true one ($f'$):
\begin{equation}\label{eq:tc}
T_f = \frac{\mathcal{O}_f}{\mathcal{O}_{f'}} - 1\,,\quad C_{f} = \frac{1}{N_{\rm pix}-1}\sum_{i}\frac{ (f_i-\bar{f}) (f'_i-\bar{f'})}{\sigma_f \sigma_{f'}}\,,
\end{equation}
where $\mathcal{O}_f$ stands for an arbitrary observable for $f$. The correlation $C_f$ is defined between reconstructed ($f$) and true fields ($f'$) with the same total number of pixels $N_{\rm pix}$. The sample mean and the standard deviation of field $f$ are denoted by $\bar{f}$ and $\sigma_f$, respectively. Both metrics provide a physical insight for comparison such that the perfect reconstruction is equivalent to $T_f=0$ and to $C_f=1$.

\subsubsection{Visual inspection and point-wise comparison}

\begin{figure*}
	\includegraphics[width=500pt]{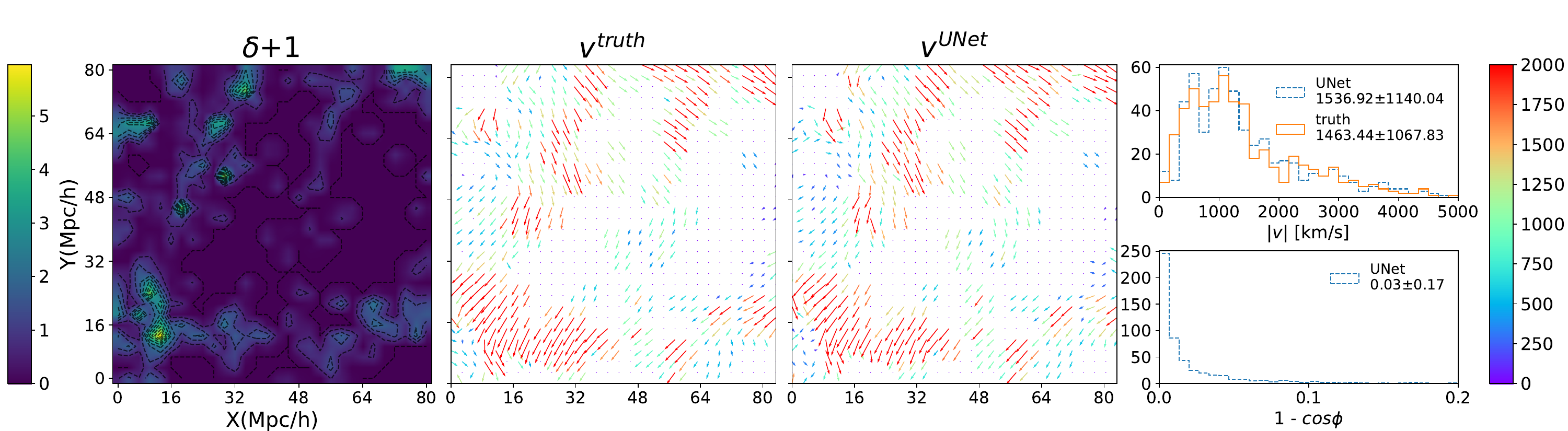}\\
    \includegraphics[width=500pt]{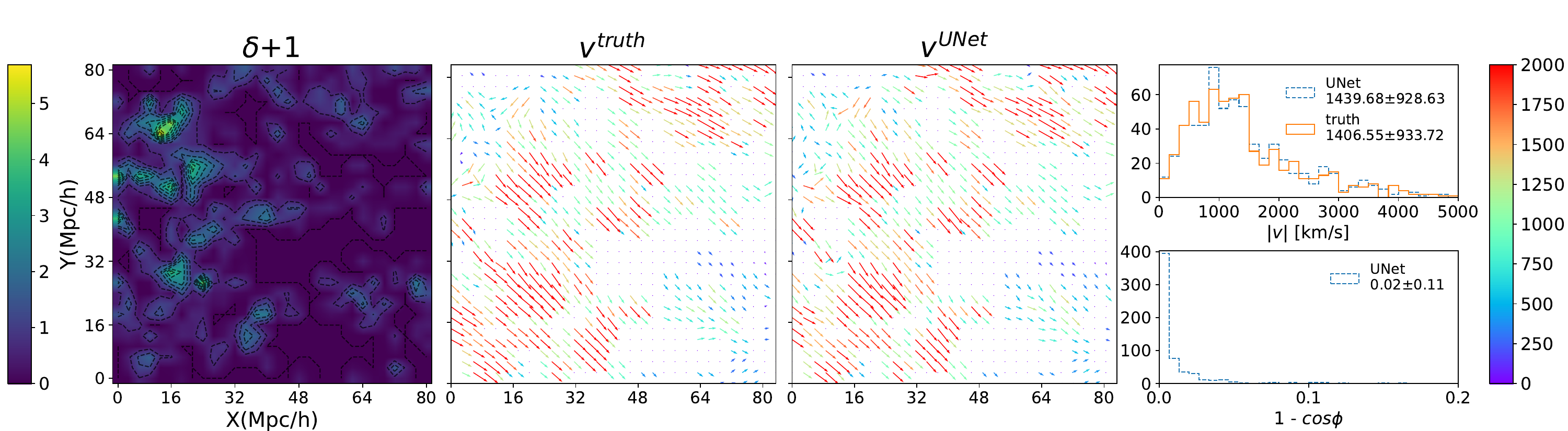}
    \caption{Point-wise comparison between the UNet-reconstructed velocity field and true one. From top to bottom, we show the results for the two slices in the test sets, each with volume of $83\times 83\times 28$ $(h^{-1}{\rm Mpc})^3$. From left to right, the fields of halo number density, the UNet-reconstructed velocity, the true velocity and the corresponding histogram are shown, respectively. The length and the orientation of the colored arrows in the velocity fields represent the velocity magnitude and direction. For each slice, the rightmost panels show the statistical histogram distributions of the velocity samples for magnitude and direction, respectively. UNet is able to reconstruct velocity well via visual inspection and the statistical analysis on the histogram.}
    \label{fig:v_point}
\end{figure*}

\begin{figure*}
    \includegraphics[width=500pt]{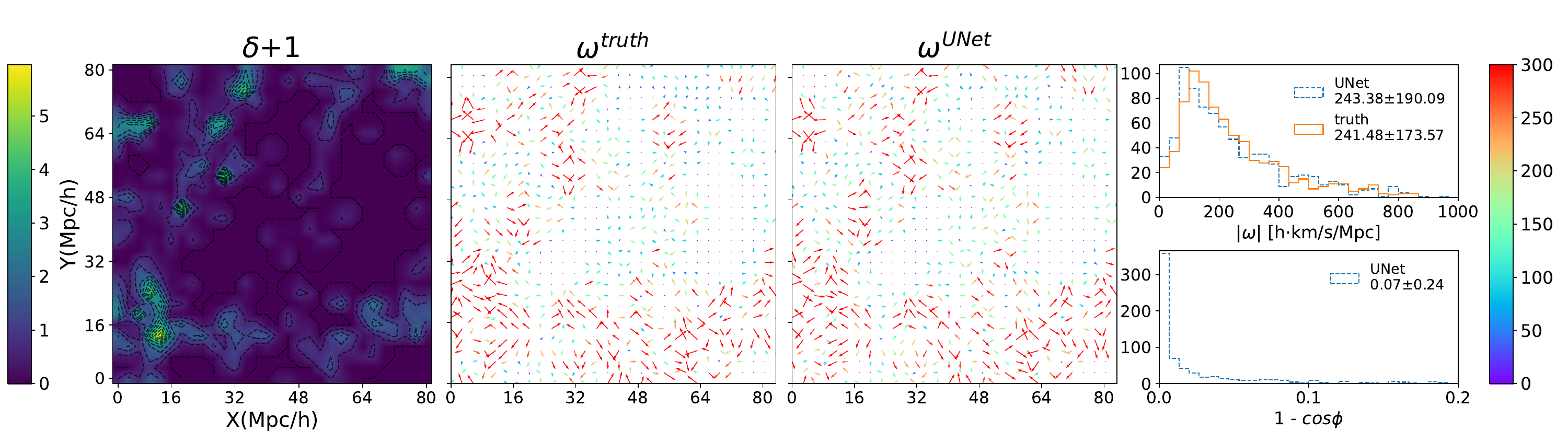}\\
    \includegraphics[width=500pt]{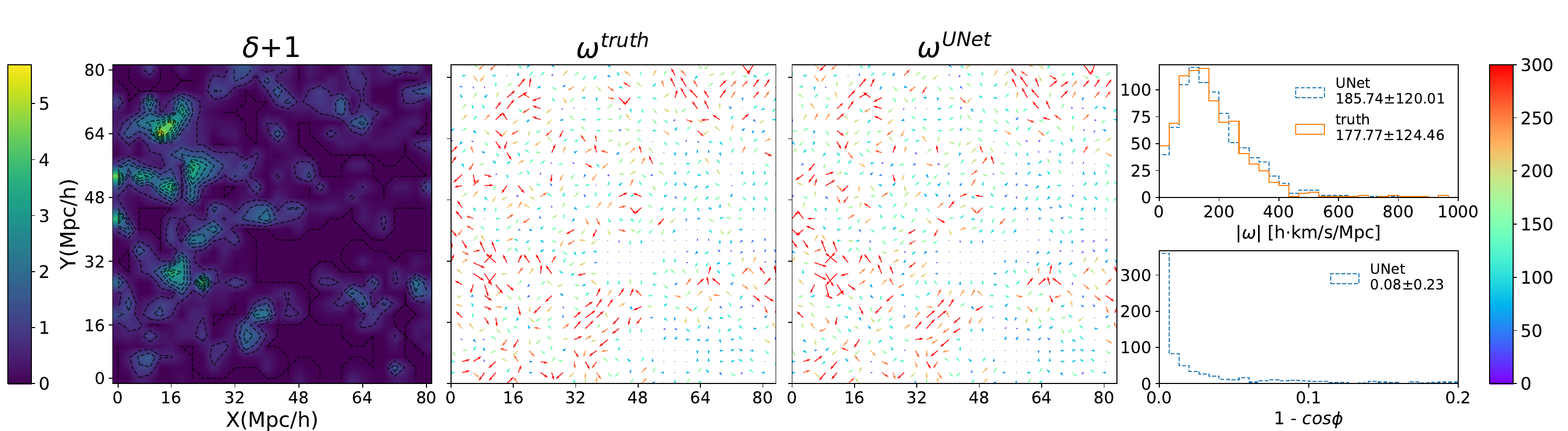}
    \caption{Same as in Fig.~\ref{fig:v_point}, but for the vorticity field of halo  velocities, $\omega$. As seen, the vorticity field, exclusively generated from the nonlinear structure formation, has a more complex distribution pattern than that of the velocity field. In spite of this, the Unet approach still performs very well in reconstruction.}
    \label{fig:v_curl_point}
\end{figure*}

\begin{figure*}
    \includegraphics[width=500pt]{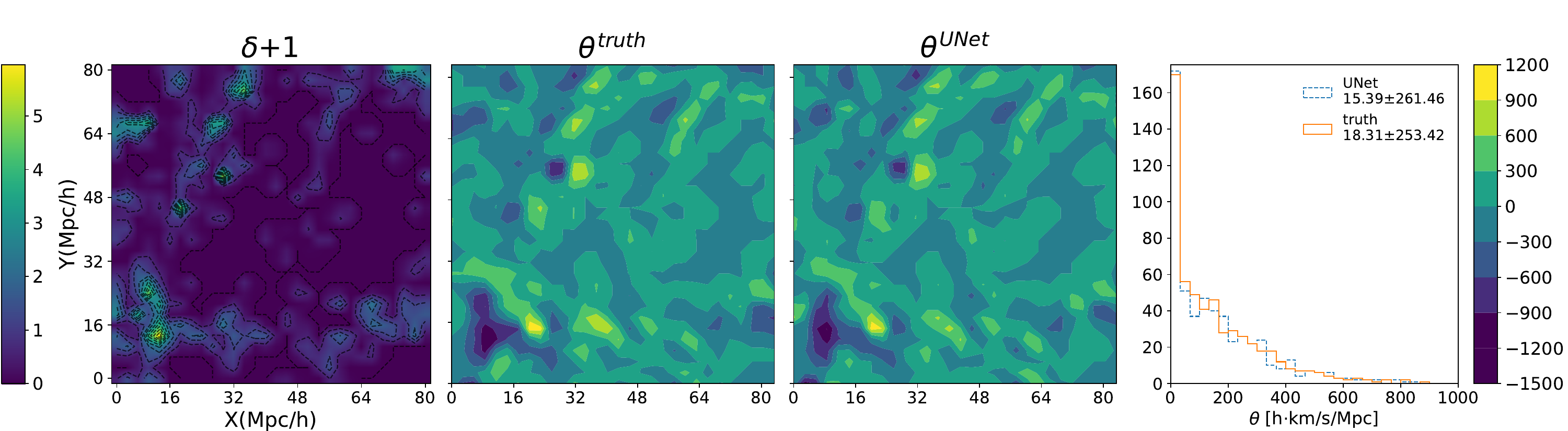}\\
    \includegraphics[width=500pt]{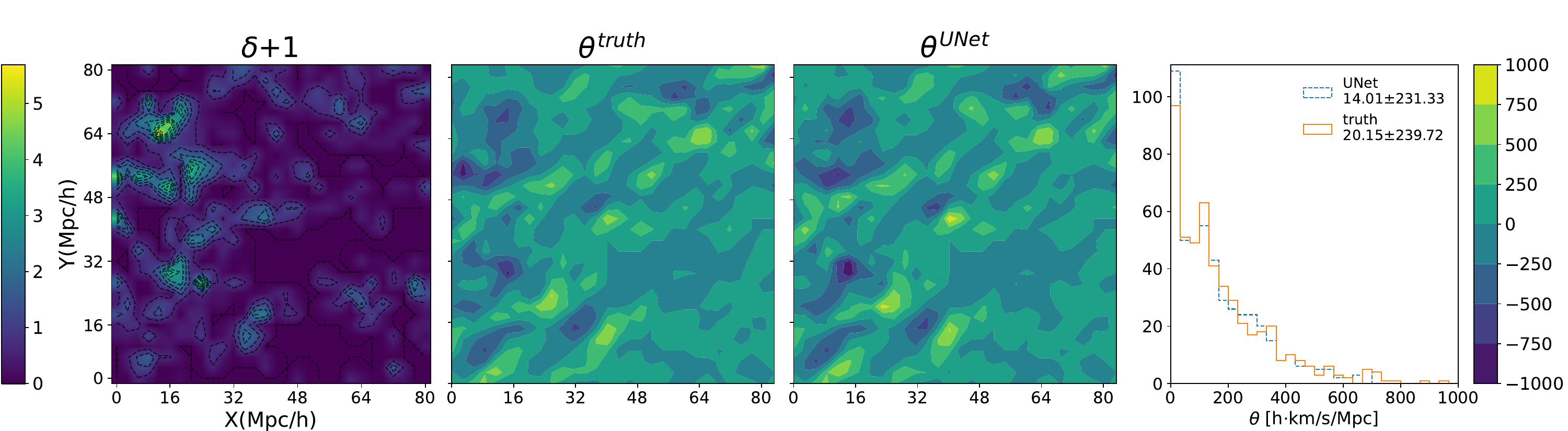}
    \caption{Same as in Fig.~\ref{fig:v_point}, but for the divergence field of halo  velocities, $\theta$. Reconstruction results show the effectiveness in reconstructing the divergence field.}
    \label{fig:v_div_point}
\end{figure*}

As a first validation, we perform a point-wise comparison between the UNet-predicted halo velocity field to the simulation truth. To do so, we randomly selected two slices in the test sets. 

Fig.~\ref{fig:v_point} visualizes the number density distribution of dark matter halos (and subhalos) and the velocity field in these two slices, zooming in to highlight small-scale nonlinear features.

As seen, there are many massive halos in the range of $M/M_{\odot}\in [10^{12}, 10^{15}]$, typically with 60 halos per slice. In the middle and right panels, we display the true and the predicted velocity fields, respectively. The colored arrows show the average velocities at the meshgrid points and are projected onto the image plane. The length and the direction of the arrow represent the magnitude and the direction of the projected velocity, respectively. To show clearly,
the projected velocity magnitude is also marked by the color from purple to red, reflecting the halo velocities from small to large. As seen, a high number density region typically leads to a larger velocity field, since the gravitational collapse and non-linear structure formation occur intensively there. Moreover, the  visualized morphology for the UNet-predicted and the true velocity fields clearly indicates the effectiveness of our neural network, as they are almost indistinguishable by eye. Interestingly, although the simulated halo velocity field is sparse, we can still reliably reconstruct it, especially for the regions with small velocities.
To quantitatively validate such reconstruction, we show  the histogram distributions (in the rightmost panel right panels) of magnitude and direction of the velocities in these two slices. We do find that, statistically, the distributions of the reconstructed velocity magnitude and direction agree well with the true values.   

Specifically, the mean value and its $1\sigma$ dispersion (with a Guassian fit) for the UNet-reconstructed velocity magnitude in each slice are 
$1536.92\pm 1140.04$ and $1439.68\pm 928.63$ ${\rm km/s}$, respectively. These mean values are consistent with the true ones at about higher than $95\%$ accuracy among all these slices. Similarly, we obtain very good results in the direction reconstruction, with $1-\cos\theta$ (defined in Eq.~\ref{eq:loss}) of 
$0.03\pm 0.17$ and $0.02\pm 0.11$
for these slices, respectively. By averaging over the slices, the deviation in the velocity direction, $\Delta \phi=|\phi^{\rm true}-\phi^{\rm UNet}|$, achieves $\Delta \phi= 14.1^\circ\pm 30.7^\circ$, implying $\Delta \phi<45^{\circ}$ at $1\sigma$ level. 

Furthermore, let us focus on the vorticity field, $\bm{\omega}$, which generally appears in high-density regions of halos and is essentially induced by non-linear structure formation. The reconstruction of the vorticity field for two randomly selected slices is present in Fig.~\ref{fig:v_curl_point}. As seen, the vorticity field is tightly concentrated on high-density regions where the nonlinear processes such as shell crossing are occurring. Thus its magnitude is tightly coupled to the local density and decays rapidly at the linear regime (low density) of structure formation. The direction of the virial motion in these high-density regions is randomized and distributed randomly, indicating a strong nonlinear process. Due to the effect of nonlinearity on small scales, the vorticity fields  are theoretically very difficult to reconstruct, especially through sparse halo samples. Here, however, we show the advantages of UNet, which can provide the reconstruction in $|\bm{\omega}|$ at very high accuracy, with the deviations in the mean and dispersion only about $|\bm{\omega}^{\rm true}-\bm{\omega}^{\rm UNet}| = 1.90\pm 16.52$ and $7.97\pm 4.45$ $h{\rm km/s}/{\rm Mpc}$ for these two slices, respectively. Also, from the histogram distribution, the reconstructed directions for the vorticity component indicate that the reconstruction is unbiased, with the mean and $1\sigma$ error of $\Delta \phi \approx 22.3^\circ\pm 39.5^\circ$. Furthermore, Fig.~\ref{fig:v_div_point} shows that the results of the reconstructed divergence field are generally good, with deviations in the mean and dispersion of $|\theta^{\rm true}-\theta^{\rm UNet}| = 2.92 \pm 8.04$ and $6.14 \pm 8.39$ $h{\rm km/s}/{\rm Mpc}$.

\begin{table}
	\centering
	\caption{Summary of the correlation coefficients $C_f$ between the reconstructed and true fields of the velocity $\bm{v}$, the divergence component $\theta$ and the vorticity $\bm{\omega}$ via Eq.~\ref{eq:tc}, estimated by averaging over 27 test sets, each with the box size of 625 ${\rm Mpc}/h$.}
		\begin{tabular}{c|ccc} 
		\hline
		field  & $\bm{v}$ & $\theta$ & $\bm{\omega}$   \\
		\hline
		 $C_f$  &0.94&0.78&0.88   \\
		\hline
	\end{tabular}\label{tab:cfv}
\end{table}

In order to quantitatively compare the reconstructed and real fields, we compute the coefficients, $C_f$, through Eq.~\ref{eq:tc} for various velocity components.
The resulting coefficients are summarized in Tab.~\ref{tab:cfv}, estimated by averaging over 27 test sets, each with the box size of side 625 ${\rm Mpc}/h$. Our proposed UNet model has excellent performance in terms of the linear correlation in real-space domain, demonstrating that the network produces high-fidelity reconstructions in $\bm{v}$, $\theta$ and $\bm{\omega}$, with $C_f$  calculated based on the magnitudes of velocity, vorticity, and divergence. in the range of [0.94, 0.78]. As seen, the reconstruction in $\bm{\omega}$ is almost as good as in the other components, which indicates that the UNet model can give a good prediction for the vorticity field with complicated morphological properties.

To further test the reconstructed velocity field with the ground truth, in Fig.~\ref{fig:v_delta}, a visual inspection for {\it the joint probability distributions of density-divergence and density-vorticity} are shown. Obviously, all predictions are in good agreement with the the true values, even in the very high density regions ($\delta \gg 1$). Furthermore, we find that, for a given $\delta$, the reconstructed distributions appear slightly narrower than the true ones. This is probably because the neural network would slightly lose some random perturbation information when learning and compressing the information in the training sets.

\begin{figure*}
\centering
\includegraphics[width=0.32\textwidth]{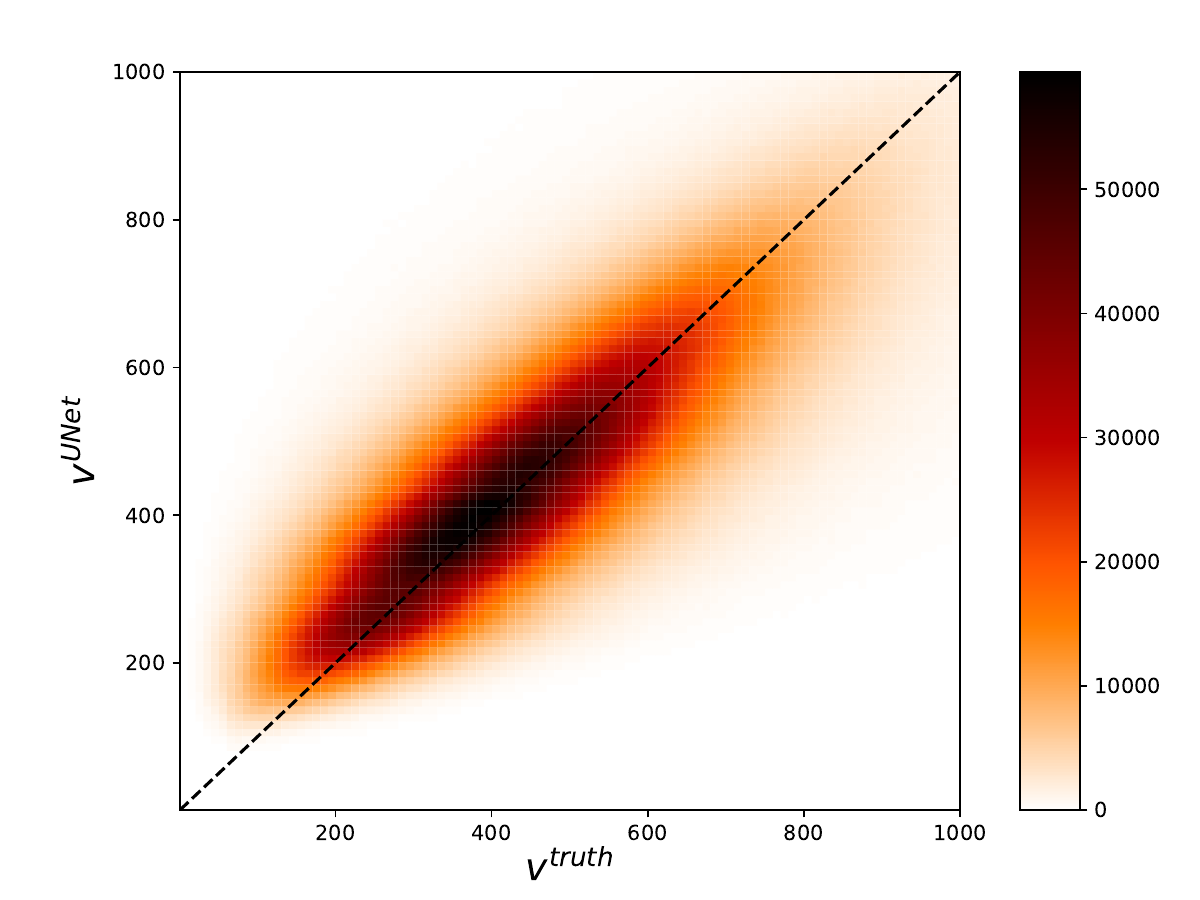}
\includegraphics[width=0.32\textwidth]{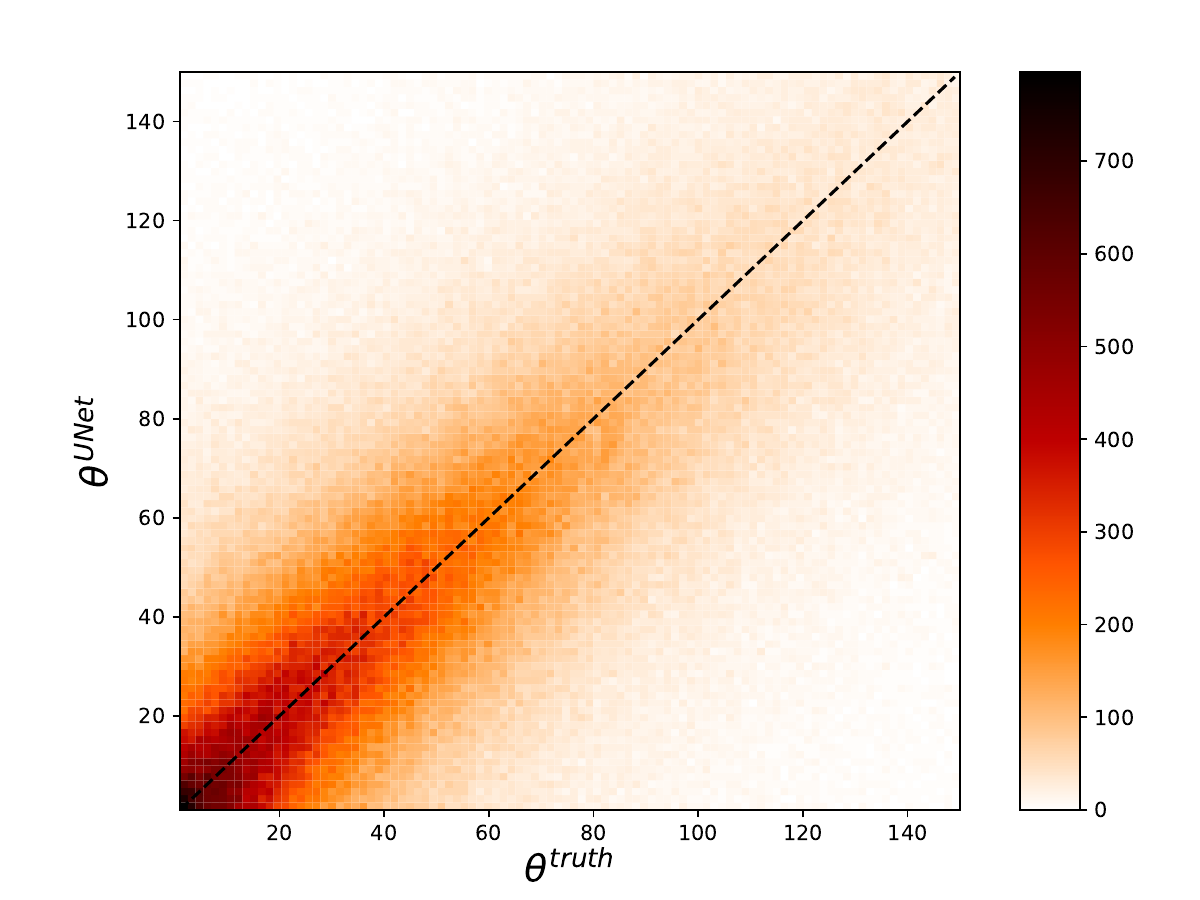}
\includegraphics[width=0.32\textwidth]{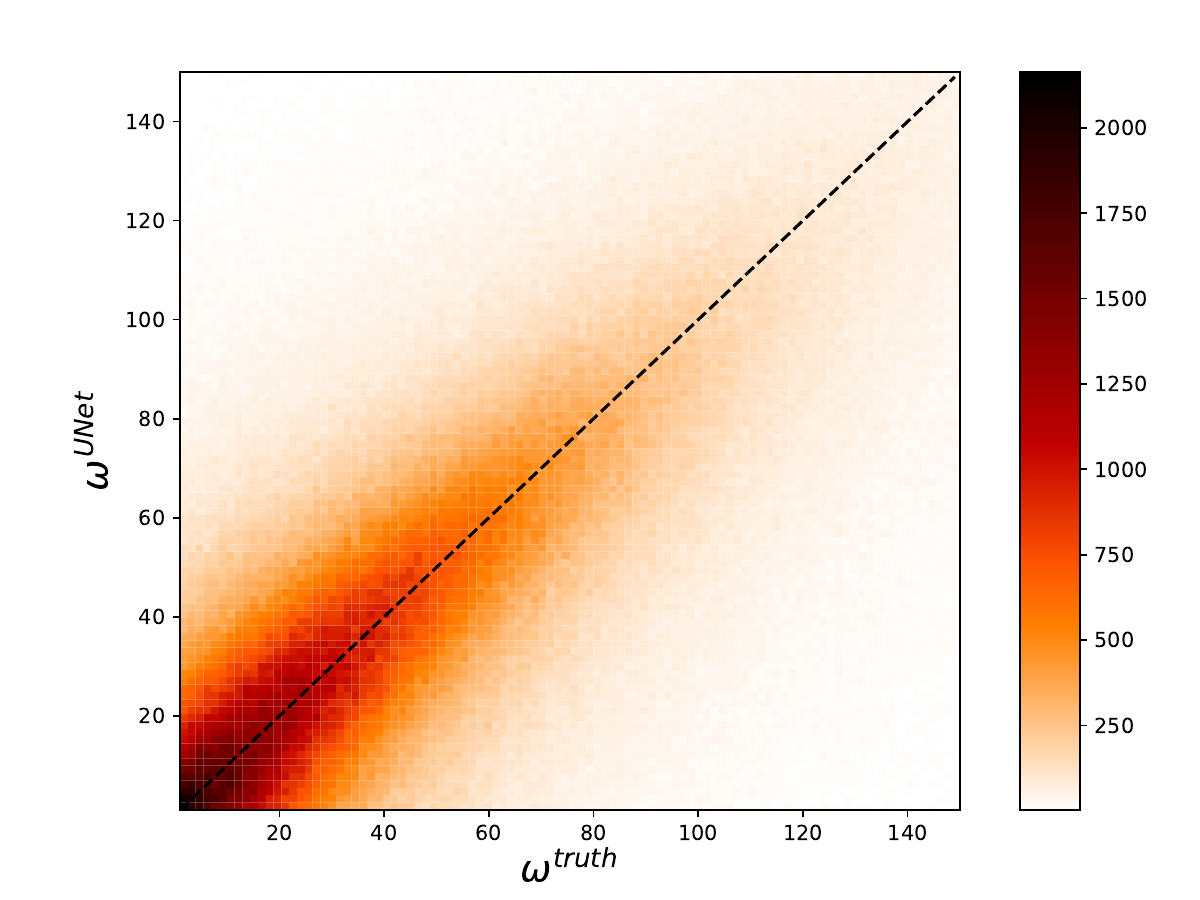}
\caption{Joint probability distribution of the UNet reconstructed field and the true field, for velocity ($v$), divergence ($\theta$), and vorticity ($\omega$) from left to right. For each case, the distribution are calculated  using 27 test sets, each with a box size of 625 $({\rm Mpc}/h)^3$. As demonstrated, the predicted distribution are strongly correlated with the true distribution.}
    \label{fig:v_joint}
\end{figure*}

\begin{figure}
        \includegraphics[width=0.45\textwidth]{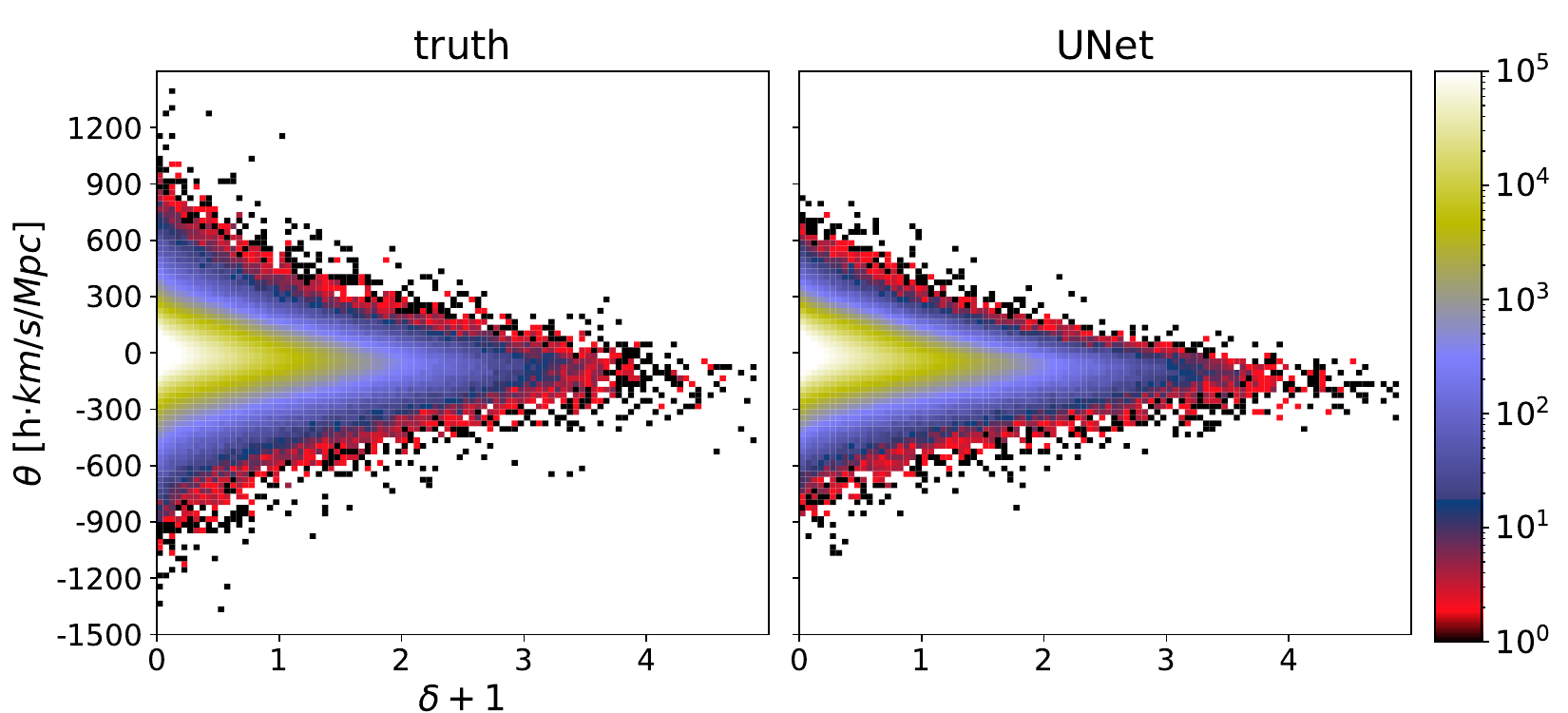}
        \includegraphics[width=0.45\textwidth]{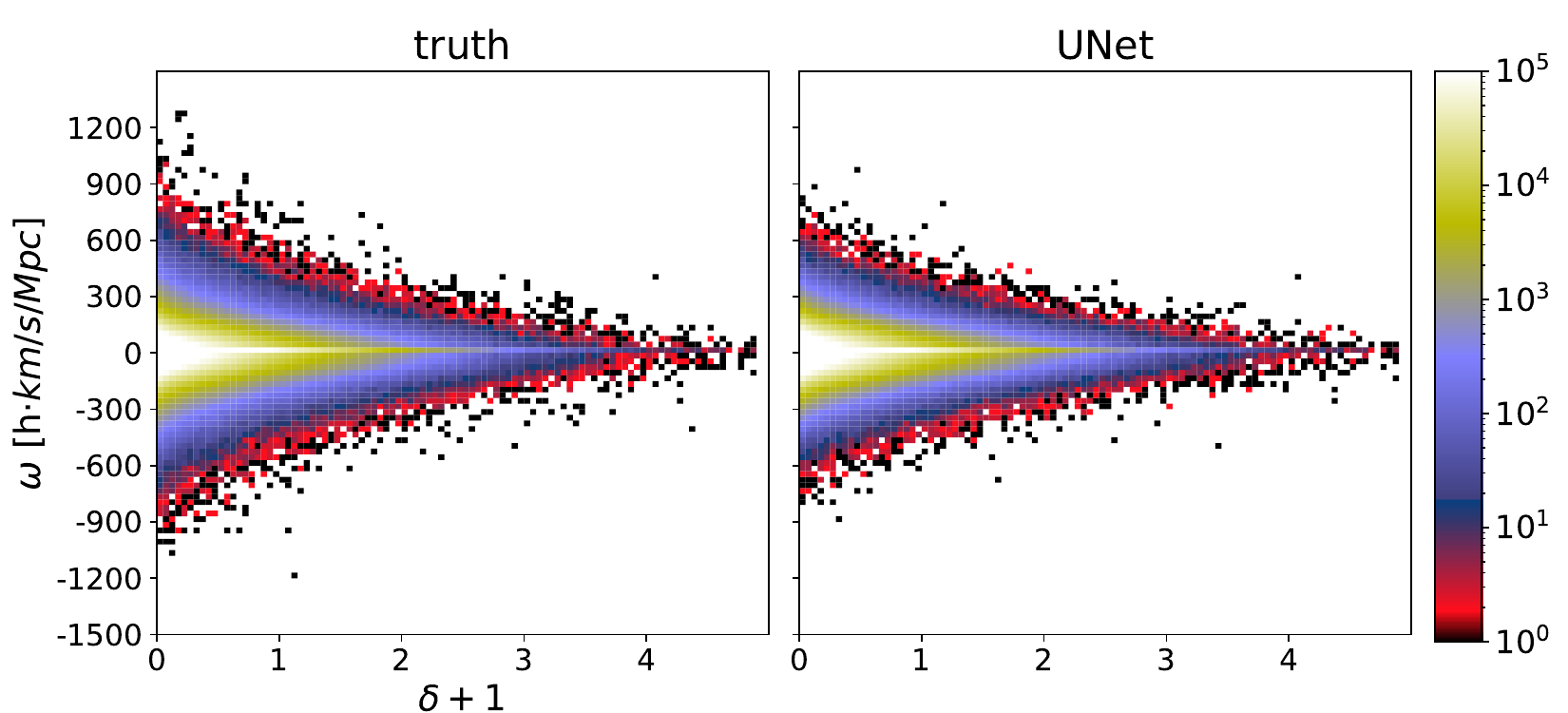}
    \caption{Joint probability distributions of density-divergence, $\rho(\delta,\theta)$ (upper), and density-vorticity, $\rho(\delta,|\bm{\omega}|)$ (lower). In each case, the distribution results are calculated from 5 test sets, each with the box size of 625 ${\rm Mpc}/h$. As seen, the predicted velocity distributions (right) agree well with the simulation truth (left) for all of the halo number densities of $\delta +1 \in [0, 5]$.}
    \label{fig:v_delta}
\end{figure}

We also present a joint distribution plot between the reconstructed velocity field and the true one in Fig.~\ref{fig:v_joint}. Our findings indicate a strong correlation between the reconstructed velocity field and the true one for all three quantities, namely $v$, $\theta$, and $\omega$.

\subsubsection{Comparison for power spectrum}

\begin{figure*}
\centering
\includegraphics[width=0.32\textwidth]{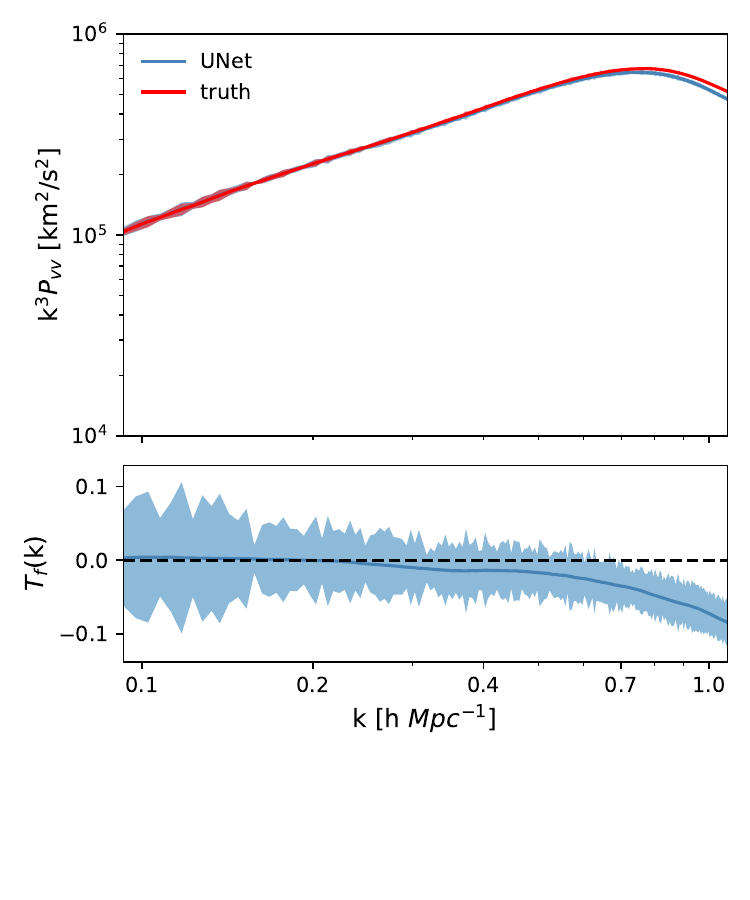}
\includegraphics[width=0.32\textwidth]{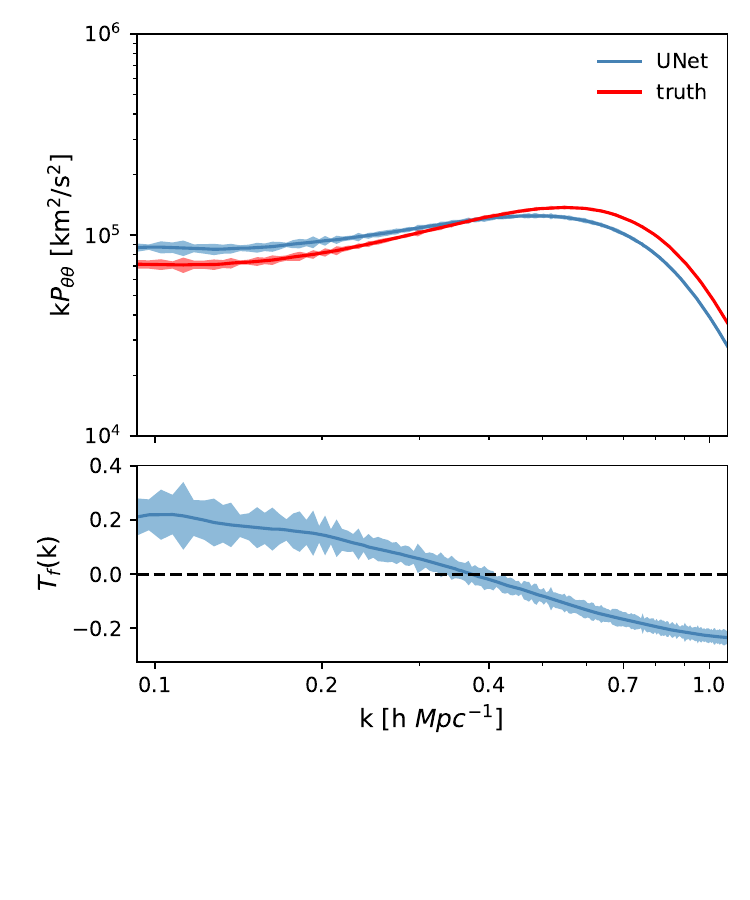}
\includegraphics[width=0.32\textwidth]{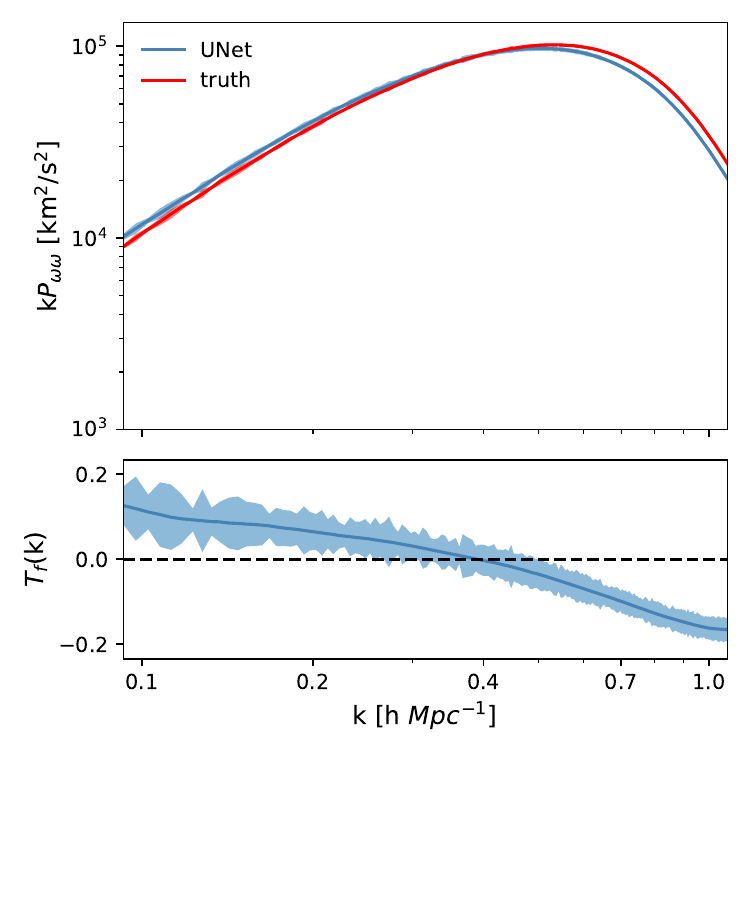}
\caption{Comparison of the UNet-predicted power spectrum and the simulation truth. From left to right, we show the the reconstruction results for the velocity, the velocity divergence and the vorticity, respectively. These results are based on the 27 test sets, each with the box of side length 625 ${\rm Mpc}/h$. The shaded area gives $1\sigma$ deviation measured from these test sets.}
    \label{fig:pk_vtheta}
\end{figure*}

Here we describe the reconstruction accuracy for the power spectra of velocity components, $P_f(\bm{k})$. For each component, we have computed the relative deviation (see Eq.~\ref{eq:tc}) in terms of the predicted power spectrum and the true one. Intuitively, by use of power spectrum as the observable, $T_f$ measures the accuracy of reconstruction in magnitude as a function of wavevector in Fourier domain. Taking the directional average, the function $T_f(k)$ represents the relative deviation with spatial averaging over $|\bm{k}|$ bins. Also, in general, $T(k)$ is not explicitly optimized during the training stage, since the training minimizes the proposed loss function (see Eq.~\ref{eq:loss}) composed of the velocity magnitude and direction in real-space domain.

As observed in Fig.~\ref{fig:pk_vtheta}, there is a small bias to slightly underestimate the power spectrum for the velocity field $\bm{v}$ over all scales, $|T_f(k)| < 0.1$. This underestimate may be due to the fact that non-linear scale is a complex process. For the divergence component $\theta$, the deviation varies for positive to negative, i.e., $T_f(k)\in[0, 0.2]$ for $k\lesssim0.4$ $h/\rm Mpc$ and $T_f(k)\in [-0.25, 0]$ otherwise. As known, the vorticity $\bm{\omega}$ is generated by nonlinear evolution, and so its reconstruction has always been a challenge. However, we find the reconstructed vorticity power spectrum successfully match the true one, yielding a similar deviation level as in $\theta$, with $|T_f(k)|\lesssim 0.20$ in all range, especially when $k\lesssim0.7$ $h/\rm Mpc$, $|T_f(k)| \lesssim 0.1$. This is remarkable considering that the UNet model performs well from the linear to deeply nonlinear scales. All these test results highlight the ability of UNet in learning various velocity components from the halo number density field, especially on the nonlinear scales.

\begin{figure*}
\centering
\includegraphics[width=0.32\textwidth]{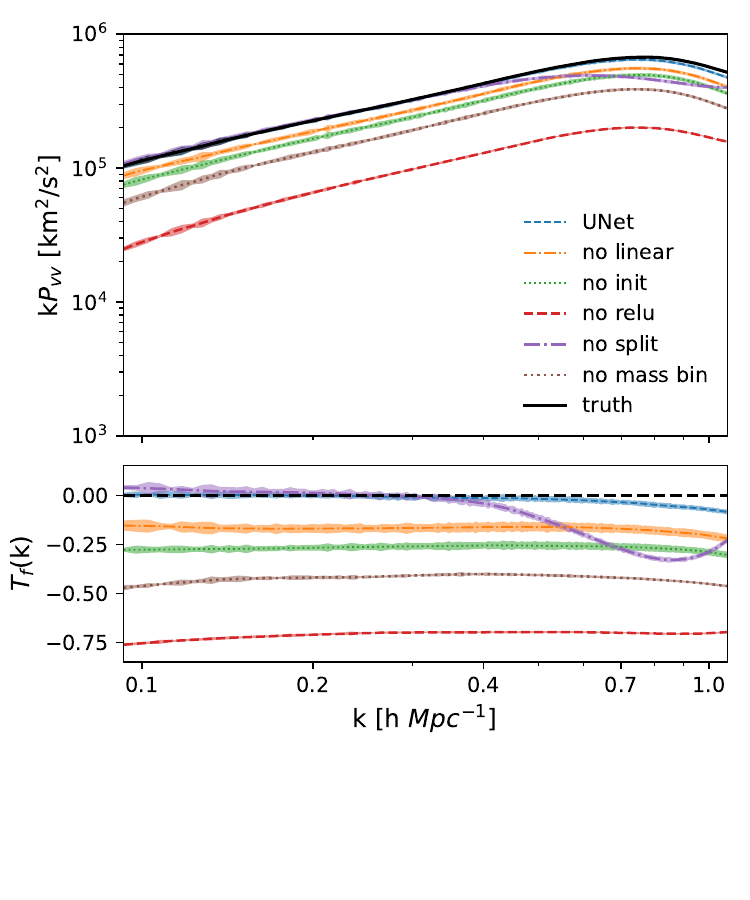}
\includegraphics[width=0.32\textwidth]{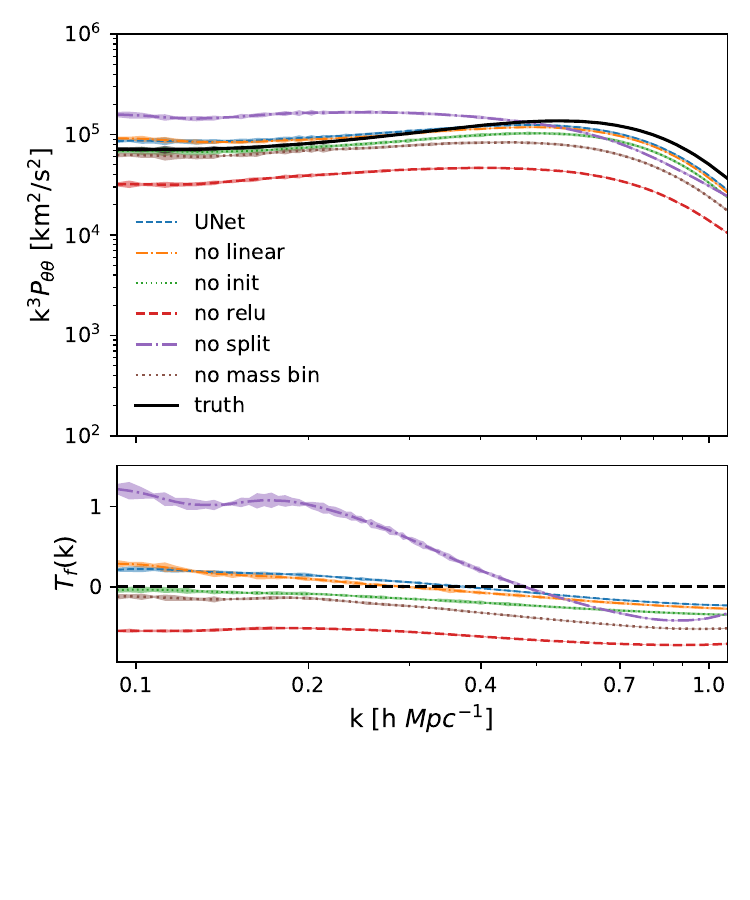}
\includegraphics[width=0.32\textwidth]{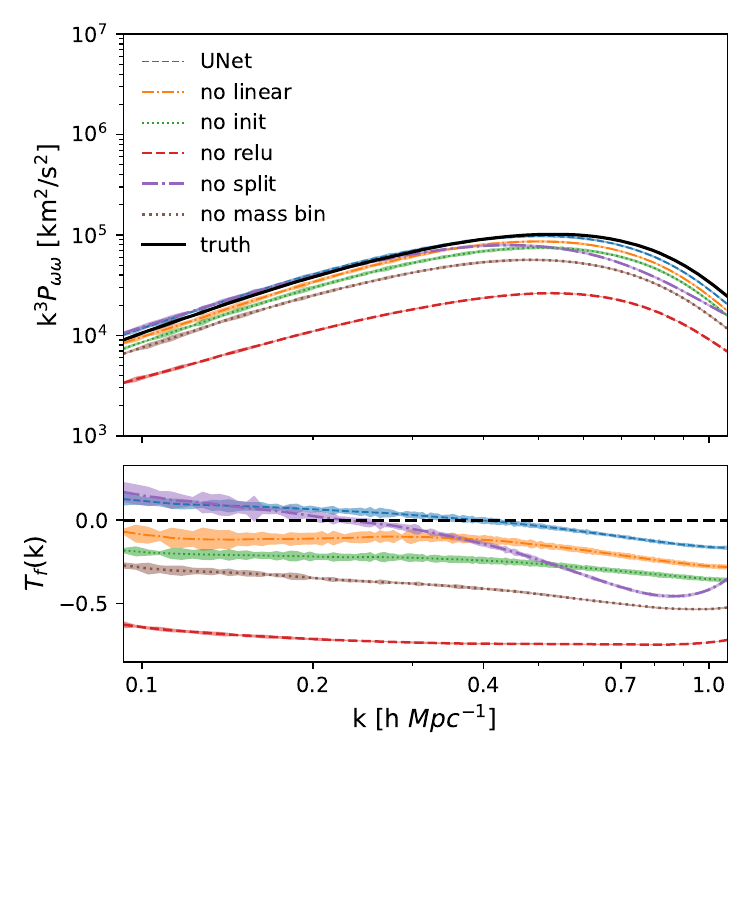}
\caption{Comparison of the power spectrum predicted by UNet with the simulation truth, as well as variations of the UNet model, including the "no init", "no relu", "no split" and "no mass bin" models. The "no init" model replaces all "init" blocks in the UNet model with "conv" blocks, while the "no relu" model eliminates all "relu" layers. The "no split" model does not split the velocity magnitude into two parts, and the "no mass bin" model does not divide different halo masses into separate bins. Our results suggest that the UNet architecture outperforms all other models.}
    \label{fig:diff_architecture}
\end{figure*}

Furthermore, in Fig.~\ref{fig:diff_architecture}, 
we present a comparison of the performance of various model architectures by calculating the power spectrum of each architecture in 5 large test boxes, each with a side length of 625 ${\rm Mpc}/h$. In this study, we use the "UNet" model as our reference architecture. The label "no linear" denotes that the input of the model does not include the linear velocity field. The label "no init" indicates that all the "init" blocks in the model have been replaced with "conv" blocks. The label "no relu" implies that we have closed all the "relu" layers in the model. The label "no split" means that the velocity magnitude has not been split into two parts. Lastly, the label "no mass bin" suggests that we have not divided the different halo masses into different bins. Our results indicate that the "UNet" architecture outperforms all other architectures, and it also demonstrates that each of these factors affects the performance of the model. Therefore, when designing a model architecture, it is essential to consider these factors carefully. Our approach aims to incorporate as much information as possible into the model, including different bins to represent the halo mass, separation of the velocity magnitude to ensure small velocity magnitudes are considered, "init" blocks to capture large-scale correlations, and linear velocity to provide large-scale information.

\subsection{RSD Corrections}

An important application of the UNet-based velocity reconstruction is to map a halo distribution from redshift to real space as well as inferring the distances of individual halos (galaxies). To do so, redshift-space distortions (RSD) are corrected by moving the halos from redshift  to real space according to their peculiar velocities reconstructed from the halo number density field using the trained UNet network. By performing a tri-linear interpolation of the reconstructed velocity field, the velocities at every halo positions can be obtained with reasonable accuracy. In the following, we will present the performance of such RSD correction.


\subsubsection{Two-point correlation function}

In redshift space, anisotropic two-point correlation function (2PCF), $\xi(\bm{r})$, provides a measurement for  halo (galaxy) clustering through the standard \cite{1993ApJ...412...64L} estimator, 
\begin{equation}
\xi(r, \mu)=\frac{D D(r, \mu)-2 D R(r, \mu)+R R(r, \mu)}{R R(r, \mu)}\,
\end{equation}
where $DD$, $DR$, and $RR$ are the normalized galaxy-galaxy, galaxy-random, and random-random number of pairs with separation $(r, \mu)$, respectively. Here the 3D separation vector between pairs of objects, $\bm{r}$, has been decomposed into ($r$, $\mu$) coordinates, where $r$ is the norm of the separation vector and $\mu$ is the cosine of the angle between the line-of-sight and separation vector directions.

It is common to expand 2PCF into Legendre polynomials as
\begin{equation}
\xi(\bm{r})=\sum_{\ell=0}^{\infty} \xi_{\ell}(r) L_{\ell}(\mu)\,,
\end{equation}
with 
\begin{equation}
\xi_{\ell}(r)=\frac{2 \ell+1}{2} \int_{-1}^1 \xi(r, \mu) L_{\ell}(\mu) d \mu \,,
\end{equation}
where $L_{\ell}$ is the Legendre polynomial of order $\ell$. Throughout this study, ignoring the more noisy subsequent orders, we only take into account $\ell = 0$, 2 and 4 multipoles, referred to as monopole, quadrupole, and hexadecapole, respectively, where $L_0 = 1$, $L_2=\left(3 \mu^2-1\right) / 2$ and $L_4=\left(35 \mu^4-30 \mu^2+3\right) / 8$. Due to the symmetry of object pairs, only even multipoles do not vanish. In practice,  
the pair counts are linearly binned with width of $\Delta r= 1.4$ Mpc in $r$ and $\Delta \mu =0.025$ in $\mu$ for the above estimation.

\begin{figure*}
 \includegraphics[width=0.35\textwidth]{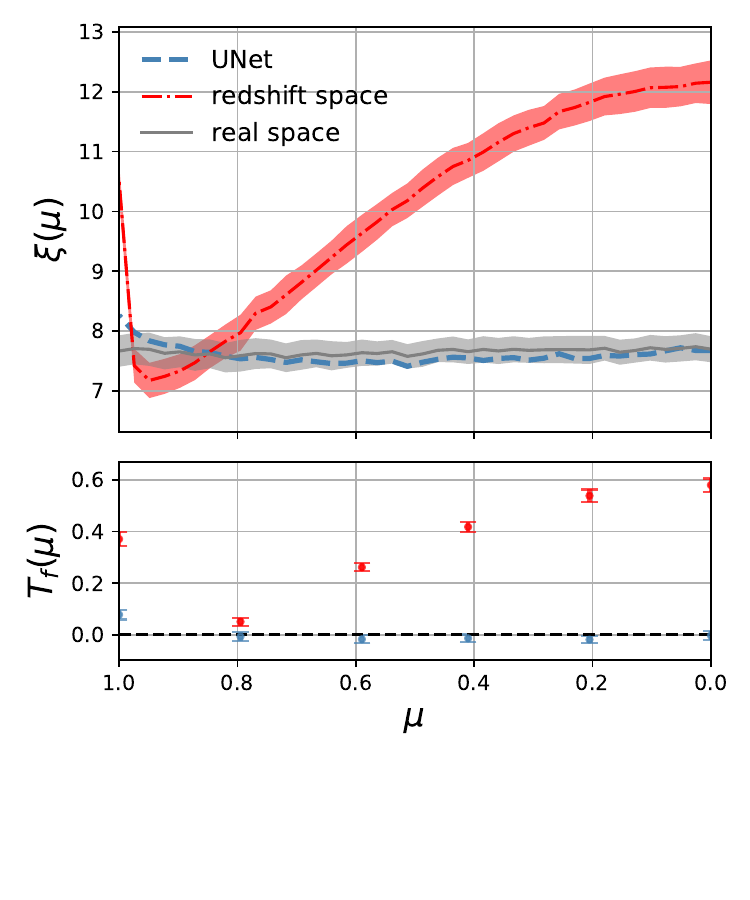}
	\includegraphics[width=0.35\textwidth]{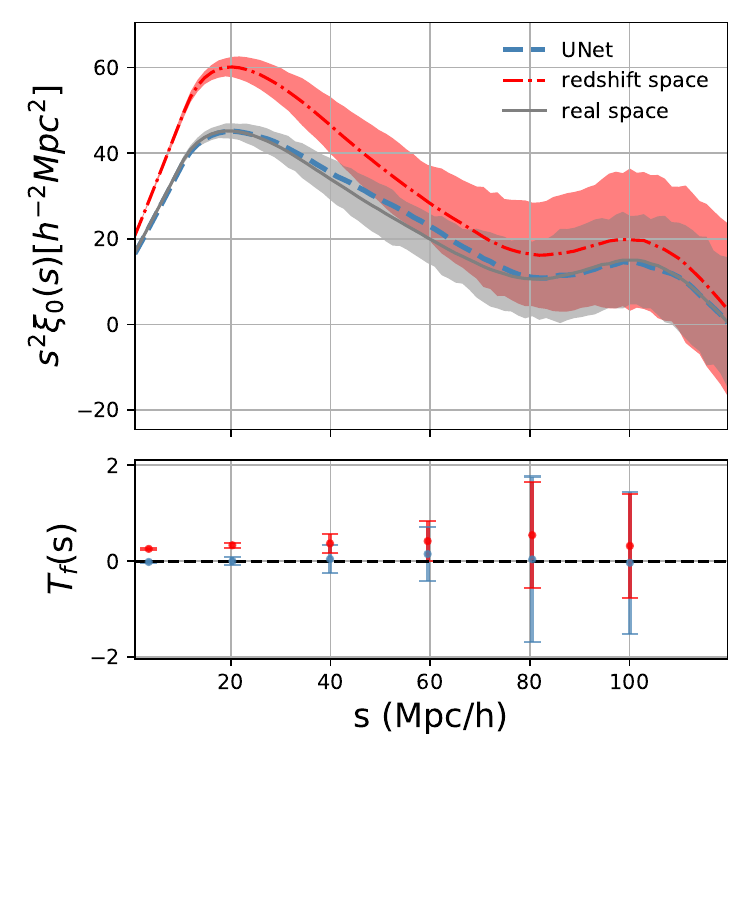}
    \includegraphics[width=0.35\textwidth]{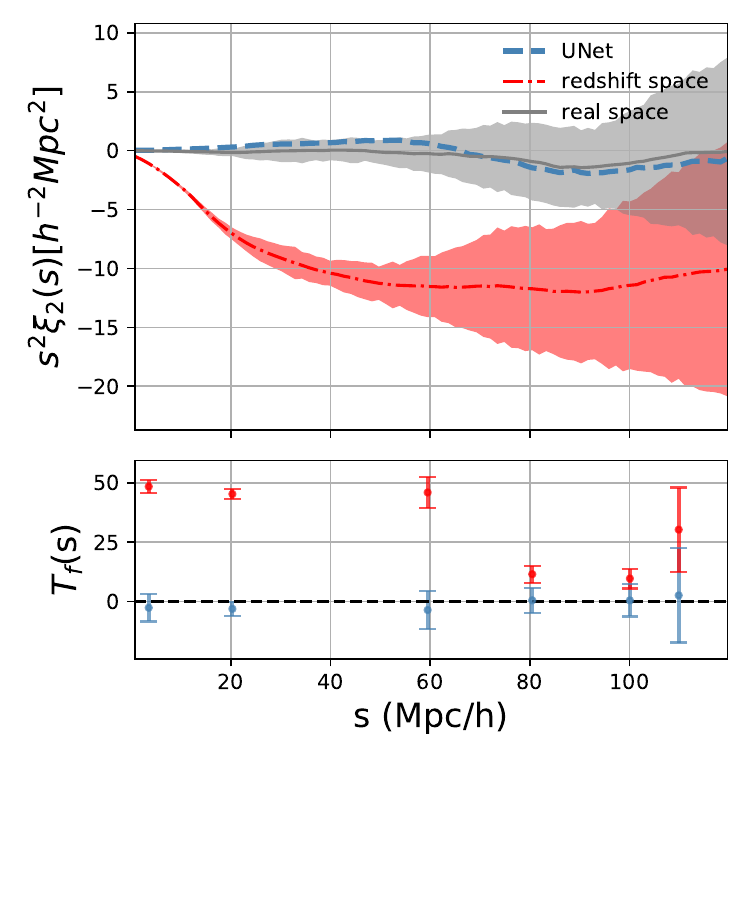}
    \includegraphics[width=0.35\textwidth]{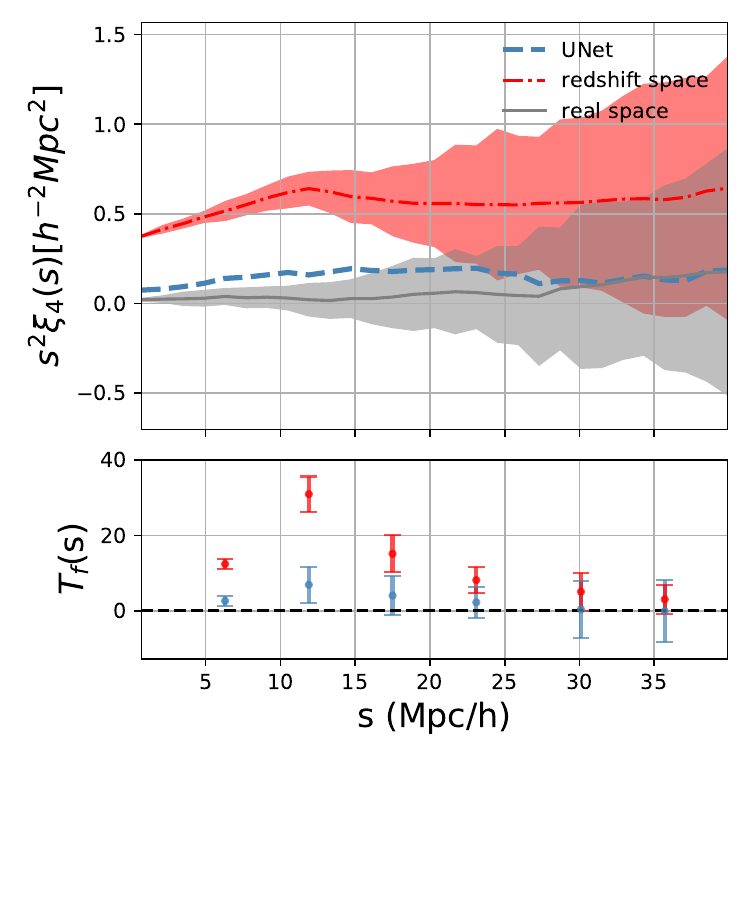}
    \caption{Comparison of the projected 2PCF, $\xi_{\rm 1D}(\mu)$ of the halo distribution (top-left) and the multipoles of 2PCF, including the monopole $\xi_0$ (top-right), the quadrupole $\xi_2$ (bottom-left) and the hexadecapole $\xi_4$ (bottom-right). The line and shaded area give the the mean and 1$\sigma$ standard deviation measured from the 27 test sets. The measurements from the halo (and subhalos) positions in real space (gray), the redshift space (pink) and the redshift space corrected by the UNet-predicted velocity field (blue) are shown, respectively.  The lower panel gives the relative deviation calculated by the ratio between the redshift-space measurements after the UNet correction and the real-space measurement. As seen, the real-space and UNet-corrected redshift-space measurements of 2PCFs are in good agreement through the relative deviation, with differences within the $1\sigma$ standard deviation.}
    \label{fig:v_2pcf}
\end{figure*}

We often measure a 1D 2PCF, $\xi(\mu)$, that projects the 2D correlation $\xi(r,\mu)$ along the $r$ axis, 
\begin{equation}\label{eq:1dxi}
\xi_{\rm 1D}(\mu) = \int_0^{\infty} dr \xi(r,\mu) dr\,.
\end{equation}

Fig.~\ref{fig:v_2pcf} shows the projected 1D 2PCF and the resultant monopole, quadrupole and hexadecapole of 2PCF. The line and the shaded area in each panel give the mean and 1$\sigma$ standard deviation, measured from 27 test sets. As seen, due to the Kaiser effect~\citep{kaiser1987clustering}, the enhancement of power over all scales is very remarkable. Meanwhile, in Tab.~\ref{tab:ximur}, we summarize the relative deviations  $T_f$ with and without UNet correction to the real-space measurements (the simulation truth), which are shown in the lower panels in Fig.~\ref{fig:v_2pcf}. 
Here, $T_f$ is evaluated using specific values from Tab.~\ref{tab:cfv}. However, since there may be some points where the calculations become inaccurate when the true value approaches zero, we estimate the uncertainty through error propagation to avoid this issue, which reads
\begin{equation}\label{eq:error}
\sigma_{T_f}^2\approx \sigma_f^2\left(\frac{\partial T_f}{\partial f}\right)_{\bar{f}}^2+\sigma^2_{f^{\rm true}}\left(\frac{\partial T_f}{\partial f^{\rm true}}\right)^2_{\bar{f}^{\rm true}}\,,
\end{equation}
where the bar denotes the average taken over samples.

Overall, the RSD effects are prominent, whereas they can be highly corrected by UNet with the differences within $1\sigma$ level. To highlight the changes due to RSD (Kasier and fingers-of-god effects), we show the 1D projected 2PCF, $\xi_{\rm 1D}$, with the variable $r$ integrated out, where the 2PCF without any correction strongly deviates from the real-space one (shown in the upper-left panel) with errors of tens of percent. However, the UNet model can accurately correct the RSD effects using the reconstructed velocity field, statistically leading to the correct clustering of halos in almost all directions with an error of 0--2\% (except for $\mu=1$ with the relative deviation of 0.08). More importantly, after the UNet correction, the results for $\xi_0$ at $\sim 100$ ${\rm Mpc}/h$ demonstrate that, the baryon acoustic oscillations (BAO) can be well recovered from redshift space, deriving a very close BAO peak to the real-space one, with about 4\% lower than the true one. Interestingly, the correction for the quadrupole leads to good agreement with the true real-space one not only on small scales,  but also on large scales. Even for $\xi_4$ with a much smaller signal-to-noise ratio than the other multipoles, the RSD effects can also be removed successfully, without any visible artificial effects such as oscillations and spikes.

\begin{table*}
	\centering
	\caption{Summary of the relative deviations with and without the UNet correction to the simulation truth (measured in real space) for 1D projected 2PCF, $\xi_{1D} (\mu)$, and the multipoles of 2PCF, $\xi_\ell (r)$ (shown in the lower panels of Fig.~\ref{fig:v_2pcf}). The mean and $1\sigma$ uncertainty are based on the 27 test sets, each with the box of side length 625 ${\rm Mpc}/h$. }
	\label{tab:ximur}
	\begin{tabular}{lcccccc} 
		\hline
		$\mu$  & 1.0 & 0.8 & 0.6  &  0.4 & 0.2 & 0.0 \\
		\hline
		$T_{\xi(\mu)}$ (UNet correction) & $0.08\pm 0.02$ & $-0.01\pm 0.02$ & $-0.02\pm 0.02$ & $-0.01\pm 0.01$ & $-0.02\pm 0.01$ & $-0.00\pm 0.02$ \\
		$T_{\xi(\mu)}$ (redshift space) & $0.37\pm 0.03$ & $0.05\pm 0.02$ & $0.26\pm 0.02$ & $0.42\pm 0.02$ & $0.54\pm 0.02$ & $0.58\pm 0.03$\\
		\hline\hline
  $r$ (${\rm Mpc}/h$)  & 5 & 20 & 40  &  60 & 80 & 100 \\
		\hline
		$T_{\xi_{0}(r)}$ (UNet correction)  & $0.03\pm 0.03$ & $0.00\pm 0.09$ & $0.04\pm 0.30$ & $0.15\pm 0.57$ & $0.04\pm 1.73$ & $-0.04\pm 1.48$\\
		$T_{\xi_{0}(r)}$ (redshift space)  & $0.26\pm0.02$ & $0.33\pm 0.06$ & $0.37\pm 0.20$ & $0.42\pm 0.42$ & $0.54\pm 1.10$ & $0.32\pm 1.09$\\
		\hline
  $r$ (${\rm Mpc}/h$)  & 5 & 20 & 60  &  80 & 100 & 110 \\
		\hline
  $T_{\xi_{2}(r)}$ (UNet correction)  & $-2.56\pm 5.78$ & $-3.02\pm 2.96$ & $-3.51\pm 8.16$ & $0.56\pm 5.30$ & $0.52\pm 6.77$ & $2.64\pm 19.83$\\
		$T_{\xi_{2}(r)}$ (redshift space)  & $48.44\pm2.90$ & $45.28\pm 1.96$ & $45.95\pm 6.49$ & $11.55\pm 3.56$ & $9.73\pm 4.20$ & $30.29 \pm 17.72$\\
		\hline
  $r$ (${\rm Mpc}/h$)  & 6 & 12 & 18  &  23 & 30 & 36 \\
		\hline
  $T_{\xi_{4}(r)}$ (UNet correction)  & $2.62\pm 1.39$ & $6.91\pm 4.81$ & $4.04\pm 5.25$ & $2.25\pm 4.08$ & $0.37\pm 7.65$ & $-0.09\pm 8.25$\\
		$T_{\xi_{4}(r)}$ (redshift space)  & $12.41\pm 1.24$ & $30.94\pm 4.69$ & $15.14\pm 4.94$ & $8.14\pm 3.43$ & $5.05\pm 5.00$ & $3.05\pm 3.78$\\
		\hline
	\end{tabular}
\end{table*}

\begin{figure*}
  \includegraphics[width=\textwidth]{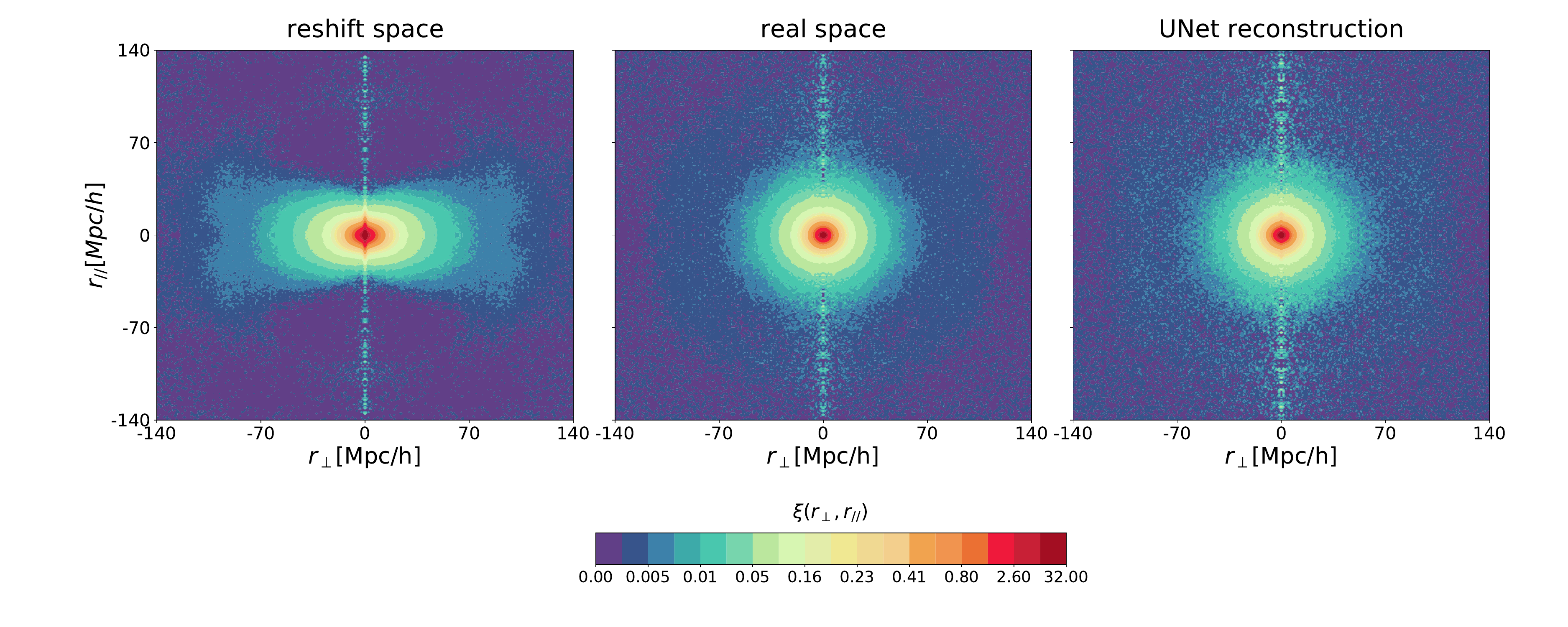}
    \caption{Contour of the anisotropic 2PCF $\xi(r_\perp, r_{\|})$. For comparison purposes, the left and middle panels display the 2PCF of redshift-space halos and the true 2PCF obtained from real space measurements, respectively. The right panel exhibits the 2PCF of redshift-space halos with UNet correction applied. The contours are produced by averaging over the results of 27 test sets in bins of size 1 ${\rm Mpc}/h$.} 
    \label{fig:v_rp}
\end{figure*}

The high-quality in the velocity reconstruction can be also appreciated in Fig.~\ref{fig:v_rp}, displaying the 2D anisotropic correlation function of redshift-space halos, $\xi (\bm{r})$, where the separation vector has been decomposed into line-of-sight and transverse separations such that $\bm{r}=(r_\perp, r_{\|})$. The contours are calculated based on the averaged result of $\xi (r_\perp, r_{\|})$ on the 27 test sets. As observed, without any RSD corrections, the anisotropic pattern is very distinctive. The Kaiser effect leads to 
galaxy clusters appearing "squashed" along the line-of-sight by a coherent infall onto galaxy clusters cancel some of the Hubble flow. Besides, the random velocities attained by galaxies in the non-linear regime produce the so-called fingers-of-god (FoG) effect, making structures elongated along the line of sight. As expected, the measured anisotropic correlation function in redshfit space present a BAO feature at $r \simeq 100$ ${\rm Mpc}/h$, as well as the impacts of the Kasier and the FoG effects. Compared with the UNet-corrected results, we find the isotropy of the correlation function is well recovered at all scales, demonstrating the effectiveness of the UNet approach.  Remarkably, our proposed method not only corrects Kasier effect on large scales, but also on small scales with $r \lesssim 10~h/{\rm Mpc}$, where the FoG effect is well removed, indicating that UNet can even accurately reconstruct the velocity field in the nonlinear regime.

\section{Conclusion}
\label{sect:conclusion}
3D velocity (and momentum) fields constructed by galaxies and clusters are very important in cosmology because they provide more information than the density field alone, and would help to improve/correct various cosmological measurements. High-fidelity reconstruction may even result in unexpected findings.

Accurate reconstruction is often a challenge for traditional reconstruction methods, typically relying on many assumptions and approximations. In this study, we have proposed an alternative scheme, a deep learning approach based on the UNet neural network to reconstruct the 3D velocity/momentum fields of halos. We find the UNet is well-suited for reconstructing such fields directly from the halo (and subhalos) density field, because the UNet is an elegant architecture that can effectively capture various features/structures of the fields at all scales and is very effective in transforming high-dimensional and structured inputs. Using multiple redshift-space halo number density fields in different mass ranges, the UNet learned how to transform halo density fields directly into velocity/momentum fields from the training data. We have performed a detailed validation with various statistics tests, and find the reconstructed velocity/momentum fields generally well agree the ground truth, but it's worth noting that there is a discernible difference in the power spectra. Further improvements could be made to address this variance by fine-tuning the UNet parameters and incorporating additional data sources to enhance the accuracy of the reconstruction.

 Furthermore, using the inferred velocity fields, the RSD effects can be well corrected by Unet. As an important application, we find that, the reconstructed velocity field directly provides a recovery of the real-space positions of individual halos, offering a reasonably well correction for the RSD effects down to a highly non-linear scale of 1.13 ${\rm Mpc}/h$, which is the Nyquist frequency ($k_{\mathrm{Ny}}=\pi N / L$) of the simulations. While there is room for improvement in some areas, this UNet-based approach is promising for many cosmological applications in terms of correcting the peculiar velocities. For example, the reconstruction of cosmic volume-weighted velocity suffers  
 severe sampling artifacts in measurements~\citep{2015PhRvD..91d3522Z,2015PhRvD..92h3527Y,2018ApJ...861...58C}. We will further extend our UNet model to tackle this long-standing problem and leave such a study for future work.   
 
 As the stage IV galaxy surveys will provide more detailed measurements of the LSS of the Universe than ever before, new computing technologies are being called upon to fully analyze these high-dimensional, massive amounts of data. Therefor, UNet-based neural networks promise to be a powerful tool to overcome the problems that traditional methods are difficult to deal with and to extract cosmological information in more depth and in a holistic manner.

\section*{Acknowledgements}
This work is supported by the National Key R\&D Program of China (2018YFA0404504, 2018YFA0404601, 2020YFC2201600), the Ministry of Science and Technology of China (2020SKA0110402, 2020SKA0110401, 2020SKA0110100), National Science Foundation of China (11890691, 11621303, 11653003, 11803094), the China Manned Space Project with No. CMS-CSST-2021 (A02, A03, B01), the Major Key Project of PCL, the 111 project No. B20019, the CAS Interdisciplinary Innovation Team (JCTD-2019-05), and the Science and Technology Program of Guangzhou, China (202002030360). We acknowledge the use of {\it Kunlun} cluster located at School of Physics and Astronomy, Sun Yat-Sen University. We also acknowledge the cosmology simulation database (CSD) in the National Basic Science Data Center (NBSDC) and its funds the NBSDC-DB-10.

\section*{Data Availability}

The BigMD simulation used in this paper is available in the CosmoSim data base (https://www.cosmosim.org/). The datasets generated in the current study are available from the corresponding authors on reasonable request.



\bibliographystyle{mnras}
\bibliography{mnras_template} 

\begin{thebibliography}{}
\makeatletter
\relax
\def\mn@urlcharsother{\let\do\@makeother \do\$\do\&\do\#\do\^\do\_\do\%\do\~}
\def\mn@doi{\begingroup\mn@urlcharsother \@ifnextchar [ {\mn@doi@}
  {\mn@doi@[]}}
\def\mn@doi@[#1]#2{\def\@tempa{#1}\ifx\@tempa\@empty \href
  {http://dx.doi.org/#2} {doi:#2}\else \href {http://dx.doi.org/#2} {#1}\fi
  \endgroup}
\def\mn@eprint#1#2{\mn@eprint@#1:#2::\@nil}
\def\mn@eprint@arXiv#1{\href {http://arxiv.org/abs/#1} {{\tt arXiv:#1}}}
\def\mn@eprint@dblp#1{\href {http://dblp.uni-trier.de/rec/bibtex/#1.xml}
  {dblp:#1}}
\def\mn@eprint@#1:#2:#3:#4\@nil{\def\@tempa {#1}\def\@tempb {#2}\def\@tempc
  {#3}\ifx \@tempc \@empty \let \@tempc \@tempb \let \@tempb \@tempa \fi \ifx
  \@tempb \@empty \def\@tempb {arXiv}\fi \@ifundefined
  {mn@eprint@\@tempb}{\@tempb:\@tempc}{\expandafter \expandafter \csname
  mn@eprint@\@tempb\endcsname \expandafter{\@tempc}}}

\bibitem[\protect\citeauthoryear{Aghanim et~al.}{Aghanim
  et~al.}{2020}]{Planck:2018vyg}
Aghanim N.,  et~al., 2020, \mn@doi [Astron. Astrophys.]
  {10.1051/0004-6361/201833910}, 641, A6

\bibitem[\protect\citeauthoryear{Alcock \& Paczy{\'n}ski}{Alcock \&
  Paczy{\'n}ski}{1979}]{ap}
Alcock C.,  Paczy{\'n}ski B.,  1979, Nature, 281, 358

\bibitem[\protect\citeauthoryear{{Ata} et~al.,}{{Ata}
  et~al.}{2017}]{2017MNRAS.467.3993A}
{Ata} M.,  et~al., 2017, \mn@doi [\mnras] {10.1093/mnras/stx178}, \href
  {https://ui.adsabs.harvard.edu/abs/2017MNRAS.467.3993A} {467, 3993}

\bibitem[\protect\citeauthoryear{{Bardeen}, {Bond}, {Kaiser}  \&
  {Szalay}}{{Bardeen} et~al.}{1986}]{1986Bardeen}
{Bardeen} J.~M.,  {Bond} J.~R.,  {Kaiser} N.,   {Szalay} A.~S.,  1986, \mn@doi
  [\apj] {10.1086/164143}, \href
  {https://ui.adsabs.harvard.edu/abs/1986ApJ...304...15B} {304, 15}

\bibitem[\protect\citeauthoryear{{Behroozi}, {Wechsler}  \& {Wu}}{{Behroozi}
  et~al.}{2013}]{2013ApJ...762..109B}
{Behroozi} P.~S.,  {Wechsler} R.~H.,   {Wu} H.-Y.,  2013, \mn@doi [\apj]
  {10.1088/0004-637X/762/2/109}, \href
  {https://ui.adsabs.harvard.edu/abs/2013ApJ...762..109B} {762, 109}

\bibitem[\protect\citeauthoryear{Berger \& Stein}{Berger \&
  Stein}{2019}]{Berger:2018aey}
Berger P.,  Stein G.,  2019, \mn@doi [Mon. Not. Roy. Astron. Soc.]
  {10.1093/mnras/sty2949}, 482, 2861

\bibitem[\protect\citeauthoryear{{Bernardeau}}{{Bernardeau}}{1992}]{VelocityRecon...1992ApJ...390L..61B}
{Bernardeau} F.,  1992, \mn@doi [\apjl] {10.1086/186372}, \href
  {https://ui.adsabs.harvard.edu/abs/1992ApJ...390L..61B} {390, L61}

\bibitem[\protect\citeauthoryear{{Bernardeau}, {Chodorowski}, {{\L}okas},
  {Stompor}  \& {Kudlicki}}{{Bernardeau}
  et~al.}{1999}]{VelocityRecon...1999MNRAS.309..543B}
{Bernardeau} F.,  {Chodorowski} M.~J.,  {{\L}okas} E.~L.,  {Stompor} R.,
  {Kudlicki} A.,  1999, \mn@doi [\mnras] {10.1046/j.1365-8711.1999.02856.x},
  \href {https://ui.adsabs.harvard.edu/abs/1999MNRAS.309..543B} {309, 543}

\bibitem[\protect\citeauthoryear{{Bilicki} \& {Chodorowski}}{{Bilicki} \&
  {Chodorowski}}{2008}]{VelocityRecon...2008MNRAS.391.1796B}
{Bilicki} M.,  {Chodorowski} M.~J.,  2008, \mn@doi [\mnras]
  {10.1111/j.1365-2966.2008.13988.x}, \href
  {https://ui.adsabs.harvard.edu/abs/2008MNRAS.391.1796B} {391, 1796}

\bibitem[\protect\citeauthoryear{{Branchini}, {Eldar}  \& {Nusser}}{{Branchini}
  et~al.}{2002}]{VelocityRecon...2002MNRAS.335...53B}
{Branchini} E.,  {Eldar} A.,   {Nusser} A.,  2002, \mn@doi [\mnras]
  {10.1046/j.1365-8711.2002.05611.x}, \href
  {https://ui.adsabs.harvard.edu/abs/2002MNRAS.335...53B} {335, 53}

\bibitem[\protect\citeauthoryear{{Caldeira}, {Wu}, {Nord}, {Avestruz},
  {Trivedi}  \& {Story}}{{Caldeira} et~al.}{2019}]{Caldeira:2018ojb}
{Caldeira} J.,  {Wu} W.~L.~K.,  {Nord} B.,  {Avestruz} C.,  {Trivedi} S.,
  {Story} K.~T.,  2019, \mn@doi [Astronomy and Computing]
  {10.1016/j.ascom.2019.100307}, \href
  {https://ui.adsabs.harvard.edu/abs/2019A&C....2800307C} {28, 100307}

\bibitem[\protect\citeauthoryear{{Carleo}, {Cirac}, {Cranmer}, {Daudet},
  {Schuld}, {Tishby}, {Vogt-Maranto}  \& {Zdeborov{\'a}}}{{Carleo}
  et~al.}{2019}]{Carleo:2019ptp}
{Carleo} G.,  {Cirac} I.,  {Cranmer} K.,  {Daudet} L.,  {Schuld} M.,  {Tishby}
  N.,  {Vogt-Maranto} L.,   {Zdeborov{\'a}} L.,  2019, \mn@doi [Reviews of
  Modern Physics] {10.1103/RevModPhys.91.045002}, \href
  {https://ui.adsabs.harvard.edu/abs/2019RvMP...91d5002C} {91, 045002}

\bibitem[\protect\citeauthoryear{{Carrick}, {Turnbull}, {Lavaux}  \&
  {Hudson}}{{Carrick} et~al.}{2015}]{2015MNRAS.450..317C}
{Carrick} J.,  {Turnbull} S.~J.,  {Lavaux} G.,   {Hudson} M.~J.,  2015, \mn@doi
  [\mnras] {10.1093/mnras/stv547}, \href
  {https://ui.adsabs.harvard.edu/abs/2015MNRAS.450..317C} {450, 317}

\bibitem[\protect\citeauthoryear{{Chardin}, {Uhlrich}, {Aubert}, {Deparis},
  {Gillet}, {Ocvirk}  \& {Lewis}}{{Chardin} et~al.}{2019}]{Chardin:2019euc}
{Chardin} J.,  {Uhlrich} G.,  {Aubert} D.,  {Deparis} N.,  {Gillet} N.,
  {Ocvirk} P.,   {Lewis} J.,  2019, \mn@doi [\mnras] {10.1093/mnras/stz2605},
  \href {https://ui.adsabs.harvard.edu/abs/2019MNRAS.490.1055C} {490, 1055}

\bibitem[\protect\citeauthoryear{{Chen}, {Zhang}, {Zheng}, {Yu}  \&
  {Jing}}{{Chen} et~al.}{2018}]{2018ApJ...861...58C}
{Chen} J.,  {Zhang} P.,  {Zheng} Y.,  {Yu} Y.,   {Jing} Y.,  2018, \mn@doi
  [\apj] {10.3847/1538-4357/aaca2f}, \href
  {https://ui.adsabs.harvard.edu/abs/2018ApJ...861...58C} {861, 58}

\bibitem[\protect\citeauthoryear{{Crittenden} \& {Turok}}{{Crittenden} \&
  {Turok}}{1996}]{1996PhRvL..76..575C}
{Crittenden} R.~G.,  {Turok} N.,  1996, \mn@doi [\prl]
  {10.1103/PhysRevLett.76.575}, \href
  {https://ui.adsabs.harvard.edu/abs/1996PhRvL..76..575C} {76, 575}

\bibitem[\protect\citeauthoryear{{Croft} \& {Gaztanaga}}{{Croft} \&
  {Gaztanaga}}{1997}]{VelocityRecon...1997MNRAS.285..793C}
{Croft} R. A.~C.,  {Gaztanaga} E.,  1997, \mn@doi [\mnras]
  {10.1093/mnras/285.4.793}, \href
  {https://ui.adsabs.harvard.edu/abs/1997MNRAS.285..793C} {285, 793}

\bibitem[\protect\citeauthoryear{{Djorgovski} \& {Davis}}{{Djorgovski} \&
  {Davis}}{1987}]{FundPlan...1987ApJ...313...59D}
{Djorgovski} S.,  {Davis} M.,  1987, \mn@doi [\apj] {10.1086/164948}, \href
  {https://ui.adsabs.harvard.edu/abs/1987ApJ...313...59D} {313, 59}

\bibitem[\protect\citeauthoryear{Dreissigacker, Sharma, Messenger, Zhao  \&
  Prix}{Dreissigacker et~al.}{2019}]{Dreissigacker:2019edy}
Dreissigacker C.,  Sharma R.,  Messenger C.,  Zhao R.,   Prix R.,  2019,
  \mn@doi [Phys. Rev.] {10.1103/PhysRevD.100.044009}, D100, 044009

\bibitem[\protect\citeauthoryear{{Dressler}, {Lynden-Bell}, {Burstein},
  {Davies}, {Faber}, {Terlevich}  \& {Wegner}}{{Dressler}
  et~al.}{1987}]{FundPlan...1987ApJ...313...42D}
{Dressler} A.,  {Lynden-Bell} D.,  {Burstein} D.,  {Davies} R.~L.,  {Faber}
  S.~M.,  {Terlevich} R.,   {Wegner} G.,  1987, \mn@doi [\apj]
  {10.1086/164947}, \href
  {https://ui.adsabs.harvard.edu/abs/1987ApJ...313...42D} {313, 42}

\bibitem[\protect\citeauthoryear{{Eisenstein} et~al.,}{{Eisenstein}
  et~al.}{2005}]{Eisenstein:2005su}
{Eisenstein} D.~J.,  et~al., 2005, \mn@doi [\apj] {10.1086/466512}, \href
  {https://ui.adsabs.harvard.edu/abs/2005ApJ...633..560E} {633, 560}

\bibitem[\protect\citeauthoryear{{Eisenstein}, {Seo}, {Sirko}  \&
  {Spergel}}{{Eisenstein} et~al.}{2007}]{Eisenstein2007BAOReconstruction}
{Eisenstein} D.~J.,  {Seo} H.-J.,  {Sirko} E.,   {Spergel} D.~N.,  2007,
  \mn@doi [\apj] {10.1086/518712}, \href
  {https://ui.adsabs.harvard.edu/abs/2007ApJ...664..675E} {664, 675}

\bibitem[\protect\citeauthoryear{{Fang}, {Forero-Romero}, {Rossi}, {Li}  \&
  {Feng}}{{Fang} et~al.}{2019}]{Fang2019}
{Fang} F.,  {Forero-Romero} J.,  {Rossi} G.,  {Li} X.-D.,   {Feng} L.-L.,
  2019, \mn@doi [\mnras] {10.1093/mnras/stz773}, \href
  {https://ui.adsabs.harvard.edu/abs/2019MNRAS.485.5276F} {485, 5276}

\bibitem[\protect\citeauthoryear{{Fluri}, {Kacprzak}, {Lucchi}, {Refregier},
  {Amara}, {Hofmann}  \& {Schneider}}{{Fluri} et~al.}{2019}]{Fluri:2019qtp}
{Fluri} J.,  {Kacprzak} T.,  {Lucchi} A.,  {Refregier} A.,  {Amara} A.,
  {Hofmann} T.,   {Schneider} A.,  2019, \mn@doi [\prd]
  {10.1103/PhysRevD.100.063514}, \href
  {https://ui.adsabs.harvard.edu/abs/2019PhRvD.100f3514F} {100, 063514}

\bibitem[\protect\citeauthoryear{Forero-Romero, Hoffman, Gottl{\"o}ber, Klypin
  \& Yepes}{Forero-Romero et~al.}{2009}]{forero2009dynamical}
Forero-Romero J.,  Hoffman Y.,  Gottl{\"o}ber S.,  Klypin A.,   Yepes G.,
  2009, Monthly Notices of the Royal Astronomical Society, 396, 1815

\bibitem[\protect\citeauthoryear{Forero-Romero, Contreras  \&
  Padilla}{Forero-Romero et~al.}{2014}]{forero2014cosmic}
Forero-Romero J.~E.,  Contreras S.,   Padilla N.,  2014, Monthly Notices of the
  Royal Astronomical Society, 443, 1090

\bibitem[\protect\citeauthoryear{{Ganeshaiah Veena}, {Lilow}  \&
  {Nusser}}{{Ganeshaiah Veena} et~al.}{2022}]{2022arXiv221206439G}
{Ganeshaiah Veena} P.,  {Lilow} R.,   {Nusser} A.,  2022, \mn@doi [arXiv
  e-prints] {10.48550/arXiv.2212.06439}, \href
  {https://ui.adsabs.harvard.edu/abs/2022arXiv221206439G} {p. arXiv:2212.06439}

\bibitem[\protect\citeauthoryear{{Gebhard}, {Kilbertus}, {Harry}  \&
  {Sch{\"o}lkopf}}{{Gebhard} et~al.}{2019}]{Gebhard:2019ldz}
{Gebhard} T.~D.,  {Kilbertus} N.,  {Harry} I.,   {Sch{\"o}lkopf} B.,  2019,
  \mn@doi [\prd] {10.1103/PhysRevD.100.063015}, \href
  {https://ui.adsabs.harvard.edu/abs/2019PhRvD.100f3015G} {100, 063015}

\bibitem[\protect\citeauthoryear{Gillet, Mesinger, Greig, Liu  \& Ucci}{Gillet
  et~al.}{2019}]{Gillet:2018fgb}
Gillet N.,  Mesinger A.,  Greig B.,  Liu A.,   Ucci G.,  2019, \mn@doi [Mon.
  Not. Roy. Astron. Soc.] {10.1093/mnras/stz010}, 484, 282

\bibitem[\protect\citeauthoryear{Gupta, Matilla, Hsu  \& Haiman}{Gupta
  et~al.}{2018}]{Gupta:2018eev}
Gupta A.,  Matilla J. M.~Z.,  Hsu D.,   Haiman Z.,  2018, \mn@doi [Phys. Rev.]
  {10.1103/PhysRevD.97.103515}, D97, 103515

\bibitem[\protect\citeauthoryear{Hahn, Porciani, Carollo  \& Dekel}{Hahn
  et~al.}{2007}]{hahn2007properties}
Hahn O.,  Porciani C.,  Carollo C.~M.,   Dekel A.,  2007, Monthly Notices of
  the Royal Astronomical Society, 375, 489

\bibitem[\protect\citeauthoryear{Hassan, Liu, Kohn  \& La~Plante}{Hassan
  et~al.}{2019}]{Hassan:2018bbm}
Hassan S.,  Liu A.,  Kohn S.,   La~Plante P.,  2019, \mn@doi [Mon. Not. Roy.
  Astron. Soc.] {10.1093/mnras/sty3282}, 483, 2524

\bibitem[\protect\citeauthoryear{{Hassan}, {Andrianomena}  \&
  {Doughty}}{{Hassan} et~al.}{2020}]{Hassan:2019cal}
{Hassan} S.,  {Andrianomena} S.,   {Doughty} C.,  2020, \mn@doi [\mnras]
  {10.1093/mnras/staa1151}, \href
  {https://ui.adsabs.harvard.edu/abs/2020MNRAS.494.5761H} {494, 5761}

\bibitem[\protect\citeauthoryear{He, Li, Feng, Ho, Ravanbakhsh, Chen  \&
  Póczos}{He et~al.}{2019}]{He:2018ggn}
He S.,  Li Y.,  Feng Y.,  Ho S.,  Ravanbakhsh S.,  Chen W.,   Póczos B.,
  2019, \mn@doi [Proc. Nat. Acad. Sci.] {10.1073/pnas.1821458116}, 116, 13825

\bibitem[\protect\citeauthoryear{Hoffman, Metuki, Yepes, Gottl{\"o}ber,
  Forero-Romero, Libeskind  \& Knebe}{Hoffman
  et~al.}{2012}]{hoffman2012kinematic}
Hoffman Y.,  Metuki O.,  Yepes G.,  Gottl{\"o}ber S.,  Forero-Romero J.~E.,
  Libeskind N.~I.,   Knebe A.,  2012, Monthly Notices of the Royal Astronomical
  Society, 425, 2049

\bibitem[\protect\citeauthoryear{{Hong}, {Jeong}, {Hwang}  \& {Kim}}{{Hong}
  et~al.}{2021}]{2021ApJ...913...76H}
{Hong} S.~E.,  {Jeong} D.,  {Hwang} H.~S.,   {Kim} J.,  2021, \mn@doi [\apj]
  {10.3847/1538-4357/abf040}, \href
  {https://ui.adsabs.harvard.edu/abs/2021ApJ...913...76H} {913, 76}

\bibitem[\protect\citeauthoryear{Ishida et~al.}{Ishida
  et~al.}{2019}]{Ishida:2018uqu}
Ishida E. E.~O.,  et~al., 2019, \mn@doi [Mon. Not. Roy. Astron. Soc.]
  {10.1093/mnras/sty3015}, 483, 2

\bibitem[\protect\citeauthoryear{Jackson}{Jackson}{1972}]{jackson1972critique}
Jackson J.,  1972, Monthly Notices of the Royal Astronomical Society, 156, 1P

\bibitem[\protect\citeauthoryear{{Jasche} \& {Lavaux}}{{Jasche} \&
  {Lavaux}}{2019}]{2019A&A...625A..64J}
{Jasche} J.,  {Lavaux} G.,  2019, \mn@doi [\aap] {10.1051/0004-6361/201833710},
  \href {https://ui.adsabs.harvard.edu/abs/2019A&A...625A..64J} {625, A64}

\bibitem[\protect\citeauthoryear{{Jeffrey}, {Lanusse}, {Lahav}  \&
  {Starck}}{{Jeffrey} et~al.}{2020}]{Jeffrey:2019fag}
{Jeffrey} N.,  {Lanusse} F.,  {Lahav} O.,   {Starck} J.-L.,  2020, \mn@doi
  [\mnras] {10.1093/mnras/staa127}, \href
  {https://ui.adsabs.harvard.edu/abs/2020MNRAS.492.5023J} {492, 5023}

\bibitem[\protect\citeauthoryear{{Jennings} \& {Jennings}}{{Jennings} \&
  {Jennings}}{2015}]{VelocityRecon...2015MNRAS.449.3407J}
{Jennings} E.,  {Jennings} D.,  2015, \mn@doi [\mnras] {10.1093/mnras/stv535},
  \href {https://ui.adsabs.harvard.edu/abs/2015MNRAS.449.3407J} {449, 3407}

\bibitem[\protect\citeauthoryear{Jennings, Watkinson, Abdalla  \&
  McEwen}{Jennings et~al.}{2019}]{Jennings:2018eko}
Jennings W.~D.,  Watkinson C.~A.,  Abdalla F.~B.,   McEwen J.~D.,  2019,
  \mn@doi [Mon. Not. Roy. Astron. Soc.] {10.1093/mnras/sty3168}, 483, 2907

\bibitem[\protect\citeauthoryear{{Kaiser}}{{Kaiser}}{1987}]{kaiser1987clustering}
{Kaiser} N.,  1987, \mn@doi [\mnras] {10.1093/mnras/227.1.1}, \href
  {https://ui.adsabs.harvard.edu/abs/1987MNRAS.227....1K} {227, 1}

\bibitem[\protect\citeauthoryear{{Kingma} \& {Ba}}{{Kingma} \&
  {Ba}}{2014}]{2014arXiv1412.6980K}
{Kingma} D.~P.,  {Ba} J.,  2014, arXiv e-prints, \href
  {https://ui.adsabs.harvard.edu/abs/2014arXiv1412.6980K} {p. arXiv:1412.6980}

\bibitem[\protect\citeauthoryear{{Kitaura}, {Angulo}, {Hoffman}  \&
  {Gottl{\"o}ber}}{{Kitaura}
  et~al.}{2012}]{VelocityRecon...2012MNRAS.425.2422K}
{Kitaura} F.-S.,  {Angulo} R.~E.,  {Hoffman} Y.,   {Gottl{\"o}ber} S.,  2012,
  \mn@doi [\mnras] {10.1111/j.1365-2966.2012.21589.x}, \href
  {https://ui.adsabs.harvard.edu/abs/2012MNRAS.425.2422K} {425, 2422}

\bibitem[\protect\citeauthoryear{{Klypin}, {Yepes}, {Gottl{\"o}ber}, {Prada}
  \& {He{\ss}}}{{Klypin} et~al.}{2016}]{2016MNRAS.457.4340K}
{Klypin} A.,  {Yepes} G.,  {Gottl{\"o}ber} S.,  {Prada} F.,   {He{\ss}} S.,
  2016, \mn@doi [\mnras] {10.1093/mnras/stw248}, \href
  {https://ui.adsabs.harvard.edu/abs/2016MNRAS.457.4340K} {457, 4340}

\bibitem[\protect\citeauthoryear{{Kudlicki}, {Chodorowski}, {Plewa}  \&
  {R{\'o}{\.z}yczka}}{{Kudlicki}
  et~al.}{2000}]{VelocityRecon...2000MNRAS.316..464K}
{Kudlicki} A.,  {Chodorowski} M.,  {Plewa} T.,   {R{\'o}{\.z}yczka} M.,  2000,
  \mn@doi [\mnras] {10.1046/j.1365-8711.2000.03463.x}, \href
  {https://ui.adsabs.harvard.edu/abs/2000MNRAS.316..464K} {316, 464}

\bibitem[\protect\citeauthoryear{La~Plante \& Ntampaka}{La~Plante \&
  Ntampaka}{2018}]{LaPlante:2018pst}
La~Plante P.,  Ntampaka M.,  2018, \mn@doi [Astrophys. J.]
  {10.3847/1538-4357/ab2983}, 810, 110

\bibitem[\protect\citeauthoryear{{Landy} \& {Szalay}}{{Landy} \&
  {Szalay}}{1993}]{1993ApJ...412...64L}
{Landy} S.~D.,  {Szalay} A.~S.,  1993, \mn@doi [\apj] {10.1086/172900}, \href
  {https://ui.adsabs.harvard.edu/abs/1993ApJ...412...64L} {412, 64}

\bibitem[\protect\citeauthoryear{{Lavaux}, {Mohayaee}, {Colombi}, {Tully},
  {Bernardeau}  \& {Silk}}{{Lavaux}
  et~al.}{2008}]{VelocityRecon...2008MNRAS.383.1292L}
{Lavaux} G.,  {Mohayaee} R.,  {Colombi} S.,  {Tully} R.~B.,  {Bernardeau} F.,
  {Silk} J.,  2008, \mn@doi [\mnras] {10.1111/j.1365-2966.2007.12539.x}, \href
  {https://ui.adsabs.harvard.edu/abs/2008MNRAS.383.1292L} {383, 1292}

\bibitem[\protect\citeauthoryear{{Lazanu}}{{Lazanu}}{2021}]{2021JCAP...09..039L}
{Lazanu} A.,  2021, \mn@doi [\jcap] {10.1088/1475-7516/2021/09/039}, \href
  {https://ui.adsabs.harvard.edu/abs/2021JCAP...09..039L} {2021, 039}

\bibitem[\protect\citeauthoryear{{Li}, {Park}, {Forero-Romero}  \& {Kim}}{{Li}
  et~al.}{2014}]{Li2014}
{Li} X.-D.,  {Park} C.,  {Forero-Romero} J.~E.,   {Kim} J.,  2014, \mn@doi
  [\apj] {10.1088/0004-637X/796/2/137}, \href
  {http://adsabs.harvard.edu/abs/2014ApJ...796..137L} {796, 137}

\bibitem[\protect\citeauthoryear{Li, Park, Sabiu  \& Kim}{Li
  et~al.}{2015}]{Li2015}
Li X.-D.,  Park C.,  Sabiu C.~G.,   Kim J.,  2015, Monthly Notices of the Royal
  Astronomical Society, 450, 807

\bibitem[\protect\citeauthoryear{{Li}, {Park}, {Sabiu}, {Park}, {Weinberg},
  {Schneider}, {Kim}  \& {Hong}}{{Li} et~al.}{2016}]{Li2016}
{Li} X.-D.,  {Park} C.,  {Sabiu} C.~G.,  {Park} H.,  {Weinberg} D.~H.,
  {Schneider} D.~P.,  {Kim} J.,   {Hong} S.~E.,  2016, \mn@doi [\apj]
  {10.3847/0004-637X/832/2/103}, \href
  {https://ui.adsabs.harvard.edu/abs/2016ApJ...832..103L} {832, 103}

\bibitem[\protect\citeauthoryear{{Li}, {Li}  \& {Zhang}}{{Li}
  et~al.}{2019}]{Li:2019ybe}
{Li} S.-Y.,  {Li} Y.-L.,   {Zhang} T.-J.,  2019, \mn@doi [Research in Astronomy
  and Astrophysics] {10.1088/1674-4527/19/9/137}, \href
  {https://ui.adsabs.harvard.edu/abs/2019RAA....19..137L} {19, 137}

\bibitem[\protect\citeauthoryear{Li, Ni, Croft, Matteo, Bird  \& Feng}{Li
  et~al.}{2021}]{Li_2021}
Li Y.,  Ni Y.,  Croft R. A.~C.,  Matteo T.~D.,  Bird S.,   Feng Y.,  2021,
  \mn@doi [Proceedings of the National Academy of Sciences]
  {10.1073/pnas.2022038118}, 118

\bibitem[\protect\citeauthoryear{{Lilow} \& {Nusser}}{{Lilow} \&
  {Nusser}}{2021}]{2021MNRAS.507.1557L}
{Lilow} R.,  {Nusser} A.,  2021, \mn@doi [\mnras] {10.1093/mnras/stab2009},
  \href {https://ui.adsabs.harvard.edu/abs/2021MNRAS.507.1557L} {507, 1557}

\bibitem[\protect\citeauthoryear{{Linder}}{{Linder}}{2005}]{2005PhRvD..72d3529L}
{Linder} E.~V.,  2005, \mn@doi [\prd] {10.1103/PhysRevD.72.043529}, \href
  {https://ui.adsabs.harvard.edu/abs/2005PhRvD..72d3529L} {72, 043529}

\bibitem[\protect\citeauthoryear{Lochner, McEwen, Peiris, Lahav  \&
  Winter}{Lochner et~al.}{2016}]{Lochner:2016hbn}
Lochner M.,  McEwen J.~D.,  Peiris H.~V.,  Lahav O.,   Winter M.~K.,  2016,
  \mn@doi [Astrophys. J. Suppl.] {10.3847/0067-0049/225/2/31}, 225, 31

\bibitem[\protect\citeauthoryear{Lucie-Smith, Peiris, Pontzen  \&
  Lochner}{Lucie-Smith et~al.}{2018}]{Lucie-Smith:2018smo}
Lucie-Smith L.,  Peiris H.~V.,  Pontzen A.,   Lochner M.,  2018, \mn@doi [Mon.
  Not. Roy. Astron. Soc.] {10.1093/mnras/sty1719}, 479, 3405

\bibitem[\protect\citeauthoryear{{Lucie-Smith}, {Peiris}  \&
  {Pontzen}}{{Lucie-Smith} et~al.}{2019}]{Lucie-Smith:2019hdl}
{Lucie-Smith} L.,  {Peiris} H.~V.,   {Pontzen} A.,  2019, \mn@doi [\mnras]
  {10.1093/mnras/stz2599}, \href
  {https://ui.adsabs.harvard.edu/abs/2019MNRAS.490..331L} {490, 331}

\bibitem[\protect\citeauthoryear{Makinen, Lancaster, Villaescusa-Navarro,
  Melchior, Ho, Perreault-Levasseur  \& Spergel}{Makinen
  et~al.}{2021}]{Makinen:2020gvh}
Makinen T.~L.,  Lancaster L.,  Villaescusa-Navarro F.,  Melchior P.,  Ho S.,
  Perreault-Levasseur L.,   Spergel D.~N.,  2021, \mn@doi [JCAP]
  {10.1088/1475-7516/2021/04/081}, 04, 081

\bibitem[\protect\citeauthoryear{{Mao}, {Wang}, {Li}, {Cai}, {Falck},
  {Neyrinck}  \& {Szalay}}{{Mao}
  et~al.}{2020}]{AIBAORecon...2020arXiv200210218M}
{Mao} T.-X.,  {Wang} J.,  {Li} B.,  {Cai} Y.-C.,  {Falck} B.,  {Neyrinck} M.,
  {Szalay} A.,  2020, arXiv e-prints, \href
  {https://ui.adsabs.harvard.edu/abs/2020arXiv200210218M} {p. arXiv:2002.10218}

\bibitem[\protect\citeauthoryear{{Masters}, {Springob}, {Haynes}  \&
  {Giovanelli}}{{Masters} et~al.}{2006}]{TullyFisher...2006ApJ...653..861M}
{Masters} K.~L.,  {Springob} C.~M.,  {Haynes} M.~P.,   {Giovanelli} R.,  2006,
  \mn@doi [\apj] {10.1086/508924}, \href
  {https://ui.adsabs.harvard.edu/abs/2006ApJ...653..861M} {653, 861}

\bibitem[\protect\citeauthoryear{{Masters}, {Springob}  \& {Huchra}}{{Masters}
  et~al.}{2008}]{TullyFisher...2008AJ....135.1738M}
{Masters} K.~L.,  {Springob} C.~M.,   {Huchra} J.~P.,  2008, \mn@doi [\aj]
  {10.1088/0004-6256/135/5/1738}, \href
  {https://ui.adsabs.harvard.edu/abs/2008AJ....135.1738M} {135, 1738}

\bibitem[\protect\citeauthoryear{{Mathews}, {Rose}, {Garnavich}, {Yamazaki}  \&
  {Kajino}}{{Mathews} et~al.}{2016}]{SNIaflow...2016ApJ...827...60M}
{Mathews} G.~J.,  {Rose} B.~M.,  {Garnavich} P.~M.,  {Yamazaki} D.~G.,
  {Kajino} T.,  2016, \mn@doi [\apj] {10.3847/0004-637X/827/1/60}, \href
  {https://ui.adsabs.harvard.edu/abs/2016ApJ...827...60M} {827, 60}

\bibitem[\protect\citeauthoryear{Mehta, Bukov, Wang, Day, Richardson, Fisher
  \& Schwab}{Mehta et~al.}{2019}]{Mehta:2018dln}
Mehta P.,  Bukov M.,  Wang C.-H.,  Day A. G.~R.,  Richardson C.,  Fisher C.~K.,
    Schwab D.~J.,  2019, \mn@doi [Phys. Rept.] {10.1016/j.physrep.2019.03.001},
  810, 1

\bibitem[\protect\citeauthoryear{Merten, Giocoli, Baldi, Meneghetti, Peel,
  Lalande, Starck  \& Pettorino}{Merten et~al.}{2019}]{Merten:2018bgr}
Merten J.,  Giocoli C.,  Baldi M.,  Meneghetti M.,  Peel A.,  Lalande F.,
  Starck J.-L.,   Pettorino V.,  2019, \mn@doi [Mon. Not. Roy. Astron. Soc.]
  {10.1093/mnras/stz972}, 487, 104

\bibitem[\protect\citeauthoryear{{Mishra}, {Reddy}  \& {Nigam}}{{Mishra}
  et~al.}{2019}]{Mishra:2019sep}
{Mishra} A.,  {Reddy} P.,   {Nigam} R.,  2019, arXiv e-prints, \href
  {https://ui.adsabs.harvard.edu/abs/2019arXiv190804682M} {p. arXiv:1908.04682}

\bibitem[\protect\citeauthoryear{Modi, Feng  \& Seljak}{Modi
  et~al.}{2018}]{Modi:2018cfi}
Modi C.,  Feng Y.,   Seljak U.,  2018, \mn@doi [JCAP]
  {10.1088/1475-7516/2018/10/028}, 1810, 028

\bibitem[\protect\citeauthoryear{{Mohayaee} \& {Tully}}{{Mohayaee} \&
  {Tully}}{2005}]{VelocityRecon...2005ApJ...635L.113M}
{Mohayaee} R.,  {Tully} R.~B.,  2005, \mn@doi [\apjl] {10.1086/499774}, \href
  {https://ui.adsabs.harvard.edu/abs/2005ApJ...635L.113M} {635, L113}

\bibitem[\protect\citeauthoryear{{Moss}}{{Moss}}{2018}]{Moss:2018tug}
{Moss} A.,  2018, arXiv e-prints, \href
  {https://ui.adsabs.harvard.edu/abs/2018arXiv181006441M} {p. arXiv:1810.06441}

\bibitem[\protect\citeauthoryear{{M{\"u}nchmeyer} \& {Smith}}{{M{\"u}nchmeyer}
  \& {Smith}}{2019}]{Munchmeyer:2019kng}
{M{\"u}nchmeyer} M.,  {Smith} K.~M.,  2019, arXiv e-prints, \href
  {https://ui.adsabs.harvard.edu/abs/2019arXiv190505846M} {p. arXiv:1905.05846}

\bibitem[\protect\citeauthoryear{{Muthukrishna}, {Parkinson}  \&
  {Tucker}}{{Muthukrishna} et~al.}{2019}]{Muthukrishna:2019wpf}
{Muthukrishna} D.,  {Parkinson} D.,   {Tucker} B.~E.,  2019, \mn@doi [\apj]
  {10.3847/1538-4357/ab48f4}, \href
  {https://ui.adsabs.harvard.edu/abs/2019ApJ...885...85M} {885, 85}

\bibitem[\protect\citeauthoryear{{Ni}, {Li}, {Lachance}, {Croft}, {Di Matteo},
  {Bird}  \& {Feng}}{{Ni} et~al.}{2021}]{2021MNRAS.507.1021N}
{Ni} Y.,  {Li} Y.,  {Lachance} P.,  {Croft} R. A.~C.,  {Di Matteo} T.,  {Bird}
  S.,   {Feng} Y.,  2021, \mn@doi [\mnras] {10.1093/mnras/stab2113}, \href
  {https://ui.adsabs.harvard.edu/abs/2021MNRAS.507.1021N} {507, 1021}

\bibitem[\protect\citeauthoryear{{Ntampaka} et~al.,}{{Ntampaka}
  et~al.}{2019}]{Ntampaka:2019udw}
{Ntampaka} M.,  et~al., 2019, \baas, \href
  {https://ui.adsabs.harvard.edu/abs/2019BAAS...51c..14N} {51, 14}

\bibitem[\protect\citeauthoryear{{Nusser} \& {Davis}}{{Nusser} \&
  {Davis}}{1994}]{1994ApJ...421L...1N}
{Nusser} A.,  {Davis} M.,  1994, \mn@doi [\apjl] {10.1086/187172}, \href
  {https://ui.adsabs.harvard.edu/abs/1994ApJ...421L...1N} {421, L1}

\bibitem[\protect\citeauthoryear{{Nusser}, {Dekel}, {Bertschinger}  \&
  {Blumenthal}}{{Nusser} et~al.}{1991}]{VelocityRecon...1991ApJ...379....6N}
{Nusser} A.,  {Dekel} A.,  {Bertschinger} E.,   {Blumenthal} G.~R.,  1991,
  \mn@doi [\apj] {10.1086/170480}, \href
  {https://ui.adsabs.harvard.edu/abs/1991ApJ...379....6N} {379, 6}

\bibitem[\protect\citeauthoryear{{Okumura}, {Seljak}, {Vlah}  \&
  {Desjacques}}{{Okumura} et~al.}{2014}]{2014JCAP...05..003O}
{Okumura} T.,  {Seljak} U.,  {Vlah} Z.,   {Desjacques} V.,  2014, \mn@doi
  [\jcap] {10.1088/1475-7516/2014/05/003}, \href
  {https://ui.adsabs.harvard.edu/abs/2014JCAP...05..003O} {2014, 003}

\bibitem[\protect\citeauthoryear{{Pan}, {Liu}, {Forero-Romero}, {Sabiu}, {Li},
  {Miao}  \& {Li}}{{Pan} et~al.}{2020}]{Li2020...ML...2020SCPMA..63k0412P}
{Pan} S.,  {Liu} M.,  {Forero-Romero} J.,  {Sabiu} C.~G.,  {Li} Z.,  {Miao} H.,
    {Li} X.-D.,  2020, \mn@doi [Science China Physics, Mechanics, and
  Astronomy] {10.1007/s11433-020-1586-3}, \href
  {https://ui.adsabs.harvard.edu/abs/2020SCPMA..63k0412P} {63, 110412}

\bibitem[\protect\citeauthoryear{Peel, Lalande, Starck, Pettorino, Merten,
  Giocoli, Meneghetti  \& Baldi}{Peel et~al.}{2019}]{Peel:2018aei}
Peel A.,  Lalande F.,  Starck J.-L.,  Pettorino V.,  Merten J.,  Giocoli C.,
  Meneghetti M.,   Baldi M.,  2019, \mn@doi [Phys. Rev.]
  {10.1103/PhysRevD.100.023508}, D100, 023508

\bibitem[\protect\citeauthoryear{Perraudin, Defferrard, Kacprzak  \&
  Sgier}{Perraudin et~al.}{2019}]{Perraudin:2018rbt}
Perraudin N.,  Defferrard M.,  Kacprzak T.,   Sgier R.,  2019, \mn@doi [Astron.
  Comput.] {10.1016/j.ascom.2019.03.004}, 27, 130

\bibitem[\protect\citeauthoryear{{Pfeffer}, {Breysse}  \& {Stein}}{{Pfeffer}
  et~al.}{2019}]{Pfeffer:2019pca}
{Pfeffer} D.~N.,  {Breysse} P.~C.,   {Stein} G.,  2019, arXiv e-prints, \href
  {https://ui.adsabs.harvard.edu/abs/2019arXiv190510376P} {p. arXiv:1905.10376}

\bibitem[\protect\citeauthoryear{{Phillips}}{{Phillips}}{1993}]{1993ApJ...413L.105P}
{Phillips} M.~M.,  1993, \mn@doi [\apjl] {10.1086/186970}, \href
  {https://ui.adsabs.harvard.edu/abs/1993ApJ...413L.105P} {413, L105}

\bibitem[\protect\citeauthoryear{{Radburn-Smith}, {Lucey}  \&
  {Hudson}}{{Radburn-Smith} et~al.}{2004}]{SNIaflow...2004MNRAS.355.1378R}
{Radburn-Smith} D.~J.,  {Lucey} J.~R.,   {Hudson} M.~J.,  2004, \mn@doi
  [\mnras] {10.1111/j.1365-2966.2004.08420.x}, \href
  {https://ui.adsabs.harvard.edu/abs/2004MNRAS.355.1378R} {355, 1378}

\bibitem[\protect\citeauthoryear{{Ramanah}, {Lavaux}, {Jasche}  \&
  {Wandelt}}{{Ramanah} et~al.}{2019a}]{KR2018}
{Ramanah} D.~K.,  {Lavaux} G.,  {Jasche} J.,   {Wandelt} B.~D.,  2019a, \mn@doi
  [\aap] {10.1051/0004-6361/201834117}, \href
  {https://ui.adsabs.harvard.edu/abs/2019A%26A...621A..69R} {621, A69}

\bibitem[\protect\citeauthoryear{Ramanah, Charnock  \& Lavaux}{Ramanah
  et~al.}{2019b}]{Ramanah:2019cbm}
Ramanah D.~K.,  Charnock T.,   Lavaux G.,  2019b, \mn@doi [Phys. Rev.]
  {10.1103/PhysRevD.100.043515}, D100, 043515

\bibitem[\protect\citeauthoryear{{Ravanbakhsh}, {Oliva}, {Fromenteau}, {Price},
  {Ho}, {Schneider}  \& {Poczos}}{{Ravanbakhsh}
  et~al.}{2017}]{Ravanbakhsh:2017bbi}
{Ravanbakhsh} S.,  {Oliva} J.,  {Fromenteau} S.,  {Price} L.~C.,  {Ho} S.,
  {Schneider} J.,   {Poczos} B.,  2017, arXiv e-prints, \href
  {https://ui.adsabs.harvard.edu/abs/2017arXiv171102033R} {p. arXiv:1711.02033}

\bibitem[\protect\citeauthoryear{{Rees} \& {Sciama}}{{Rees} \&
  {Sciama}}{1968}]{1968Natur.217..511R}
{Rees} M.~J.,  {Sciama} D.~W.,  1968, \mn@doi [\nat] {10.1038/217511a0}, \href
  {https://ui.adsabs.harvard.edu/abs/1968Natur.217..511R} {217, 511}

\bibitem[\protect\citeauthoryear{{Riess}, {Davis}, {Baker}  \&
  {Kirshner}}{{Riess} et~al.}{1997}]{SNIaflow...1997ApJ...488L...1R}
{Riess} A.~G.,  {Davis} M.,  {Baker} J.,   {Kirshner} R.~P.,  1997, \mn@doi
  [\apjl] {10.1086/310917}, \href
  {https://ui.adsabs.harvard.edu/abs/1997ApJ...488L...1R} {488, L1}

\bibitem[\protect\citeauthoryear{Rodriguez, Kacprzak, Lucchi, Amara, Sgier,
  Fluri, Hofmann  \& Réfrégier}{Rodriguez et~al.}{2018}]{Rodriguez:2018mjb}
Rodriguez A.~C.,  Kacprzak T.,  Lucchi A.,  Amara A.,  Sgier R.,  Fluri J.,
  Hofmann T.,   Réfrégier A.,  2018, \mn@doi [Comput. Astrophys. Cosmol.]
  {10.1186/s40668-018-0026-4}, 5, 4

\bibitem[\protect\citeauthoryear{{Sachs} \& {Wolfe}}{{Sachs} \&
  {Wolfe}}{1967}]{1967ApJ...147...73S}
{Sachs} R.~K.,  {Wolfe} A.~M.,  1967, \mn@doi [\apj] {10.1086/148982}, \href
  {https://ui.adsabs.harvard.edu/abs/1967ApJ...147...73S} {147, 73}

\bibitem[\protect\citeauthoryear{{Schmelzle}, {Lucchi}, {Kacprzak}, {Amara},
  {Sgier}, {R{\'e}fr{\'e}gier}  \& {Hofmann}}{{Schmelzle}
  et~al.}{2017}]{Schmelzle:2017vwd}
{Schmelzle} J.,  {Lucchi} A.,  {Kacprzak} T.,  {Amara} A.,  {Sgier} R.,
  {R{\'e}fr{\'e}gier} A.,   {Hofmann} T.,  2017, arXiv e-prints, \href
  {https://ui.adsabs.harvard.edu/abs/2017arXiv170705167S} {p. arXiv:1707.05167}

\bibitem[\protect\citeauthoryear{{Shallue} \& {Eisenstein}}{{Shallue} \&
  {Eisenstein}}{2022}]{2022arXiv220712511S}
{Shallue} C.~J.,  {Eisenstein} D.~J.,  2022, arXiv e-prints, \href
  {https://ui.adsabs.harvard.edu/abs/2022arXiv220712511S} {p. arXiv:2207.12511}

\bibitem[\protect\citeauthoryear{{Springel}}{{Springel}}{2005}]{2005MNRAS.364.1105S}
{Springel} V.,  2005, \mn@doi [\mnras] {10.1111/j.1365-2966.2005.09655.x},
  \href {https://ui.adsabs.harvard.edu/abs/2005MNRAS.364.1105S} {364, 1105}

\bibitem[\protect\citeauthoryear{{Springer}, {Ofek}, {Weiss}  \&
  {Merten}}{{Springer} et~al.}{2020}]{Springer:2018aak}
{Springer} O.~M.,  {Ofek} E.~O.,  {Weiss} Y.,   {Merten} J.,  2020, \mn@doi
  [\mnras] {10.1093/mnras/stz2991}, \href
  {https://ui.adsabs.harvard.edu/abs/2020MNRAS.491.5301S} {491, 5301}

\bibitem[\protect\citeauthoryear{{Springob}, {Masters}, {Haynes}, {Giovanelli}
  \& {Marinoni}}{{Springob} et~al.}{2007}]{FundPlan...2007ApJS..172..599S}
{Springob} C.~M.,  {Masters} K.~L.,  {Haynes} M.~P.,  {Giovanelli} R.,
  {Marinoni} C.,  2007, \mn@doi [\apjs] {10.1086/519527}, \href
  {https://ui.adsabs.harvard.edu/abs/2007ApJS..172..599S} {172, 599}

\bibitem[\protect\citeauthoryear{{Sunyaev} \& {Zeldovich}}{{Sunyaev} \&
  {Zeldovich}}{1972}]{1972CoASP...4..173S}
{Sunyaev} R.~A.,  {Zeldovich} Y.~B.,  1972, Comments on Astrophysics and Space
  Physics, \href {https://ui.adsabs.harvard.edu/abs/1972CoASP...4..173S} {4,
  173}

\bibitem[\protect\citeauthoryear{{Sunyaev} \& {Zeldovich}}{{Sunyaev} \&
  {Zeldovich}}{1980}]{1980MNRAS.190..413S}
{Sunyaev} R.~A.,  {Zeldovich} Y.~B.,  1980, \mn@doi [\mnras]
  {10.1093/mnras/190.3.413}, \href
  {https://ui.adsabs.harvard.edu/abs/1980MNRAS.190..413S} {190, 413}

\bibitem[\protect\citeauthoryear{{Tanimura}, {Aghanim}, {Bonjean}  \&
  {Zaroubi}}{{Tanimura} et~al.}{2022}]{2022A&A...662A..48T}
{Tanimura} H.,  {Aghanim} N.,  {Bonjean} V.,   {Zaroubi} S.,  2022, \mn@doi
  [\aap] {10.1051/0004-6361/202243046}, \href
  {https://ui.adsabs.harvard.edu/abs/2022A&A...662A..48T} {662, A48}

\bibitem[\protect\citeauthoryear{Tewes, Kuntzer, Nakajima, Courbin, Hildebrandt
   \& Schrabback}{Tewes et~al.}{2019}]{Tewes:2018she}
Tewes M.,  Kuntzer T.,  Nakajima R.,  Courbin F.,  Hildebrandt H.,   Schrabback
  T.,  2019, \mn@doi [Astron. Astrophys.] {10.1051/0004-6361/201833775}, 621,
  A36

\bibitem[\protect\citeauthoryear{Tröster, Ferguson, Harnois-Déraps  \&
  McCarthy}{Tröster et~al.}{2019}]{Troster:2019mys}
Tröster T.,  Ferguson C.,  Harnois-Déraps J.,   McCarthy I.~G.,  2019,
  \mn@doi [Mon. Not. Roy. Astron. Soc.] {10.1093/mnrasl/slz075}, 487, L24

\bibitem[\protect\citeauthoryear{{Tully} \& {Fisher}}{{Tully} \&
  {Fisher}}{1977}]{TullyFisher...1977A&A....54..661T}
{Tully} R.~B.,  {Fisher} J.~R.,  1977, \aap, \href
  {https://ui.adsabs.harvard.edu/abs/1977A&A....54..661T} {500, 105}

\bibitem[\protect\citeauthoryear{{Turnbull}, {Hudson}, {Feldman}, {Hicken},
  {Kirshner}  \& {Watkins}}{{Turnbull}
  et~al.}{2012}]{SNIaflow...2012MNRAS.420..447T}
{Turnbull} S.~J.,  {Hudson} M.~J.,  {Feldman} H.~A.,  {Hicken} M.,  {Kirshner}
  R.~P.,   {Watkins} R.,  2012, \mn@doi [\mnras]
  {10.1111/j.1365-2966.2011.20050.x}, \href
  {https://ui.adsabs.harvard.edu/abs/2012MNRAS.420..447T} {420, 447}

\bibitem[\protect\citeauthoryear{{Wang}, {Mo}, {Yang}  \& {van den
  Bosch}}{{Wang} et~al.}{2012}]{VelocityRecon...2012MNRAS.420.1809W}
{Wang} H.,  {Mo} H.~J.,  {Yang} X.,   {van den Bosch} F.~C.,  2012, \mn@doi
  [\mnras] {10.1111/j.1365-2966.2011.20174.x}, \href
  {https://ui.adsabs.harvard.edu/abs/2012MNRAS.420.1809W} {420, 1809}

\bibitem[\protect\citeauthoryear{Wu et~al.,}{Wu et~al.}{2021}]{Wu:2021jsy}
Wu Z.,  et~al., 2021, \mn@doi [Astrophys. J.] {10.3847/1538-4357/abf3bb}, 913,
  2

\bibitem[\protect\citeauthoryear{{Yu}, {Zhang}, {Jing}  \& {Zhang}}{{Yu}
  et~al.}{2015}]{2015PhRvD..92h3527Y}
{Yu} Y.,  {Zhang} J.,  {Jing} Y.,   {Zhang} P.,  2015, \mn@doi [\prd]
  {10.1103/PhysRevD.92.083527}, \href
  {https://ui.adsabs.harvard.edu/abs/2015PhRvD..92h3527Y} {92, 083527}

\bibitem[\protect\citeauthoryear{{Zaroubi}, {Hoffman}, {Fisher}  \&
  {Lahav}}{{Zaroubi} et~al.}{1995}]{VelocityRecon...1995ApJ...449..446Z}
{Zaroubi} S.,  {Hoffman} Y.,  {Fisher} K.~B.,   {Lahav} O.,  1995, \mn@doi
  [\apj] {10.1086/176070}, \href
  {https://ui.adsabs.harvard.edu/abs/1995ApJ...449..446Z} {449, 446}

\bibitem[\protect\citeauthoryear{{Zhang}, {Zheng}  \& {Jing}}{{Zhang}
  et~al.}{2015}]{2015PhRvD..91d3522Z}
{Zhang} P.,  {Zheng} Y.,   {Jing} Y.,  2015, \mn@doi [\prd]
  {10.1103/PhysRevD.91.043522}, \href
  {https://ui.adsabs.harvard.edu/abs/2015PhRvD..91d3522Z} {91, 043522}

\bibitem[\protect\citeauthoryear{{Zhang}, {Wang}, {Zhang}, {Sun}, {He},
  {Contardo}, {Villaescusa-Navarro}  \& {Ho}}{{Zhang}
  et~al.}{2019}]{Zhang:2019ryt}
{Zhang} X.,  {Wang} Y.,  {Zhang} W.,  {Sun} Y.,  {He} S.,  {Contardo} G.,
  {Villaescusa-Navarro} F.,   {Ho} S.,  2019, arXiv e-prints, \href
  {https://ui.adsabs.harvard.edu/abs/2019arXiv190205965Z} {p. arXiv:1902.05965}

\makeatother
\end{thebibliography}




\appendix
\section{Momentum field reconstruction from Unet}
\label{app:m}

\begin{table}
	\centering
	\caption{Same as in Tab.~\ref{tab:cfv}, but for correlation coefficients for the momentum field.}
		\begin{tabular}{c|ccc} 
		\hline
		field  & $\tilde{\bm{v}}$ & $\tilde{\theta}$ &$\tilde{\bm{\omega}}$ \\
		\hline
		 $C_f$  & 0.95&0.80&0.92   \\
		\hline
	\end{tabular}\label{tab:cfm}
\end{table}

The momentum field of galaxies and clusters of galaxies is also cosmologically very important~\citep{2014JCAP...05..003O}. In the following, we will use the tilde symbol to denote the momentum field and its components , i.e., $\tilde{\bm{v}}$ for the momentum field, $\tilde{\bm{\theta}}$ and $\tilde{\bm{\omega}}$ for its divergence and vorticity, respectively. The momentum field can be defined as $\tilde{\bm{v}}(\bm{x})=[1+\delta(\bm{x})] \bm{v}(\bm{x})$, where $\delta=n / \bar{n}-1$ is the perturbation of number density field, and $\bm{v}$ is the comoving velocity field. The momentum field thus is the number-weighted velocity. We can also decompose the momentum field into the divergence and vorticity components, with
$\tilde{\theta}(\bm{k}) = i \bm{k} \cdot \tilde{\bm{v}}(\bm{k})$ and $\tilde{\bm{\omega}}(\bm{k}) = i \bm{k} \times \tilde{\bm{v}}(\bm{k})$, very similar to the velocity field. The corresponding momentum power spectra can be defined in the same way (as defined in Eq.~\ref{eq:v}), e.g.,  $P_{\tilde{\theta}}$ and $P_{\tilde{\omega}}$ for the divergence and vorticity power spectra of the momentum field, respectively.

The reconstruction results for the momentum field are summarized as follows. Overall, for the reconstruction of the momentum field, UNet can achieve similar results. From Tab.~\ref{tab:cfm}, the resulting correlation coefficients are at the level of 0.9, about 2\% larger than the ones shown in Tab.~\ref{tab:cfv} for the velocity field.

\begin{figure}
	\includegraphics[width=0.45\textwidth]{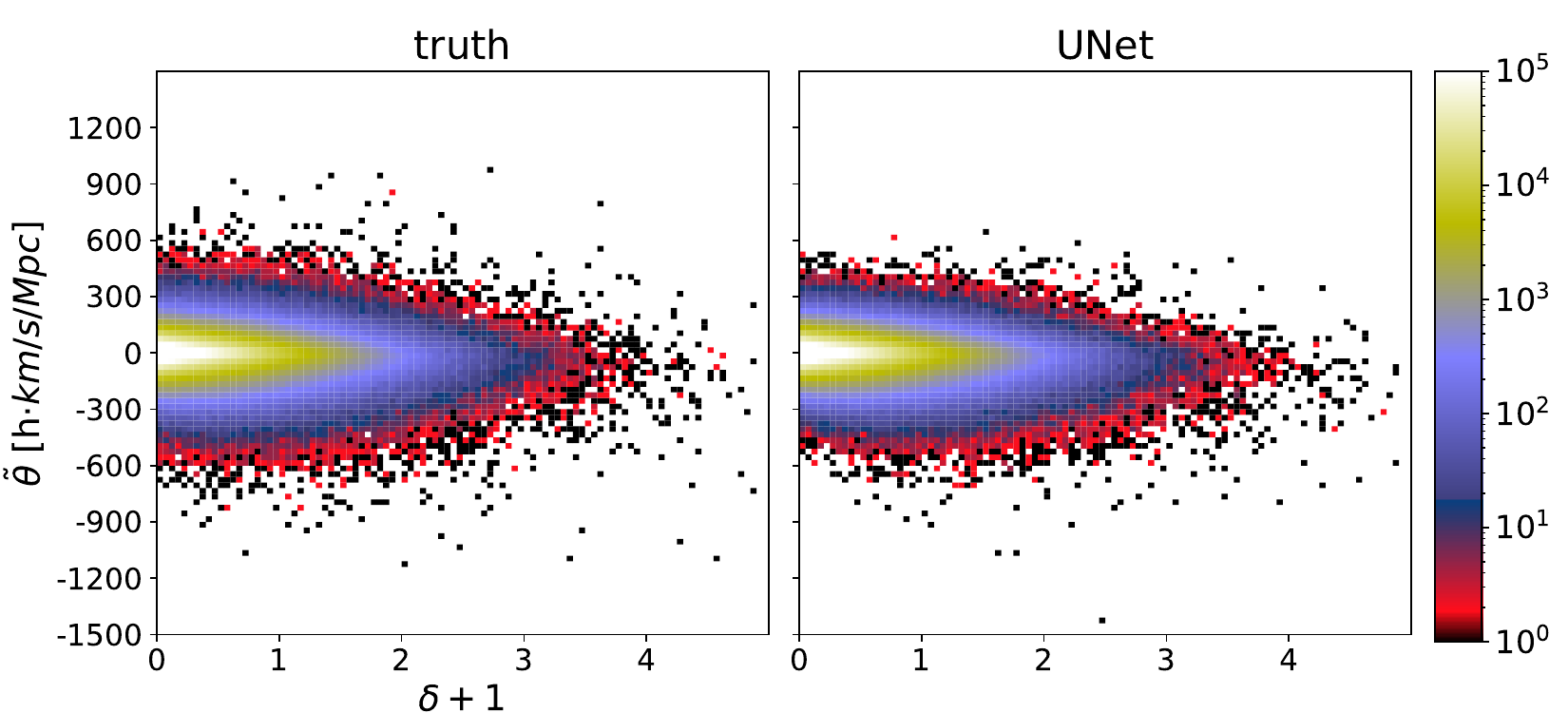}
        \includegraphics[width=0.45\textwidth]{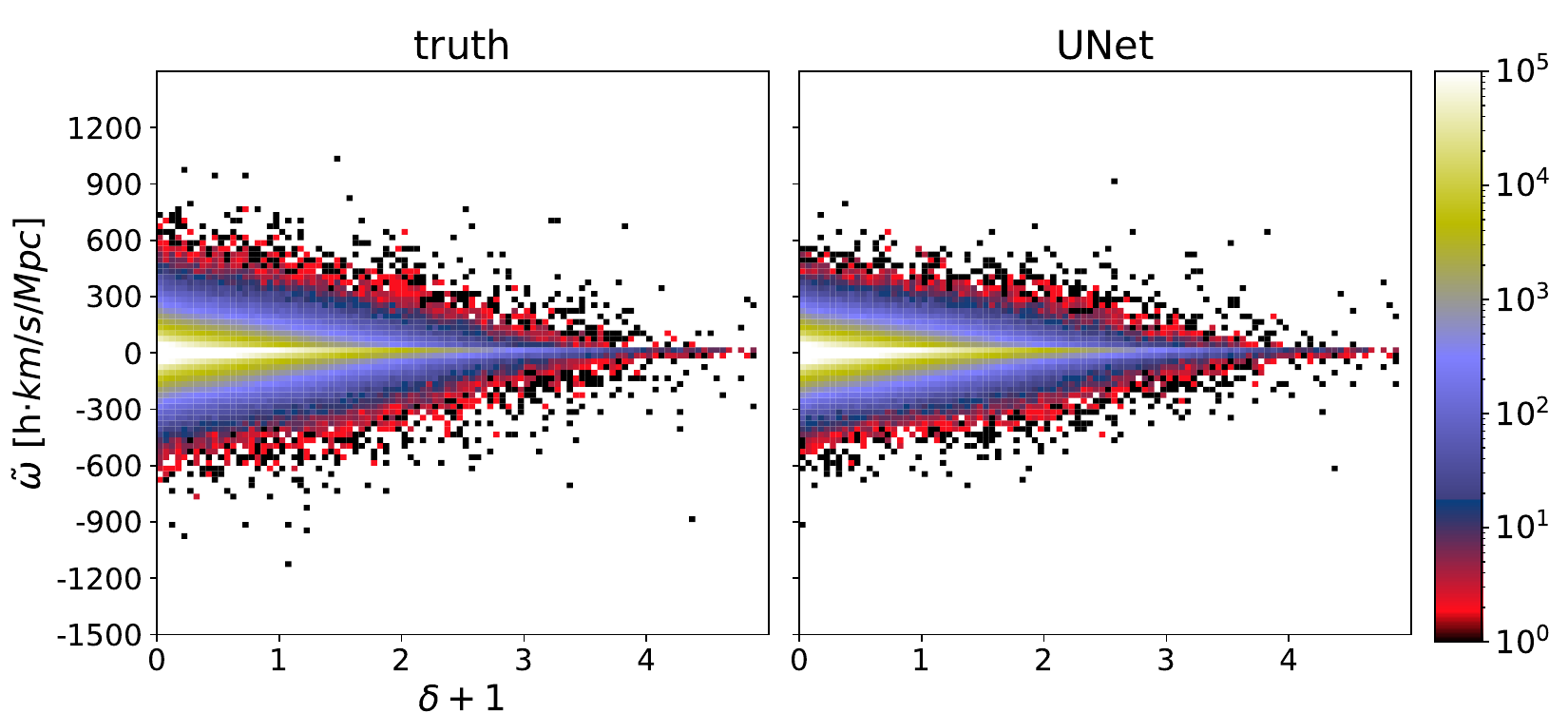}
    \caption{Same as in Fig.~\ref{fig:v_delta}, but for the joint probability distributions of density-divergence (upper), and density-vorticity (lower) for the momentum field.}
    \label{fig:momentum_delta}
\end{figure}

\begin{figure*}
	\includegraphics[width=0.8\textwidth]{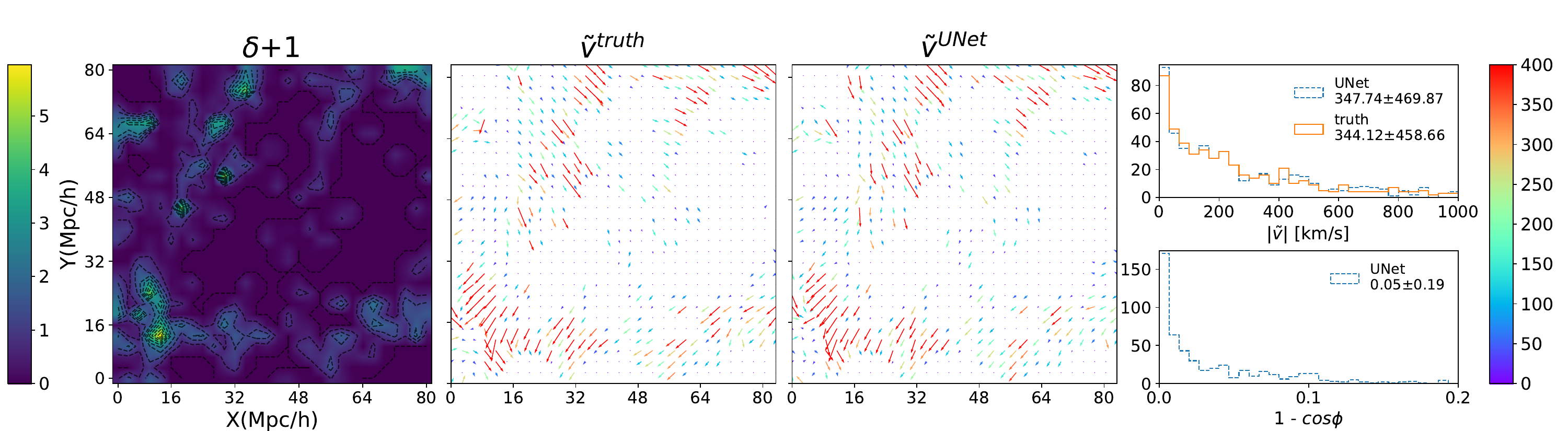}\\
    \includegraphics[width=0.8\textwidth]{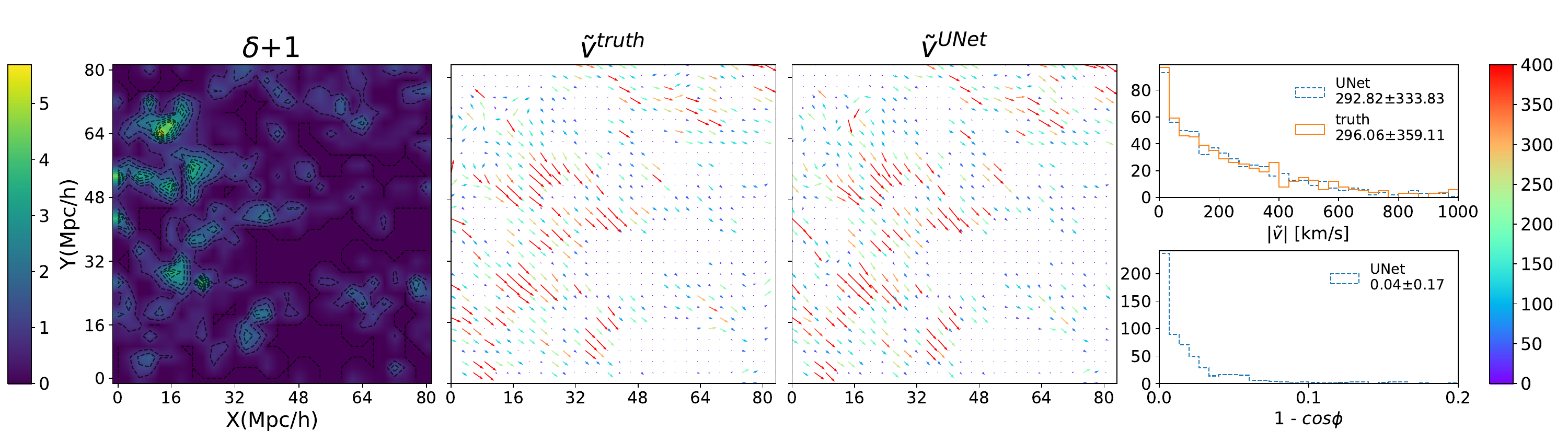}
    \caption{Same as in Fig.~\ref{fig:v_point}, but for the momentum field $\tilde{\bm v}$.}
    \label{fig:momentum_point}
\end{figure*}

\begin{figure*}
    \includegraphics[width=0.8\textwidth]{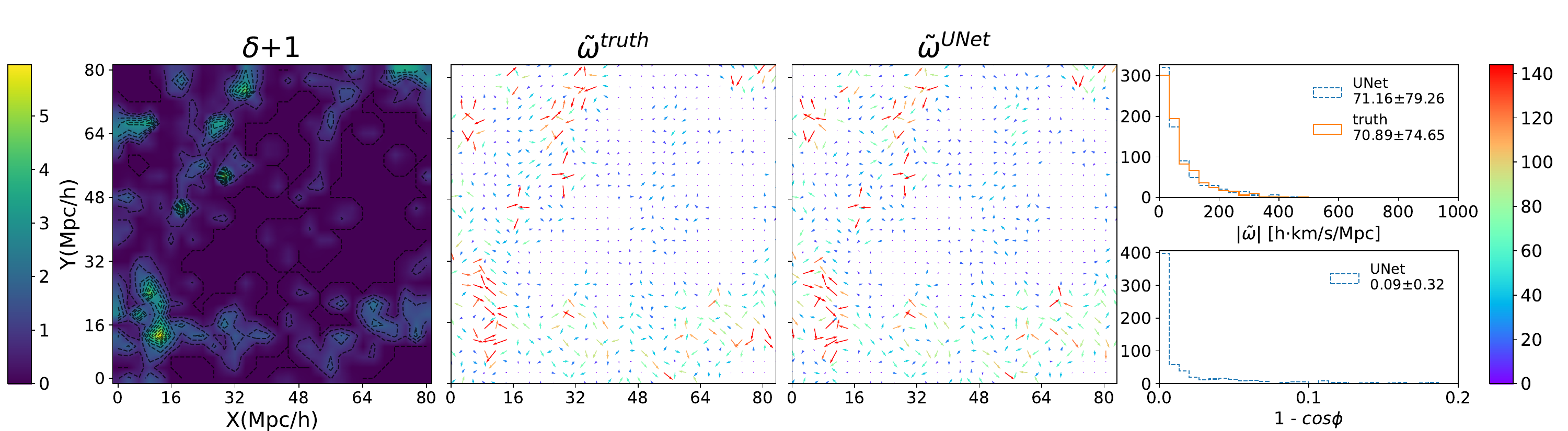}\\
    \includegraphics[width=0.8\textwidth]{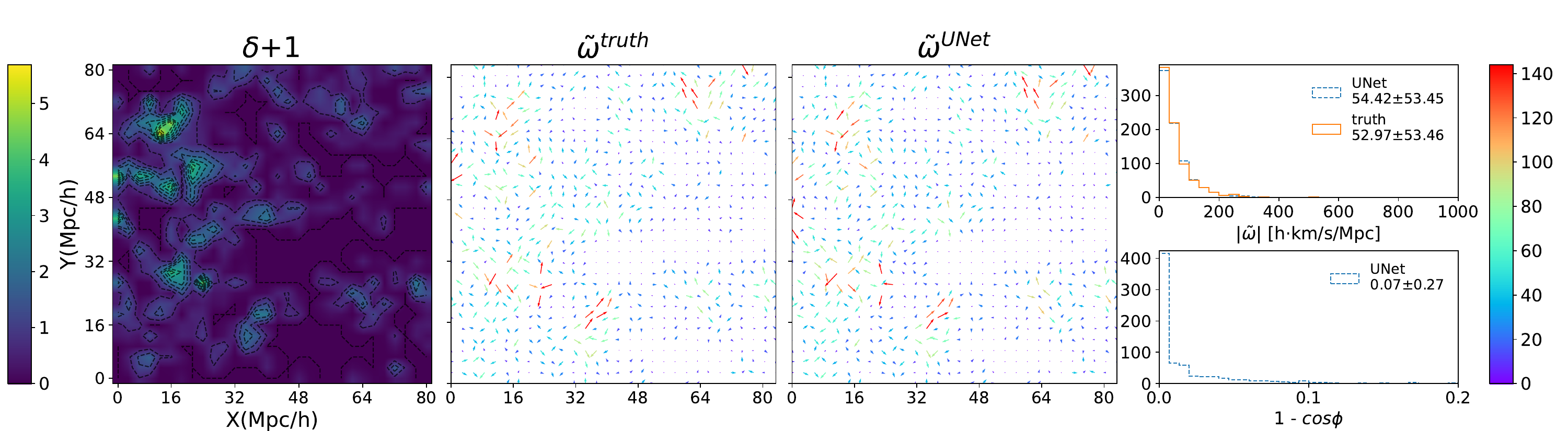}
    \caption{Same as in Fig.~\ref{fig:v_curl_point}, but for the vorticity component of the momentum field $\tilde{\bm \omega}$.}
    \label{fig:momentum_curl_point}
\end{figure*}

\begin{figure*}
	\includegraphics[width=0.8\textwidth]{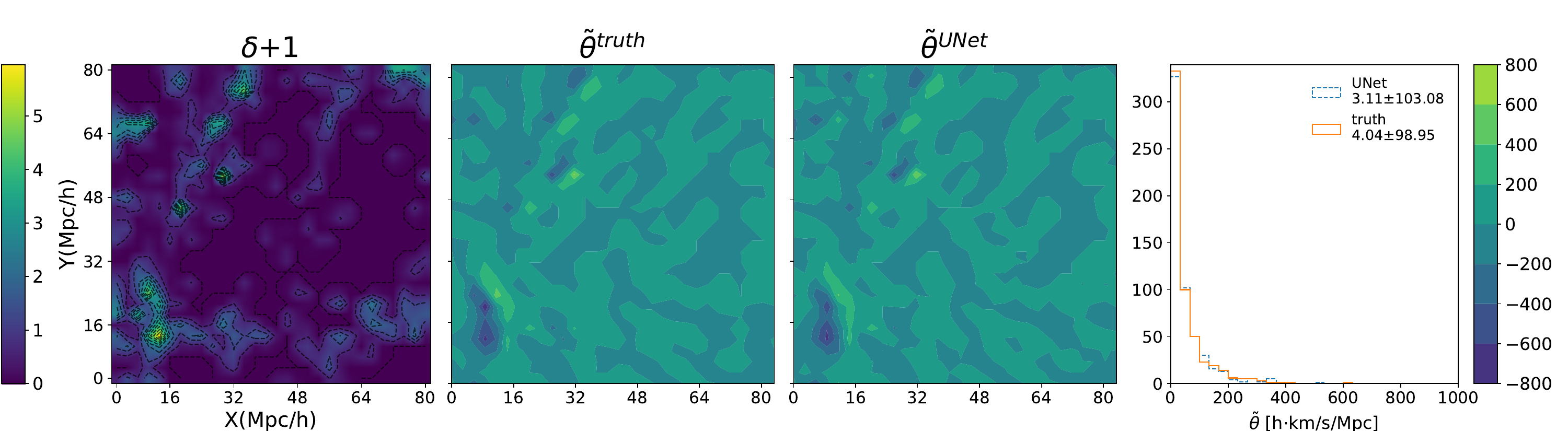}\\
    \includegraphics[width=0.8\textwidth]{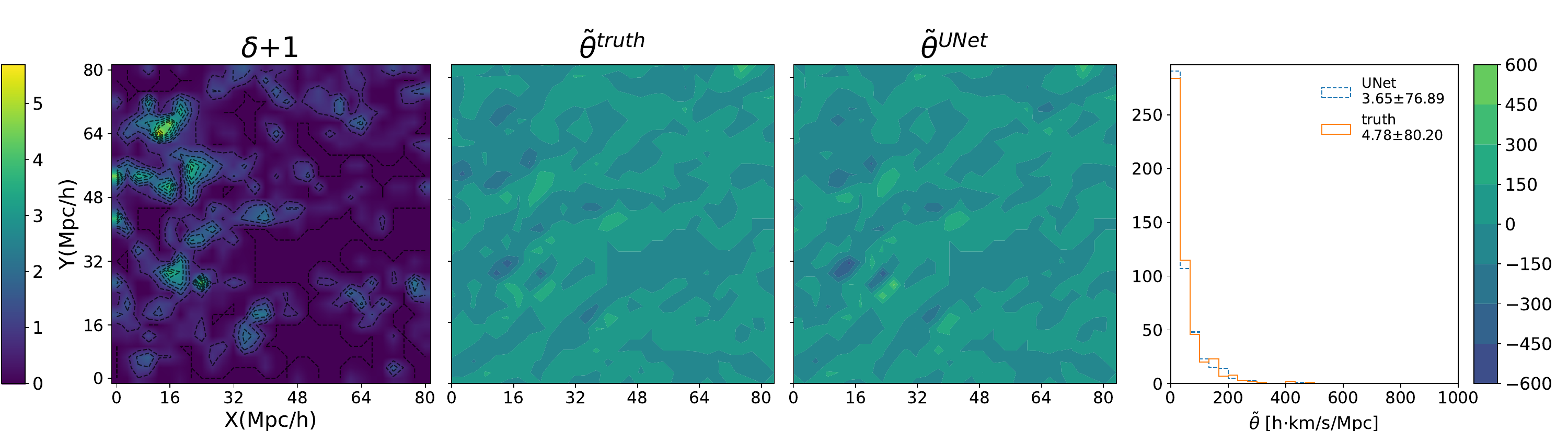}
    \caption{Same as in Fig.~\ref{fig:v_div_point}, but for the momentum field $\tilde{\bm \theta}$.}
    \label{fig:momentum_div_point}
\end{figure*}

\begin{figure*}
\centering
\includegraphics[width=0.32\textwidth]{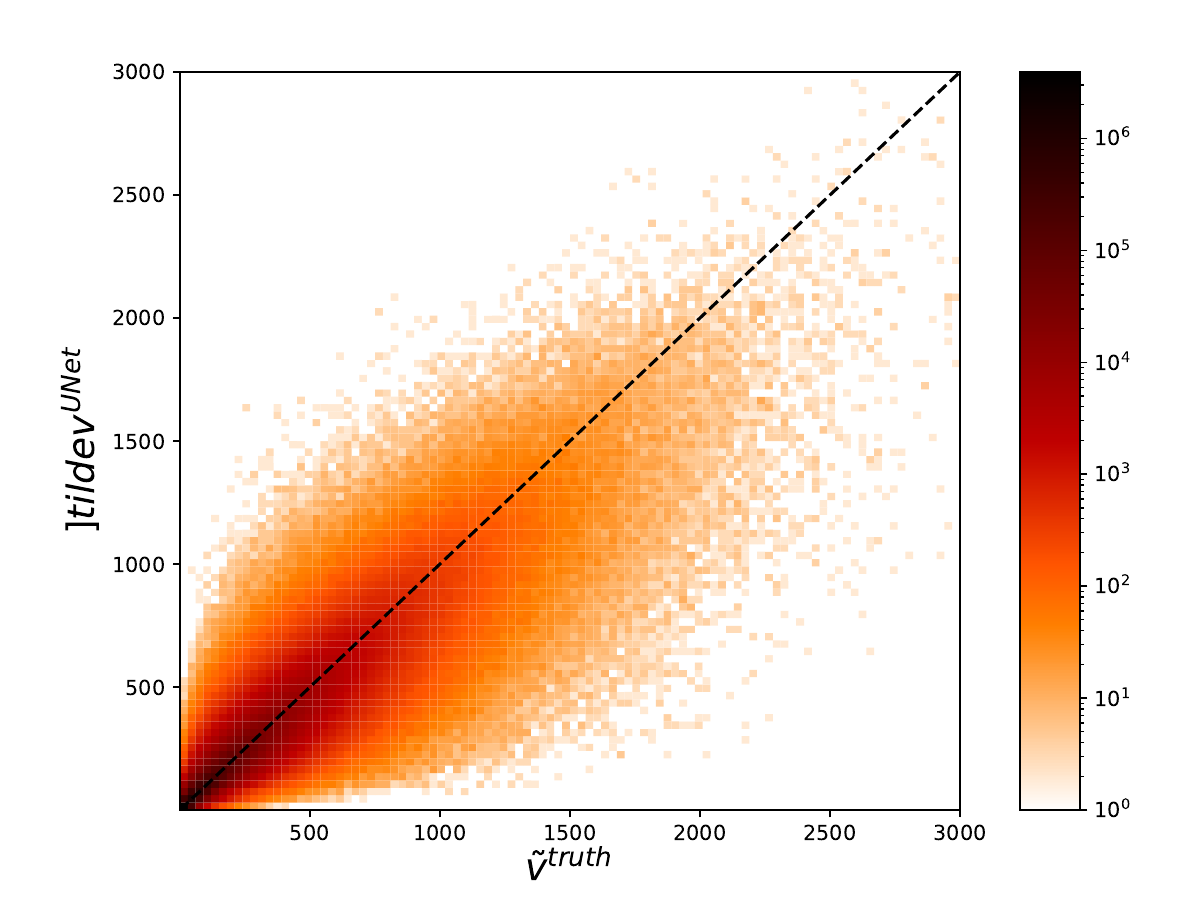}
\includegraphics[width=0.32\textwidth]{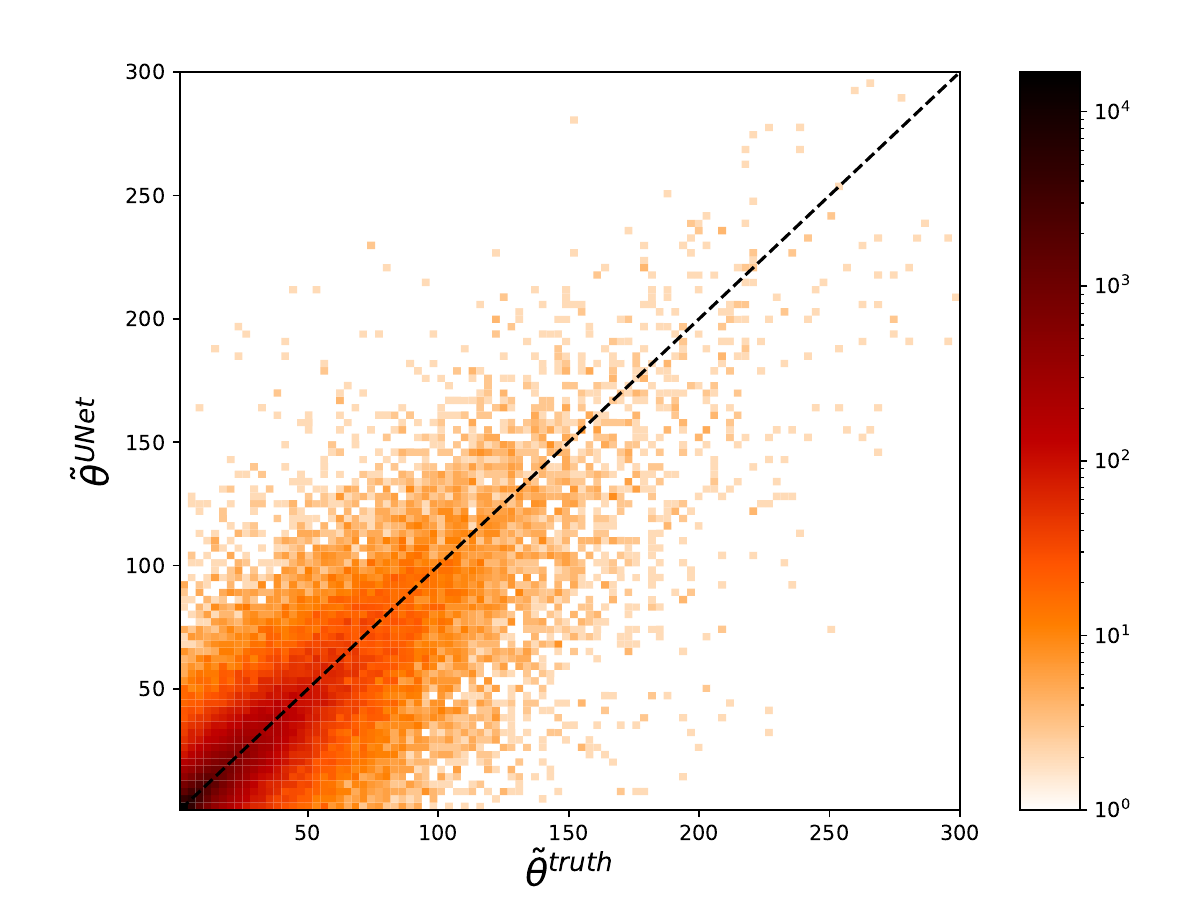}
\includegraphics[width=0.32\textwidth]{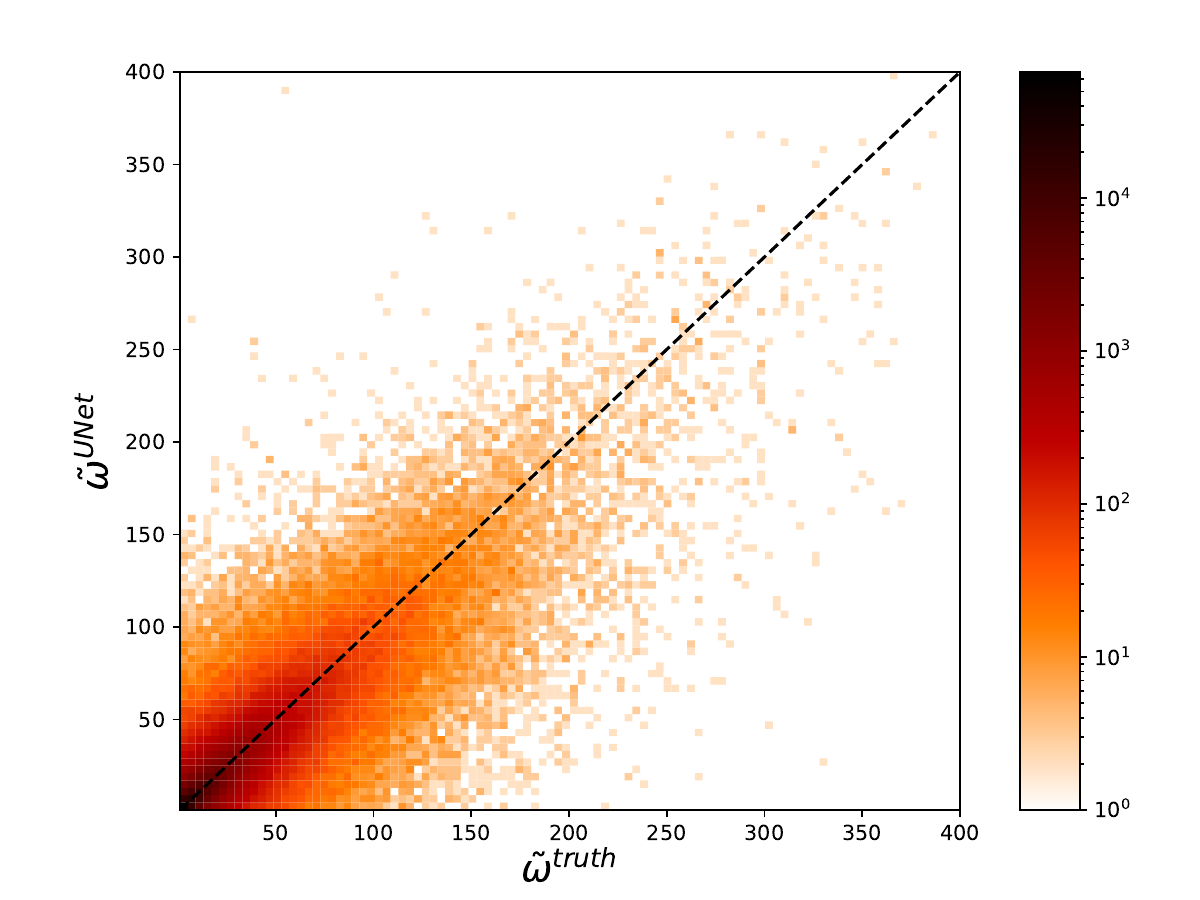}
\caption{Same as in Fig.~\ref{fig:v_joint}, but for the momentum field.}
    \label{fig:p_joint}
\end{figure*}

Furthermore, let us first compare the joint probability distributions of density-divergence, and density-vorticity for the momentum field, which are shown in Fig.~\ref{fig:momentum_delta}. We do see that the reconstructed distributions are pretty consistent with the true ones morphologically. Additionally, the reconstructions of the momentum field and its vorticity component for two randomly selected slices are present in Figs.~\ref{fig:momentum_point}~\&~\ref{fig:momentum_curl_point}\&~\ref{fig:momentum_div_point}. As seen, both of these reconstructed fields indeed correlate strongly with the true ones, providing very high reconstruction accuracy at the level of 1\%. Also, from the histogram distributions, the deviations on average are about $18^\circ$ for the direction of the momentum field, and about $23^\circ$ for the direction of the vorticity, respectively.

\begin{figure*}
\centering
\includegraphics[width=0.3\textwidth]{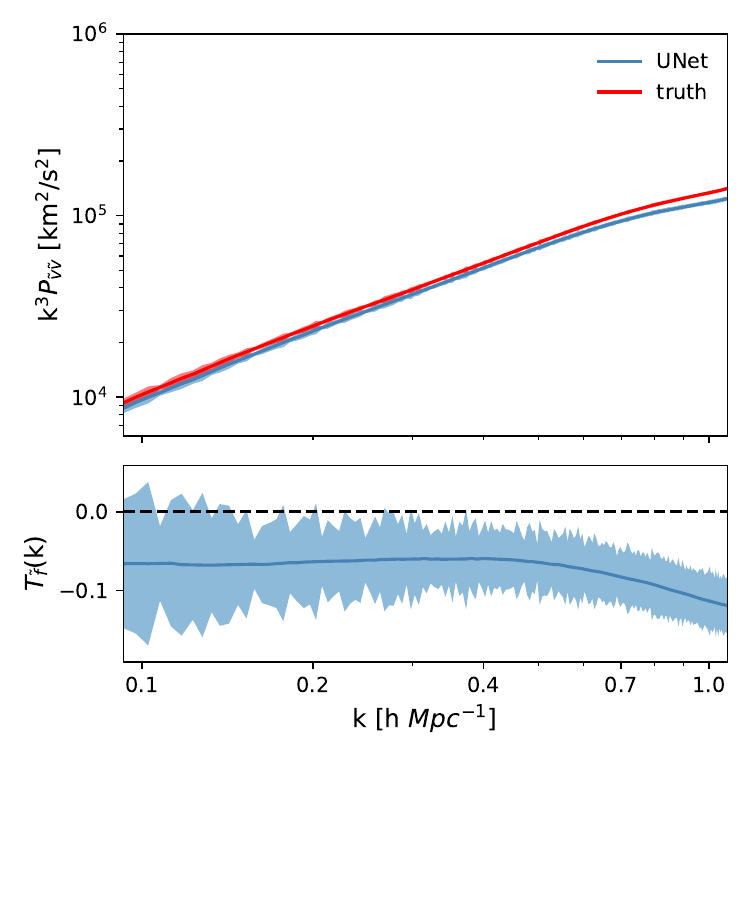}
\includegraphics[width=0.3\textwidth]{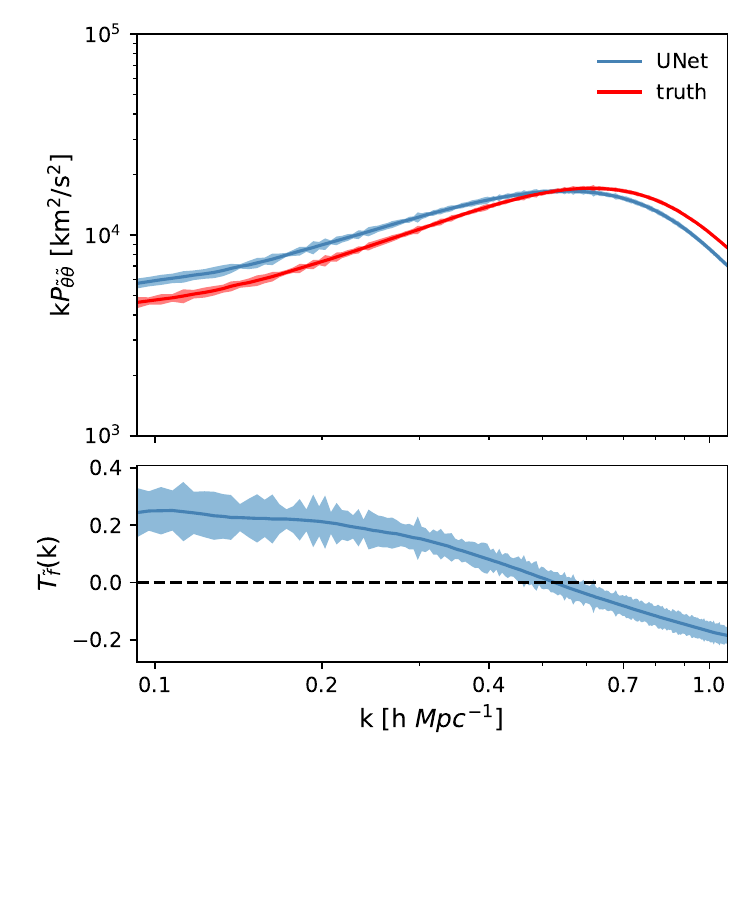}
\includegraphics[width=0.3\textwidth]{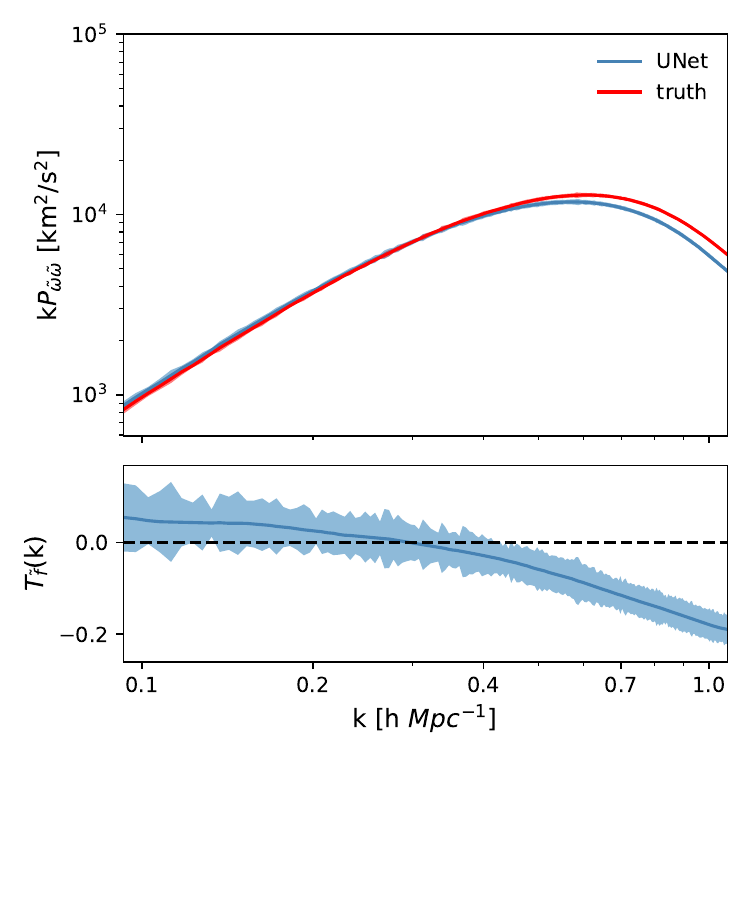}
\caption{Same as in Fig.~\ref{fig:pk_vtheta}, but for the momentum field.}
    \label{fig:pk_vtheta_m}
\end{figure*}

\begin{figure*}
    \includegraphics[width=0.35\textwidth]{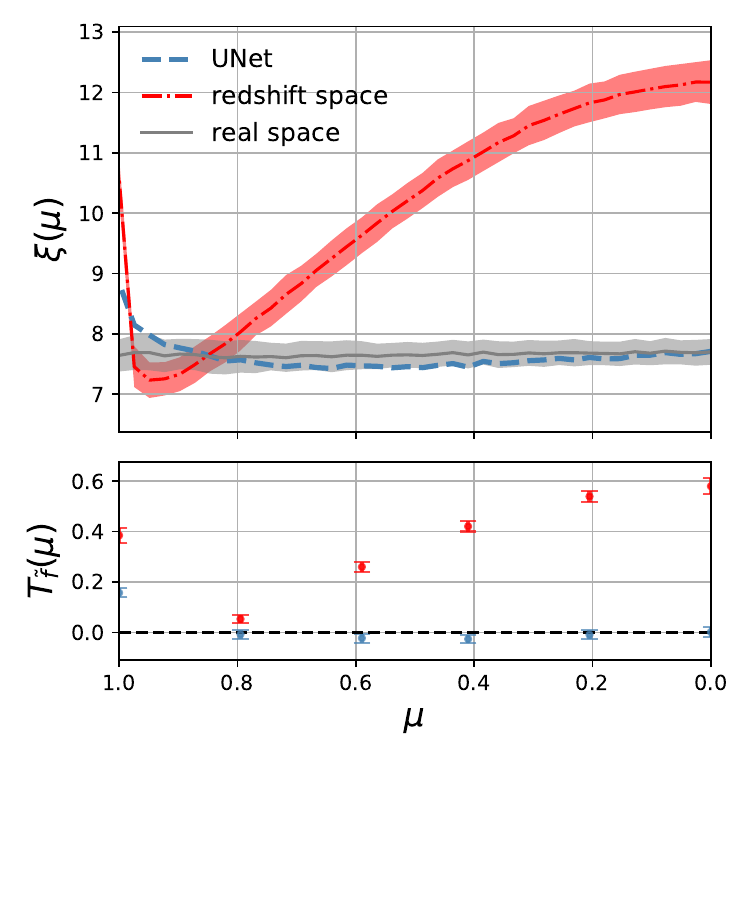}
\includegraphics[width=0.35\textwidth]{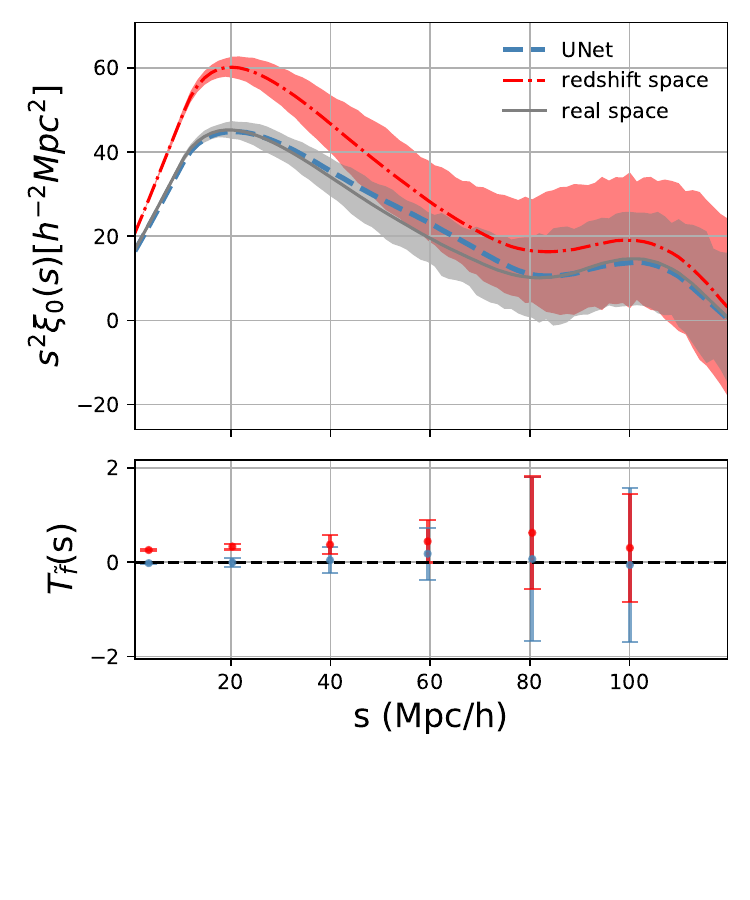}
    \includegraphics[width=0.35\textwidth]{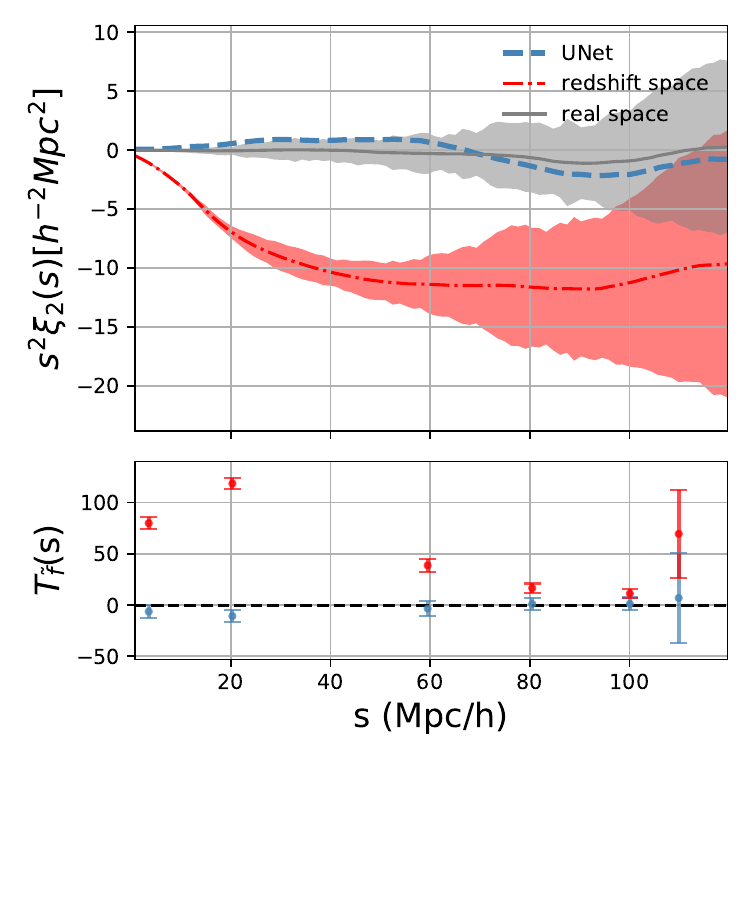}
    \includegraphics[width=0.35\textwidth]{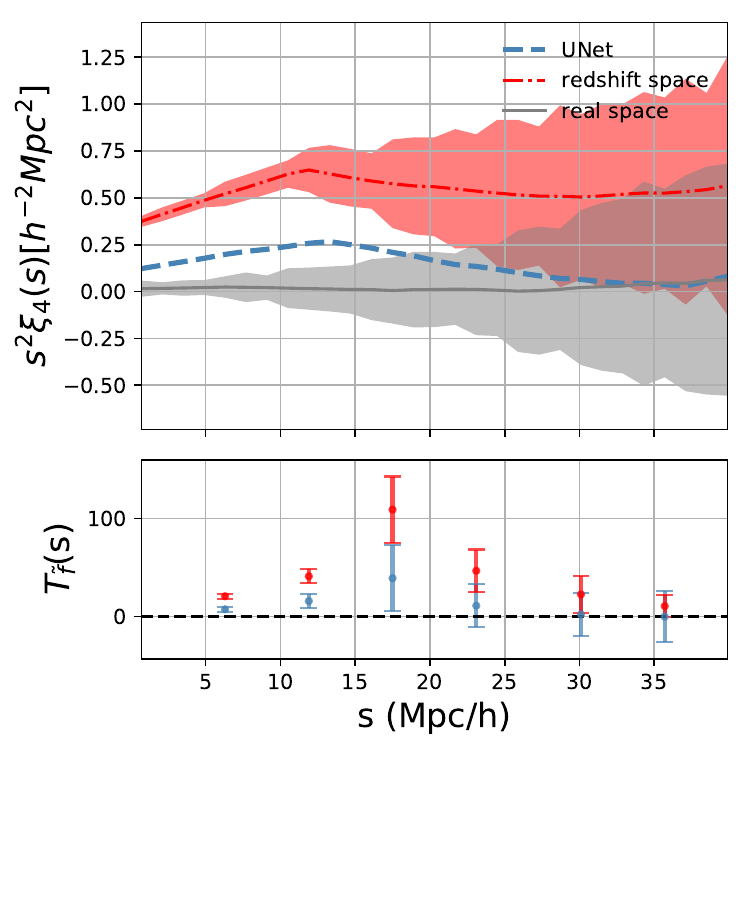}
    \caption{Same as in Fig.~\ref{fig:v_2pcf}, but for the 2PCFs reconstructed from the momentum field.}
    \label{fig:mom_2pcf}
\end{figure*}

\begin{figure*}
 \includegraphics[width=450pt]{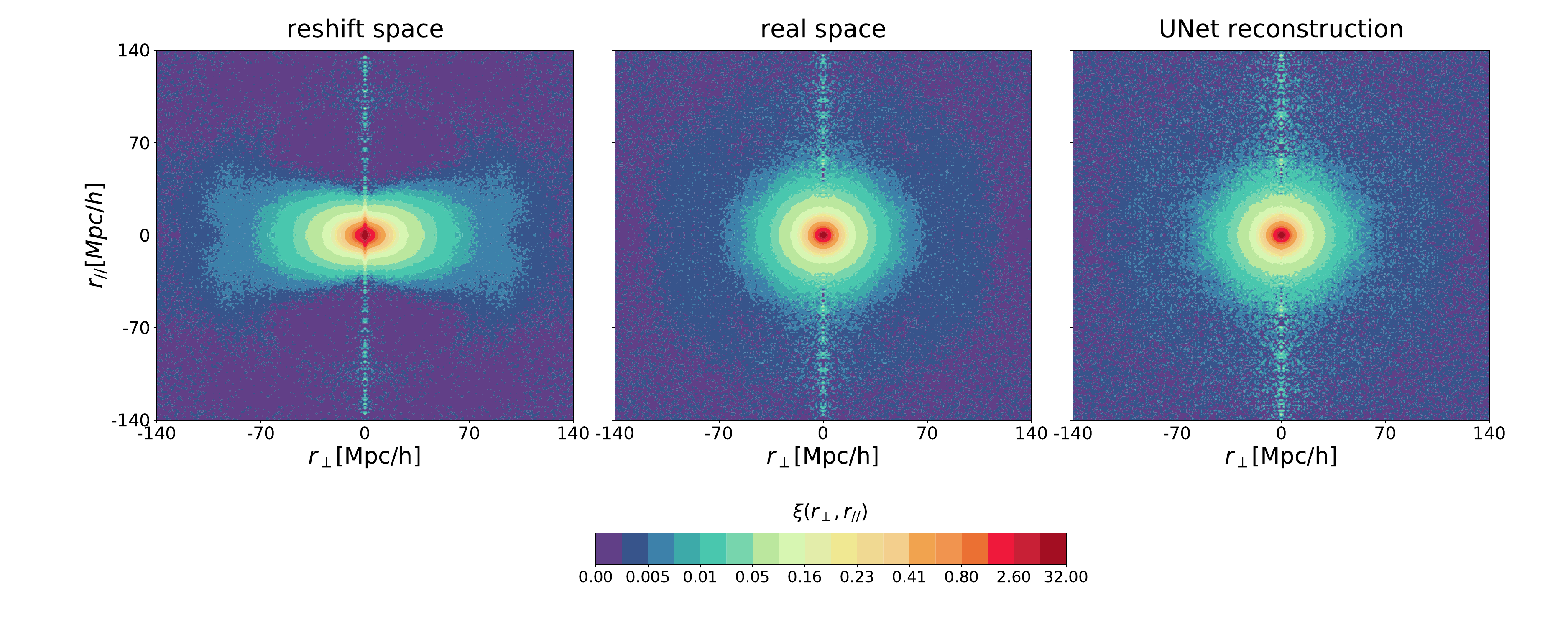}
    \caption{Same as in Fig.~\ref{fig:v_rp}, but for the contour of the anisotropic 2PCF reconstructed from the the momentum field.}
    \label{fig:momentum_rp}
\end{figure*}

Compared with the reconstruction in power spectrum, as observed in Fig.~\ref{fig:pk_vtheta_m}, the relative deviation demonstrate the UNet model yielding excellent reconstruction at all scales of $k\lesssim 1.1$ $h/\rm Mpc$, $|T(k)|\lesssim 0.15$ for the momentum field, and $|T(k)|\lesssim 0.2$ for both  momentum divergence and vorticity components. More interestingly, we can also correct the peculiar velocity of each individual halo from the UNet-reconstructed momentum field via $\bm{v} = \tilde{\bm{v}}/(1+\delta n)$, where we assume the halo number density contrast $\delta n$ is exactly known from the simulations. The projected 2PCF and the associated multipoles of 2PCF are illustrated in Fig.~\ref{fig:mom_2pcf}, the relative deviation between the reconstructed one and the truth are detailed in Tab.~\ref{tab:ximurp}. Furthermore, the comparison of the anisotropic 2PCF between the reconstruction and the true one are shown in Fig.~\ref{fig:momentum_rp}.  All of there results  obviously demonstrate a high-fidelity reconstruction of UNet.

\begin{table*}
	\centering
	\caption{Same as in Tab.~\ref{tab:ximur}, but for the momentum field.}
	\label{tab:ximurp}
	\begin{tabular}{lcccccr} 
		\hline
		$\mu$  & 1.0 & 0.8 & 0.6  &  0.4 & 0.2 & 0.0 \\
		\hline
		$T_{\xi(\mu)}$ (UNet correction) & $0.16\pm 0.02$ & $-0.01\pm 0.02$ & $-0.02\pm 0.02$ & $-0.03\pm 0.02$ & $-0.01\pm 0.02$ & $0.0\pm 0.02$ \\
		$T_{\xi(\mu)}$ (redshift space) & $0.38\pm 0.03$ & $0.05\pm 0.02$ & $0.24\pm 0.02$ & $0.42\pm 0.02$ & $0.54\pm 0.02$ & $0.58\pm 0.03$\\
		\hline\hline
  $r$ (${\rm Mpc}/h$)  & 5 & 20 & 40  &  60 & 80 & 100 \\
		\hline
		$T_{\xi_{0}(r)}$ (UNet correction)  & $-0.02\pm 0.03$ & $-0.01\pm 0.09$ & $0.04\pm  0.27$ & $0.18\pm 0.55$ & $0.07\pm 1.74$ & $-0.06\pm 1.64$\\
		$T_{\xi_{0}(r)}$ (redshift space)  & $0.26\pm0.02$ & $0.33\pm 0.06$ & $0.37\pm 0.20$ & $0.44\pm 0.45$ & $0.62\pm 1.19$ & $0.30\pm 1.14$\\
		\hline
  $r$ (${\rm Mpc}/h$)  & 5 & 20 & 60  &  80 & 100 & 110 \\
		\hline
  $T_{\xi_{2}(r)}$ (UNet correction)  & $-6.26\pm 5.96$ & $-10.7\pm 5.51$ & $-3.40\pm 7.55$ & $1.09\pm 6.22$ & $1.25\pm 6.15$ & $6.95\pm 44.0$\\
		$T_{\xi_{2}(r)}$ (redshift space)  & $79.98\pm5.63$ & $118.70\pm 5.40$ & $38.75\pm 6.08$ & $16.62\pm 4.48$ & $11.27 \pm 4.64$ & $69.55 \pm 42.99$\\
		\hline
  $r$ (${\rm Mpc}/h$)  & 6 & 12 & 18  &  23 & 30 & 36 \\
		\hline
  $T_{\xi_{4}(r)}$ (UNet correction)  & $7.22\pm 2.41$ & $ 15.66\pm 7.31$ & $39.00\pm 33.84$ & $10.91\pm 22.02$ & $1.99\pm 22.34$ & $-0.18\pm 26.21$\\
		$T_{\xi_{4}(r)}$ (redshift space)  & $20.6\pm 2.37$ & $40.97\pm 7.27$ & $109.15\pm 33.81$ & $46.63\pm 21.72$ & $22.45\pm 19.25$ & $ 10.50\pm 11.06$\\
		\hline
	\end{tabular}
\end{table*}


\bsp	
\label{lastpage}
\end{document}